\documentclass{article}
\usepackage{graphicx}
\usepackage{xcolor}
\usepackage{amsfonts}
\usepackage{amssymb}
\def\ligne#1{\hbox to\hsize{#1}}
\def\leurre{\noindent\leftskip0pt\small\baselineskip 10pt}
\newtheorem{thm}{\textbf{Theorem}}
\newtheorem{fig}{\textbf{Figure}}
\newtheorem{tab}{\textbf{Table}}
\newtheorem{lem}{\textbf{Lemma}}

\newtheorem{prop}{\textbf{Proposition}}

\author{Maurice {\sc Margenstern}}
\title{A strongly universal cellular automaton in the dodecagrid with four states}
\begin{document}
\maketitle

\begin{abstract}
In this paper, we prove that there is a strongly universal cellular automaton in the 
dodecagrid, the tessellation $\{5,3,4\}$ of the hyperbolic $3D$ space, with four states 
but, it is not rotation invariant as the automaton of~\cite{mmarXiv21a} is with five 
states. The present paper is not an improvement of~\cite{mmarXiv21a}. The reduction to 
four states of the automaton of the present paper results from a relaxation of the 
condition of rotation invariance. However, the relaxation is not complete: the rules
are invariant with respect to rotations of the dodecahedron which leave it globally
invariant which also leave invariant a couple of opposite faces of the dodecahedron.
As there are twenty five such rotations, it can be considered that the reduction
to four states is significant although the relaxation it implied.
\end{abstract}

\section{Introduction}~\label{intro}

    The present paper is a kind of continuation of~\cite{mmarXiv21a} which also deals with
strong universality of a cellular automaton in the dodecagrid, the tessellation
\hbox{$\{5,3,4\}$} of the hyperbolic $3D$-space. We remind the reader that the tessellation
is based on Poincar\'e's dodecahedron whose faces are pentagons with right-angles and 
which are mutually perpendicular when they share a side. The numbers which occurs in the
signature of the tessellation mean the following: the faces of the dodecahedron 
have five sides, at each of its vertices three faces meet and around each of its sides,
there are four dodecahedrons of the tessellation. We also remind the reader that
strong universality means that we consider computations of the cellular automaton which
start from a finite configuration. That latter condition says that all cells of the space
are in a quiescent state, finitely many of them being excepted. And the quiescent state
is state such that if a cell and all its neighbours are in that state, the new state
of the cell remains that state. We get the strong universality by reduction: we
simulate a two-register machine which is know to be strongly universal, a result
established by Coke-Minsky, see~\cite{minsky}. In the present paper, we repeat several 
pictures of~\cite{mmarXiv21a} and several corresponding explanations in order to 
facilitate the reading of the paper which we try to make as self-contained as possible.

\newcount\compterel\compterel=1
\def\numerrel{\the\compterel\global \advance\compterel by 1}
From now on, we denote by $\Delta$ any dodecahedron we consider in the dodecagrid, unless
it is otherwise indicated.

Figure~\ref{fdodecs} provides us with a representation of~$\Delta$ according to a
Schlegel representation of the solid. That representation could be used to represent 
the dodecagrid too. However, here, we shall use another representation illustrated by 
Figure~\ref{stab_fix0} which we shall again meet later in Section~\ref{scenario}. That 
figure makes use of a property we shall see with the explanations about  
Figure~\ref{fdodecs}. Before turning to that argumentation, we presently deal with the 
Schlegel representation.

\vskip 10pt
\vtop{
\ligne{\hfill
\includegraphics[scale=0.5]{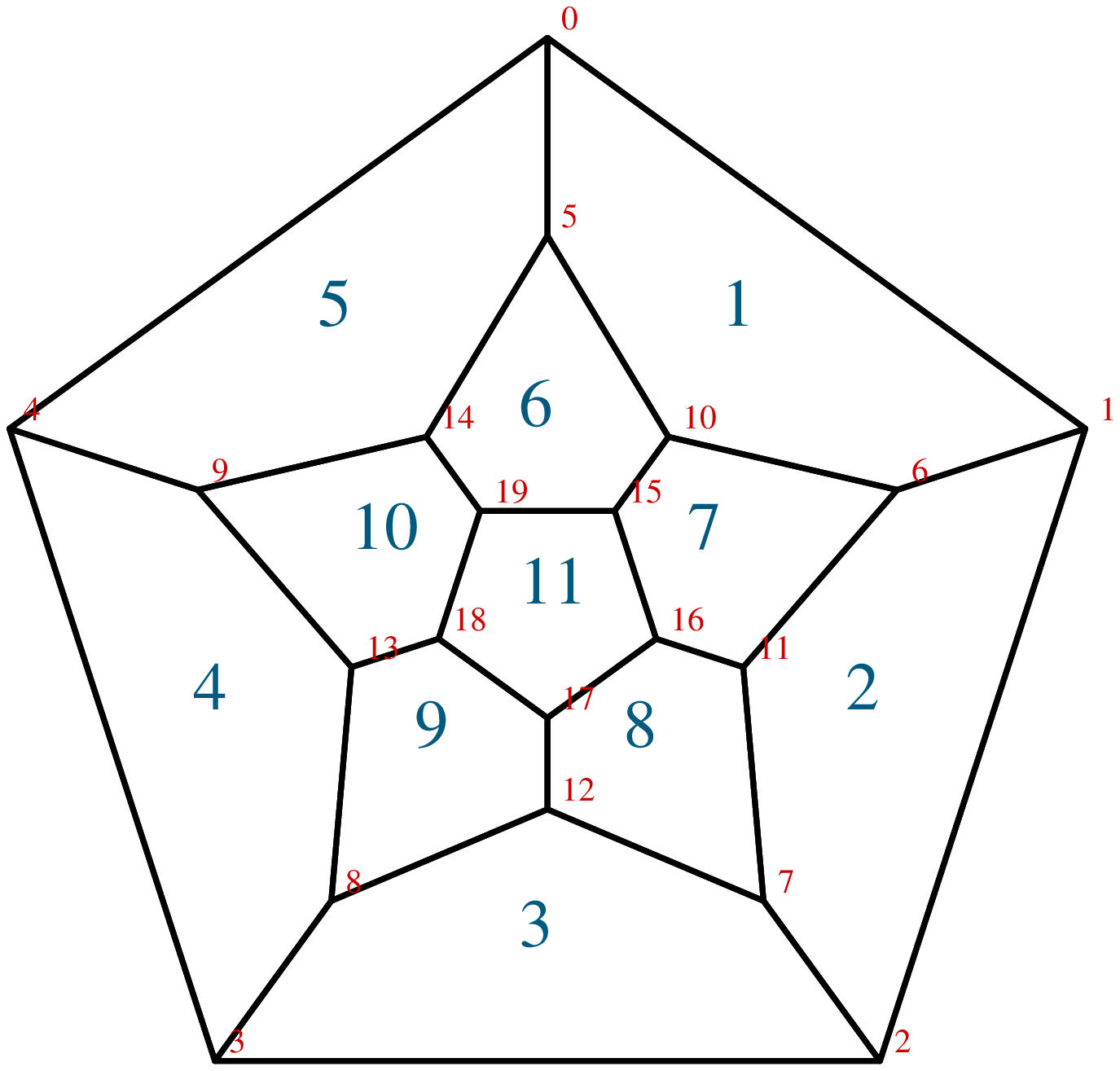}
\hfill}
\vspace{-5pt}
\begin{fig}\label{fdodecs}
\leurre
The Schlegel representation of the dodecahedron~$\Delta$.
\end{fig}
}

   Figure~\ref{fdodecs} is obtained by the projection of the vertices and the edges of
$\Delta$ on the plane of one of its faces say $F_0$. The projection is performed from a
point $P$, called the {\bf centre of projection}, which belongs to the straight line 
passing through the mid-point of~$F_0$ and orthogonal to the plane~$\Pi$ of that face. 
It is usually assumed that $P$ and $\Delta$ are both contained in the half-space 
delimited by $\Pi$ and that $P$ does not belong to $\Delta$.  Note that such 
a projection can be performed in both the Euclidean and the hyperbolic $3D$-space. 

As shown on the figure, we number the faces of~$\Delta$. 
Face~0 is the face which belongs to $\Pi$, the plane on which the projection is 
performed. We number the faces around face~0 by clockwise turning around the face when 
we look at the plane from $P$, the centre of projection. If we consider the face which 
is opposed to face~0, say face~11, we also number the remaining faces of~$\Delta$, they 
are also clockwise numbered from~6 to~10 with face~6 sharing a side with face~5 and 
another one with face~1. That numbering will also be used to number the edges and the
vertices of~$\Delta$ as follows. As far as an edge~$\sigma$ of~$\Delta$ is shared by
two faces~$F_i$ and~$F_j$ and by those faces only, we denote $\sigma$ by $i$-$j$.
Similarly, as a vertex~$v$ of~$\Delta$ belongs to three faces~$F_i$, $F_j$ and~$F_k$ and
to those faces only, we denote~$v$ by $i$-$j$-$k$.

Later, we shall consider the neighbours of a dodecahedron in the dodecagrid. Consider
such a tile, say $\Delta$ again, and assume a numbering of its faces obtained as in 
Figure~\ref{fdodecs}. Let $F_i$ be the face numbered by~$i$, with \hbox{$i\in\{0..11\}$},
we say later on {\bf face~$i$}. We say that $F_i$ and $F_j$, with \hbox{$i\not=j$} are 
{\bf contiguous} if and only if they share an edge. There is a single dodecahedron of 
the dodecagrid which shares $F_i$ with~$\Delta$, we denote that dodecahedron 
by~$\Delta_i$. We say that $\Delta_i$ is a {\bf neighbour} of~$\Delta$, more precisely
its $i$-neighbour. We shall often say that $\Delta$ and~$\Delta_i$ {\bf can see} each 
other through face~$i$ and other expressions connected with vision. Now the face~$i$
of $\Delta$ may receive another number in $\Delta_i$. Often,
we decide that face~$i$ in~$\Delta$ is face~0{} in~$\Delta_i$ and that a face 
of~$\Delta_i$ which is in the same plane as a face~$j$ of~$\Delta$ receives the 
number~$j$ in~$\Delta_i$ too. We shall do that in what follows if not otherwise specified.
The neighbours of~$\Delta$ share an important property: 

\begin{prop}\label{pneighs}
If $F_i$ and $F_j$ are contiguous faces of~$\Delta$, $\Delta_i$ and~$\Delta_j$ cannot 
see each other, but \hbox{$\Delta_{i_j}=\Delta_{j_i}$}.
\end{prop}

The proposition is an easy corollary of the following assertion:

\begin{lem}\label{lneighs}
Two tiles~$T_1$ and~$T_2$ of the dodecagrid which share and edge~$s$ can see each other 
if and only for both faces~$F_a$ and $F_b$ of $T_i$ sharing~$s$, $T_1$ and $T_2$ do not
belong to the same half-space defined by the plane of that face.
\end{lem}

\noindent
Proof of the proposition and of the lemma.\vskip 0pt
By construction, let~$\Pi_a$ be the plane containing~$F_a$. That plane defines a 
half-space $\mathcal S$ which contains~$\Delta$. Clearly, $\Delta_a$ and~$\Delta$
are not both in~$\mathcal S$: otherwise we would have \hbox{$\Delta=\Delta_a$}.
From a similar argument,  $\Delta_b$ and~$\Delta$ are not on the same side with respect
to~$\Pi_b$. By construction, as far as $F_a$ and $F_b$ are contiguous, 
\hbox{$\Pi_a\perp\Pi_b$}.
Now, if \hbox{$\ell=\Pi_a\cap\Pi_b$}, $\Pi_a$ and $\Pi_b$ define four dihedral right 
angles around~$\ell$. By construction, $\Delta$ lies inside one angle, $\Delta_a$ lies
in a second one which is in the same side as $\Delta$ with respect to~$\Pi_b$ and, 
symmetrically, $\Delta_b$ lies in a third angle which is in the same side as $\Delta$
with respect to~$\Pi_a$. The fourth angle may contain a dodecahedron which shares the 
same edge with $\Delta$, $\Delta_a$ and~$\Delta_b$: it is a neighbour of~$\Delta_a$ 
seen from its face~$b$ but also, for the same reason, a neighbour of~$\Delta_b$ seen 
from its face~$a$. Now, by construction of the dodecagrid, there can exactly be four 
dodecahedrons around an edge, so that the neighbour~$b$ 
of~$\Delta_a$ is the neighbour~$a$ of~$\Delta_b$ which can be written
\hbox{$\Delta_{a_b} = \Delta_{b_a}$} as in Proposition~\ref{pneighs}. \hfill$\Box$

In particular, if two faces of tiles~$T_1$ and~$T_2$ sharing and edge~$s$ are in the same
plane~$\Pi$, and if $T_1$ and~$T_2$ are not in the same half-space defined by~$\Pi$
the tiles cannot see each other. {\it A fortiori}, if the tiles have no common element,
they cannot see each other.

\def\HH{$\mathcal H$}
   The cellular automaton we construct in Section~\ref{scenario} evolves in the
hyperbolic $3D$ space but a large part of the construction deals with a single plane
which we shall call the {\bf horizontal plane} denoted by \HH. In fact, both sides
of~\HH{} will be used by the construction and most tiles of our construction have a face
on~\HH.

   We take advantage of that circumstance to define a particular representation which we
find more appropriate to our purpose. 

   The trace of the dodecagrid on~\HH{} is a tiling of the hyperbolic plane, namely the
tiling $\{5,4\}$ we call the {\bf pentagrid}. The left-hand side part of 
Figure~\ref{penta} illustrates the tiling and its right-hand side part illustrates a way 
to locate the cells of the pentagrid. Let us look closer at the figure whose pictures 
live in Poincar\'e's disc, a popular representation of the hyperbolic plane, 
see~\cite{mmbook1}.

\vskip 10pt
\vtop{
\ligne{\hfill
\includegraphics[scale=0.75]{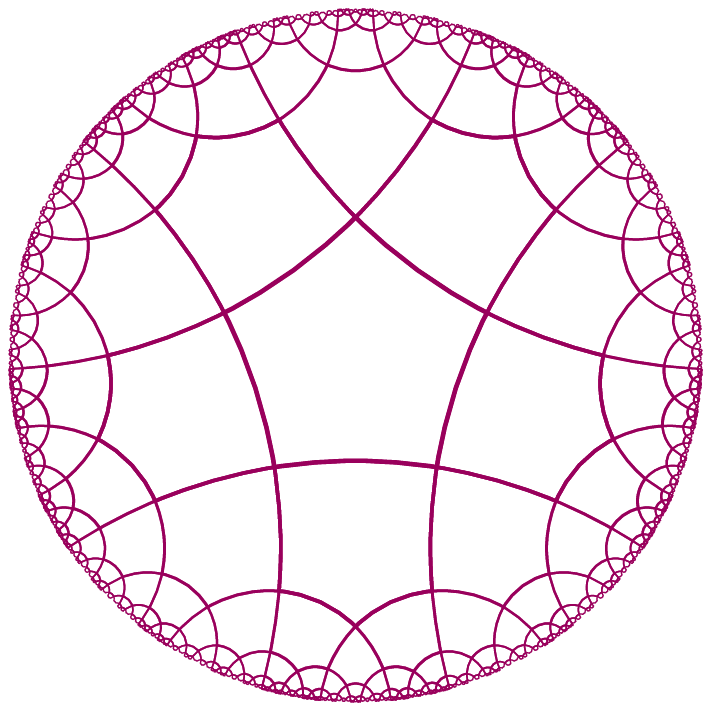}
\raise-20pt\hbox{\includegraphics[scale=0.475]{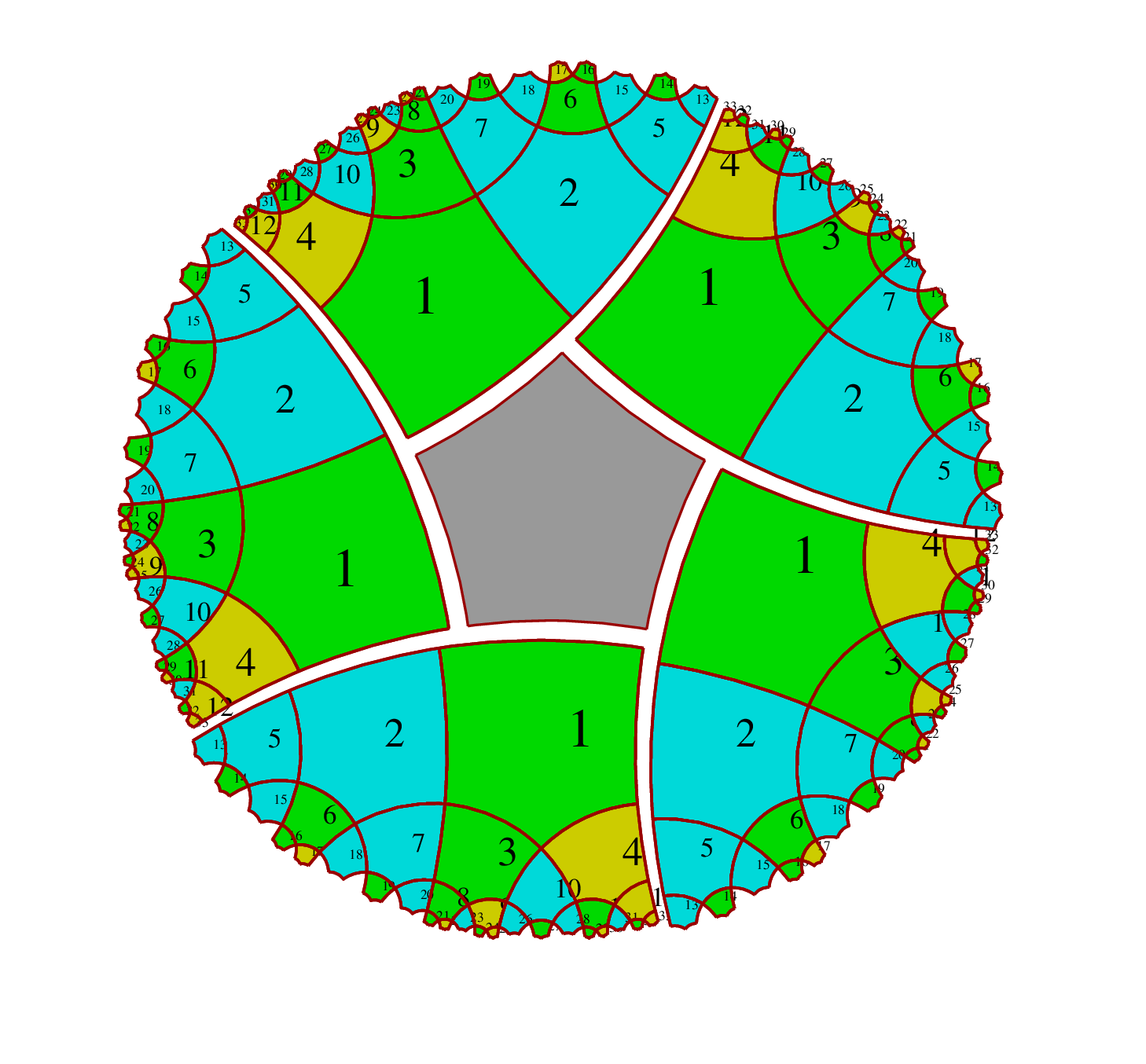}}
\hfill}
\vspace{-5pt}
\begin{fig}\label{penta}
\leurre
To left: a representation of the pentagrid in Poincar\'es disc model of the hyperbolic 
plane. To right: a splitting of the hyperbolic plane into five sectors around a once
and for all fixed central tile.
\end{fig}
}
   In the right-hand side picture, we can see five tiles which are counter-clockwise
numbered from~1 up to~5, those tiles being the neighbours of a tile which we call
the {\bf central tile} for convenience. Indeed, there is no 
central tile in the pentagrid
as there is no central point in the hyperbolic plane. We can see the disc model as a
window over the hyperbolic plane, as if we were flying over that plane in an abstract
spacecraft. The centre of the circle is the point on which are attention is focused while
the circle itself is our horizon. Accordingly, the central tile is the tile which is 
central with respect to the area under our consideration. It is also the reason to
number the central tile by~0.

The right-hand side picture shows us five blocs of tiles we call {\bf sectors}. 
Each sector is defined by a unique tile which shares and edge with the central one.
We number that tile by~1. It is a green tile on the picture. The sector is delimited
by two rays~$u$ and~$v$ issued from a vertex of tile~1: they continue two consecutive 
sides of tile~0.  Those rays are supported by straight lines in the hyperbolic plane 
and they define a right angle. Tile~1 is called the {\bf head} of the sector it defines :
the sector is the set of tiles contained in the angle defined by~$u$ and~$v$.
We also number the sectors from~1 to~5 by counter-clockwise turning around tile~0.
We also say that two tiles are {\bf neighbouring} or that they are {\bf neighbours} of 
each other if and only if they have a common side.

\vskip 10pt
\vtop{
\ligne{\hfill
\includegraphics[scale=1.5]{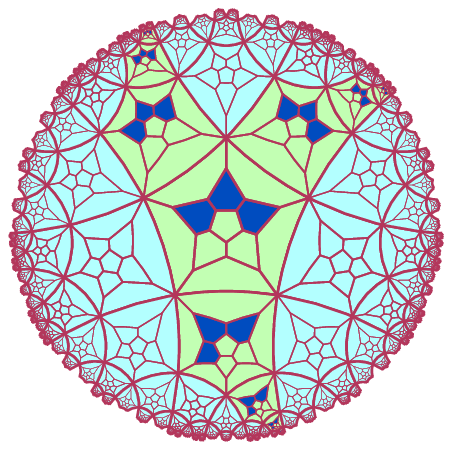}
\hfill}
\vspace{-5pt}
\begin{fig}\label{stab_fix0}
\leurre
A configuration we shall later meet in Section~{\rm\ref{scenario}}. The central tile
is, in some sense, the centre of that configuration. A dark blue face means that a
blue dodecahedron is put on that face.
\end{fig}
}

   Consider the configuration illustrated by Figure~\ref{stab_fix0}. We can see the 
Schlegel projection of a dodecahedron on each tile of the picture. A few dodecahedrons
have another light colour and on each of them, three faces have a dark blue face. We call
them the elements of a {\bf track}.  Tile~0 is the central tile of the picture. 
Around tile~0, we can see three elements of track exactly, the other neighbouring tiles
having a uniform light colour: we call them {\bf blank} tiles. We number the elements of 
tracks by the number of their sectors, here 1, 3 and~5 by counter-clockwise turning around 
tile~0. On the picture, tile~3 is just below tile~0. Let us look closer at tiles~0 and~3
of~\HH. Let $\Delta_0$ and~$\Delta_3$ be the tiles of the dodecagrid which are put on 
tiles~0 and~3 of~\HH{} respectively.  Assume that the face~1 of $\Delta_0$ and 
of~$\Delta_3$ share the side which is shared by the tiles~0 and~3 of~\HH. By construction
of the figure, tiles~0 and~3 are on the same side with respect to~\HH. As a corollary, 
the faces~1 of $\Delta_0$ and~$\Delta_3$ coincide so that,
according to Lemma~\ref{lneighs} and to Proposition~\ref{pneighs}, 
\hbox{$(\Delta_0)_1 = \Delta_3$} and \hbox{$(\Delta_3)_1=\Delta_0$}. 
From the same proposition, we can see that $(\Delta_0)_6$
and $(\Delta_0)_7$ cannot see each other but $(\Delta_0)_6$ and $(\Delta_3)_7$
can see each other and, for a similar reason, $(\Delta_0)_7$ and $(\Delta_3)_6$ can also
see each other.

Accordingly, we have to pay attention to dodecahedrons which are put 
on the faces of a dodecahedron in the representation as defined on Figure~\ref{stab_fix0}:
we call that representation the {\bf \HH-representation}. The rule is simple: a face~$F$ 
of a dodecahedron~$\Delta$ takes the colour of the neighbour that $\Delta$ can see 
through~$F$. The rule also applies to the faces of two neighbouring dodecahedrons whose
face~0 is on~\HH{} through which they see each other. Each face take the colour of its 
neighbour.

By {\it abus de langage}, we also call \HH{} the restriction of the 
dodecagrid to those which sit on that plane. When it will be needed to clarify, we denote 
by \HH$_u$, \HH$_b$ the set of dodecahedrons which are placed upon, below~\HH{} 
respectively.

   Now that the global setting is given, we shall proceed as follows: 
Section~\ref{scenario} indicates the main lines of the implementation which is precisely
described in Subsection~\ref{newrailway}. At last, Section~\ref{srules} gives us the rules
followed by the automaton. That section also contains a few figures which illustrate the 
application of the rules. Those figures were established from pieces of figures drawn by 
a computer program which applied the rules of the automaton to an appropriate window in 
each of the configurations described in Subsection~\ref{newrailway}. The computer 
program also checked that the set of rules is coherent and that rules are pairwise 
independent with respect to the restricted notion of rotation invariance which we define
a little further. As far as that latter property is important for our result, we start 
our study by a short section on the rotations of the dodecahedron, see 
Section~\ref{rotododec}.

   That allows us to prove the following property:

\begin{thm}\label{letheo}
There is a strongly universal cellular automaton in the dodecagrid, truly spatial, 
which has four states and whose rules are invariant under rotations of the dodecahedron 
around a face.
\end{thm}

\section{Rotations of the dodecahedron}\label{rotododec}

We partially repeat a similar section of~\cite{mmarXiv21a}. We do not look at all 
rotations which leave $\Delta$ invariant but at a subset of those rotations only.
It is classically known that there are sixty rotations which leave the dodecahedron
globally invariant: the image can exactly be put on the same place of the space as the 
initial dodecahedron. Using the Schlegel representation, it means that a rotation
which leaves $\Delta$ globally invariant performs a permutation on the number of the 
faces. In~\cite{mmarXiv21a}, we constructed a cellular automaton in the dodecagrid whose
rules are invariant under any rotation of~$\Delta$, we see a bit further what it exactly
means. That cellular automaton has five states and it is strongly universal. In the 
conclusion of~\cite{mmarXiv21a}, I raised the question of improving the result. A possible
improvement consists in reducing the number of states. Up to now, I do not know how to
do that. However, by relaxing a bit the condition of rotation invariance, I could
get that reduction in the number of states.

   The whole set of rotations of~$\Delta$ has sixty elements and the requirement of
invariance of the rules under that set is a huge condition. There are three kinds of
rotations which leave $\Delta$ globally invariant. The interested reader is referred to
\cite{mmarXiv21a} where that question is rather thoroughly dealt with. Let us remind the 
reader that the three kinds of rotations are the following. The identity being excepted
which belongs to all kinds, there are twenty four rotations around a couple of opposite
faces, twenty rotations around a couple of opposite vertices and fifteen rotations 
around a couple of opposite edges.

   Why the restriction to rotations around a couple of opposite faces? That restriction
is based on the main structure of the simulation which we call the tracks and which we 
soon study. A track is a finite sequence of tiles of the dodecagrid, all of them 
belonging to~$\mathcal H$$_u$ or all of them belonging to~$\mathcal H$$_b$ together with
the fact that a side of each tile belongs to a fixed line of~$\mathcal H$ and that
two consecutive elements of the sequence are neighbours of each other. In such a track
the elements have a kind of interdependence. Any rotation around vertices or edges
break that interdependence. Only a rotation around a face is able to keep the 
interdependence. As far as a cell does not know whether its neighbour belongs to a track
or not, such rotation can be around an arbitrary face. Whence the kind of restriction
we consider.

Below, Figure~\ref{frotafaces} illustrates the set of rotations we consider in the
paper.

As already mentioned, there twenty four rotations around a couple of opposite faces which
leave $\Delta$ globally invariant. That number is easy to get: the set of twelve faces 
of~$\Delta$ can be split into six couple of opposite faces, namely:
\vskip 5pt
\ligne{\hfill
$\vcenter{\hbox{
\vtop{\leftskip 0pt\parindent 0pt\hsize=80pt
0\ $-$ 11\vskip 0pt
1\ $-$ 9\vskip 0pt
2\ $-$ 10\vskip 0pt
3\ $-$ 6\vskip 0pt
4\ $-$ 7\vskip 0pt
5\ $-$ 8
}}}$
\hfill(\numerrel)\hskip 10pt}
\vskip 5pt
Each line of~(1) indicates two faces by their number which are opposite to each other
in~$\Delta$. By opposite, we mean that the faces are image of each other under the
symmetry in the central point of~$\Delta$. That property can be checked on 
Figure~\ref{fdodecs}. Consider two such faces, $F_a$ and $F_b$.
Let $C_a$, $C_b$ be the centre of $F_a$, $F_b$ respectively. Then, the centre of~$\Delta$
is the midpoint of~$[C_aC_b]$ and the rotation we consider is a rotation around the
line defined by $C_a$ and $C_b$ which leave those faces globally invariant. It is easy
to see that for each such axis of rotation, there are four possible rotations leaving 
the faces globally invariant. Whence the twenty four rotations which are illustrated
by Figure~\ref{frotafaces}.

As an example, consider the axis through the centres of faces~1 and~9. It can be checked 
that once we fix the image of~$F_0$ and that of $F_1$ under the rotation, the numbering
of all the other faces are fixed. In the line devoted to the rotations around the faces~1
and~9, we have four dodecahedrons which are the image of the first one under the four
possible rotations around that axis leaving $\Delta$ globally invariant. As face~1 is
fixed under the rotation, each face contiguous to $F_1$ in $\Delta$ is put on~$F_0$ by
such a rotation. The concerned faces are 5, 6, 7, 2 and 0, that latter one defining the
identity. If we put $F_5$ onto $F_0$, we necessarily get $F_8$ in place of~$F_{11}$ as
can be checked on the figure. Note that $F_7$ is contiguous to both~$F_1$ and~$F_8$
in $\Delta$ so that it must be the same on the image which tells us that $F_7$ is put
onto $F_6$: it is a way to fix the considered rotation around~$F_1$ and~$F_9$. It can be
checked that the considered rotation induces the following permutation on the numbering
of the faces: \hbox{\tt 5 1 0 4 10 6 7 2 3 9 11 8}, where the position, from~0 to~11
indicates the initial numbering of the faces.
\vskip 10pt
\vtop{
\ligne{\hfill
\includegraphics[scale=0.56]{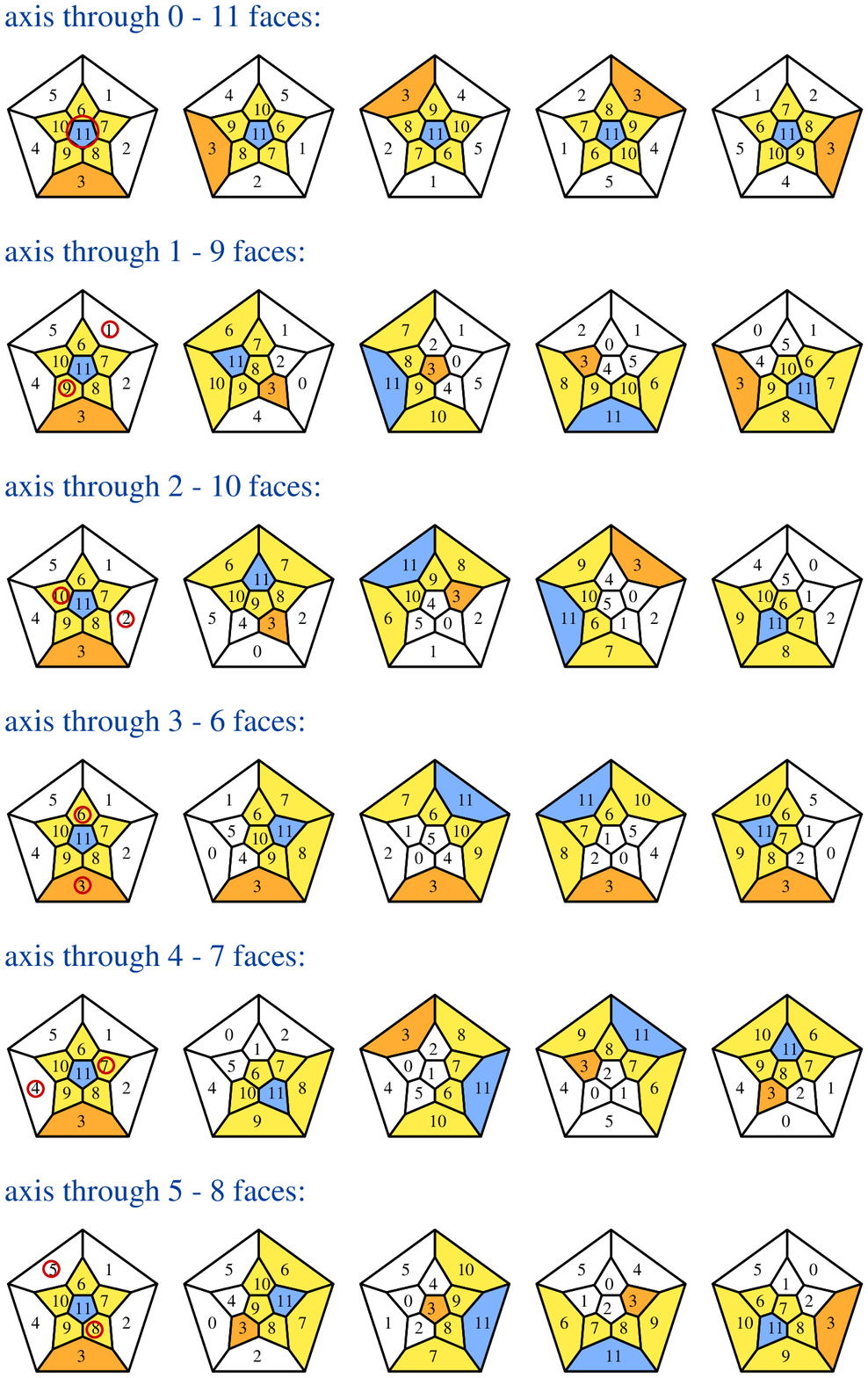}
\hfill}
\vspace{-10pt}
\begin{fig}\label{frotafaces}
\leurre
The twenty four rotations around a couple of opposite faces.
\end{fig}
}
\vskip 10pt
We leave to the reader the computations of the rotations induced by all the rotations
illustrated by Figure~\ref{frotafaces}.

Those rotations can be generated by two rotations around faces which are contiguous.
Figure~\ref{fgrotfaces} indicates how to proceed: rotations around faces~$a$ and~$e$
of the figure produce a rotation around~$f$. Clearly, once we have a rotation around
a couple of opposite faces, we can generate all the other rotations round that axis.
Figure~\ref{fgrotfaces} allows us to gradually obtain all the rotations indicated by 
Figure~\ref{frotafaces}. We also leave it to the reader. In~\cite{mmarXiv21a} we indicate
how to gradually generate the rotations around opposite vertices and around opposite edges
with the help of two rotations around fixed contiguous faces.
  
\vskip 10pt
\vtop{
\ligne{\hskip-5pt
\includegraphics[scale=0.56]{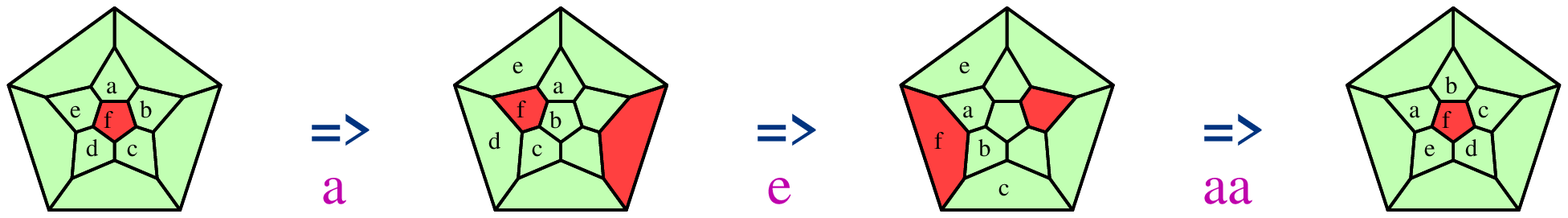}
\hfill}
\vspace{-10pt}
\begin{fig}\label{fgrotfaces}
\leurre
How to generate the rotations around opposite faces thanks to two fixed rotations 
around contiguous faces.
The letter under an arrow indicates the face around which the rotation is performed.
\end{fig}
}
\vskip 10pt 
  In order to prepare the construction of the rules explained in Section~\ref{srules}, 
we need to define a simple criterion in order to see the neighbourhood of a tile.

\vskip 10pt
\vtop{
\ligne{\hfill
\includegraphics[scale=1]{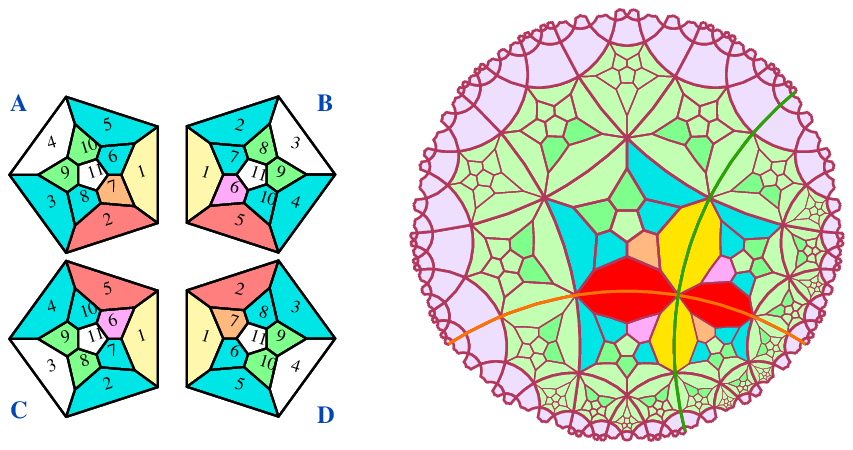}
\hfill}
\vspace{-5pt}
\begin{fig}\label{fles4}
\leurre
Correspondence between neighbours of neighbouring tiles.
\end{fig}
}
\vskip 5pt
Figure~\ref{fles4} helps us to establish that criterion. In the right-hand side part 
of the figure, we have the representation of eight tiles which share a common vertex.
The figure shows us four of them only, those which belong to \HH$_u$. Those which belong
to \HH$_b$ are not visible.
Two lines~$\ell$ and~$m$, in green and orange respectively on the right-hand 
side part of the figure, bear sides of those tiles. We fix a numbering of the sides of 
the tiles which is indicated in the left-hand side part of the figure. Our four tiles 
have their face~0 on \HH{} and in all of them, face~0 and face~1 have their common side 
on~$\ell$, the green line. The faces of a tile which are contiguous with face~0 
constitute the {\bf lower crown} while those which are contiguous with face~11 constitute the {\bf upper crown}.
It can be seen that faces~$i$ with \hbox{$i\in\{1..5\}$} belong to the lower crown
while faces~$j$ with \hbox{$j\in\{6..10\}$} constitute the upper crown.

\def\VV{\hbox{$\mathcal V$}}
Denote by {\bf A}, {\bf B}, {\bf C} and {\bf D} as indicated on the left-hand side part 
of the figure the four tiles which have an edge on both~$\ell$ and~$m$. Let~\VV{} be the 
plane which is orthogonal to~\HH, cutting it along~$m$, the orange line. Say that those 
tiles belong 
to generation~0 and denote them by {\bf L} with 
\hbox{{\bf L $\in$ $\{$A$,$ B$,$ C$,$ D$\}$}}. The neighbours of those tiles belong to 
generation~1. We denote them as {\bf L}$_i$ where $i$ is in \hbox{$\{0..11\}$},
indicating a face on which is lying the neighbour. As far as the face of {\bf A}$_6$, 
{\bf B}$_7$ on~\VV{} is orthogonal to the face~1 of {\bf A}, {\bf B} respectively,
{\bf A}$_6$ and {\bf B}$_7$ share a face on that plane so that they can see each other
according to Proposition~\ref{pneighs}.
That property is illustrated on the left-hand side part of Figure~\ref{fles4} by the fact
that the face~6 of~{\bf A} and the face~7 of~{\bf B} are both in blue. Similarly,
{\bf A}$_7$ and {\bf B}$_6$ can see each other. Denote that relationship by
\hbox{\bf A$_7$ $\Bumpeq$ B$_6$}. Here are the similar relationships which can be 
established :
\vskip 5pt
\ligne{\hfill
$\vcenter{\hbox{\vtop{\leftskip 0pt\parindent 0pt\hsize=80pt
\ligne{{\bf A$_6$ $\Bumpeq$ B$_7$}\hfill}
\ligne{{\bf A$_7$ $\Bumpeq$ B$_6$}\hfill}
\ligne{{\bf C$_6$ $\Bumpeq$ D$_7$}\hfill}
\ligne{{\bf C$_7$ $\Bumpeq$ D$_6$}\hfill}
}
\hskip 30pt
\vtop{\leftskip 0pt\parindent 0pt\hsize=80pt
\ligne{{\bf A$_7$ $\Bumpeq$ C$_6$}\hfill}
\ligne{{\bf A$_8$ $\Bumpeq$ C$_{10}$}\hfill}
\ligne{{\bf B$_6$ $\Bumpeq$ D$_7$}\hfill}
\ligne{{\bf B$_{10}$ $\Bumpeq$ D$_8$}\hfill}
}}}$
\hfill(\numerrel)\hskip 20pt}
\vskip 5pt
We can see that in (2) several neighbours can see two neighbours. Indeed, {\bf A} can see
both {\bf B} and {\bf C} but it cannot see {\bf D} as far as {\bf C} and {\bf D} are not
on the same half-space with respect to~\VV. From (2), we can see that {\bf A}$_7$ can see
both {\bf B}$_6$ and {\bf C}$_6$. Let~$s$ be the side which is shared by the faces~1 of
{\bf A}, {\bf B} {\bf C} and {\bf D}. Those tiles are the four tiles around~$s$. Now, if 
$n$ is the line which bears~$s$, we can see that {\bf A}$_7$, {\bf B}$_6$, {\bf C}$_6$
and {\bf D}$_7$ also share a common side which lies on~$n$. 

We defined {\bf L}$_i$ where {\bf L} stands for {\bf A}, {\bf B}, {\bf C} or {\bf D} as
the neighbours of generation~1 for those tiles. There is room for another generation:
indeed, as far as, already noticed, {\bf L}$_i$ and {\bf L}$_j$ do not see each other 
when $i\not=j$. But if {\bf L}$_i$ and {\bf L}$_j$ with $i\not=j$ share an edge~$\sigma$, 
they have a common neighbour~{\bf N}$_{i,j}$: around ~$\sigma$ there must be four tiles:
{\bf L}, {\bf L}$_i$, {\bf L}$_j$ and {\bf N}$_{i,j}$. Consider, for instance, 
{\bf L}$_{11}$, {\bf L}$_6$ and {\bf L}$_7$. We can define {\bf N}$_{11,6}$, 
{\bf N}$_{11,7}$ and {\bf N}$_6,7$. Consider the planes $\Pi_6$, $\Pi_7$ and $\Pi_{11}$
containing the faces~6, 7 and~11 of {\bf L} respectively. For each one, say $\Pi_i$,
{\bf L}$_i$ is in one half-space defined by $\Pi_i$ while {\bf L}$_j$ and {\bf L}$_k$
with \hbox{$\{i,j,k\}=\{6,7,1\}$} are in the other half-space. The intersections of those 
half-spaces define eight regions of the space which we call {\bf octants}. Each octant 
contains a tile touching {\bf L} as far as that situation is the same as the one we can 
see with planes giving rise to $\ell$ and $m$ together with the intersections of the 
half-spaces that they define. We already know seven tiles for seven octants: {\bf L},
{\bf L}$_i$ with $i$ in $\{6,7,11\}$, {\bf N}$_{6,7}$, {\bf N}$_{6,11}$ and 
{\bf N}$_{7,11}$. There is an eighth tile: a common neighbour to those last three tiles,
{\bf N}$_{6,7,11}$. That last tile can see all {\bf N}$_i$'s we have defined, but it
cannot see neither the {\bf L}$_i$'s nor {\bf L} itself. We say that {\bf N}$_{i,j,k}$
with, here, \hbox{$\{i,j,k\}=\{6,7,11\}$} belongs to generation~3.

Now, there is a simple criterion to distinguish those generations: the tiles of
generation~1 have a common face with {\bf L}; those of generation~2 share just a single
edge with {\bf L} while those of generation~3 share just a vertex with~{\bf L}. We have
seen that criterion for the upper crown of~{\bf L}. For the lower crown, things are a bit
different as far as $L$ has five neighbours belonging to~$\mathcal H$$_u$ and one of them
belonging to~$\mathcal H$$_b$: generation~1 is still defined by the tiles sharing a face 
with~{\bf L} and generation~2 by the tiles which share with~{\bf L} one edge only. 
Taking into account the tiles which are in~\HH$_b$, we can see that we have all the 
possible neighbours of~{\bf L}. Indeed, consider the vertex~$v$ belonging to two faces of
the lower crown and to a third one of the upper crown. Let $e$ be the edge joining~$v$ to 
a vertex of the face~0 of~{\bf L}. Then the three other tiles sharing with {\bf L} the 
edge $e$ only share also~$v$. But, as can be shown on the figure, on each tile, a 
neighbour of a face of the upper crown of a neighbour of~$L$ also shares~$v$. So that
the tile of generation~3 with respect to~$L$ at such a vertex is a tile of generation~2
with respect to a neighbour of~$L$ of generation~1. At last, consider a vertex~$w$ 
belonging to one face~$a$ of the lower crown and to two faces of the upper crown. Then 
{\bf L} with its neighbours {\bf L}$_i$ of generation~1 also sharing~$w$ defines three 
tiles sharing~$w$. The neighbour~{\bf N} of generation~2
defined by those {\bf L}$_i$'s is the fourth tile around the edge~$f$ shared by the two
faces of the upper crown defining~$w$. A similar argument can be formulated for the 
tile~{\bf K} sharing the face~$a$ with~{\bf L} so that we get the eight tiles around~$w$.
So for such a vertex too, the tiles of generation~3 belong to the generation~2 of a
neighbour of~$L$ of generation~1. A similar argument holds for a vertex belonging to
the face~0 of~$L$: its neighbour of generation~$3$ for~$L$ is a neighbour of 
generation~$2$ for the neighbour of~$L$ belonging to~$\mathcal H$$_b$.
Accordingly we proved:

\begin{prop}\label{pneighgen}
The neighbours of a tile~{\bf L} whose face~$0$ is on~\HH{} belong to generation~$1$.
The neighbours of the neighbours which share an edge only with~{\bf L} belong
to generation~$2$. There are neighbours of those of generation~$2$ which shares a vertex
only with~{\bf L}: they are the neighbours of generation~$3$. The tiles of generation~$3$
around~{\bf L} touch vertices of~{\bf L} which belong to the upper crown. The tiles of 
generation~$3$ of~$L$ touching it at its lower crown are neighbours of generation~$2$ 
for a neighbour of~$L$ belonging to~$\mathcal H$.
\end{prop}

At last but not the least, we have to see the connection of neighbours of a tile with
the distance from a tile to another one. Say that a sequence of 
tiles~$\{T_i\}_{i\in\{0..n\}}$ is a {\bf path} if $T_i$ and $T_{i+1}$ share a face for 
$0\leq i<n$. In that situation, we say that $n$ is the {\bf length} of the path. We also
say that the path {\bf joins}~$T_0$ to $T_n$ and conversely. The {\bf distance}
between two tiles~$U$ and $V$ is the length of the shortest path joining~$U$ to~$V$.
That distance is~0 if and only if $U=V$. A {\bf ball} around $T$ of radius~$k$ in the 
space is the set of tiles whose distance to~$T$ is at most~$k$. Those which are at 
distance~$k$ exactly constitute the {\bf sphere} around~$T$ of radius~$k$. Note that 
two consecutive tiles on a path are neighbours of generation~1 for one another according 
to the definition of a path. 

\begin{lem}\label{ldistneigh}
Let $T$ be a tile at distance~$k$ from a tile~$A$, with $k>0$. Let $U$ be the tile on a 
shortest path from $A$ to~$T$ which is at the distance~$k$$-$1 from~$A$. Define a 
numbering of $T$ under which the face~$0$ of $T$ is the face it shares with~$U$. Then all 
neighbours $T_i$ of $T$ with $i>0$ are at the distance~$k$+1 from~$A$. Let $\Pi_i$ be 
the plane supporting a face~$i$ of $T$ belonging to its upper crown. Then a shortest 
path from~$A$ to $T$ is completely contained in the half-space defined by~$\Pi_i$ which 
contains~$T$.
\end{lem}

\noindent
Proof of the lemma. We proceed by a complete induction on the length of the path.
As the dodecahedron is a convex set, the lemma is true for~$k=0$. Assume that it is
true for~$k$. Let $T$ be a tile at the distance~$k$+1 from~$A$ and let $U$
be a tile on a shortest path from~$A$ to~$T$ which is at distance~$k$. Define the face
$F_0$ shared by~$T$ and~$U$ as the face~0 of~$T$. Let $\Pi_0$ be the plane supporting 
$F_0$. That half-space defined by $\Pi_0$ which contains~$U$ also contains the path 
from $A$ to~$U$ by the induction hypothesis. Let $F_j$ be a face of the upper crown 
of~$T$ and let $\Pi_j$ be the plane supporting it. If $F_j$ is opposed to~$F_0$, the 
line joining the centre of~$F_0$ to that of $F_j$ is a common perpendicular to $\Pi_0$ 
and to~$\Pi_j$. Accordingly, those planes are not secant, so that the half-space 
of~$\Pi_j$ containing~$T$ also contains the half-space of~$\Pi_0$ containing~$U$. 
Consequently, that half-space also completely contains the path from~$T_0$ to~$U$. 
If $F_j$ is not opposed to~$F_0$ there is a unique edge~$s$ joining $F_0$ to~$F_j$. As 
$s$ is the intersection of two faces of the lower crown which are, by construction, 
mutually perpendicular and both perpendicular to~$F_0$, $s$ is the common perpendicular 
line to both~$F_0$ and $F_j$ so that $\Pi_0$ and $\Pi_j$ are non-secant. 
Accordingly me may repeat the previous argument which allows us to conclude that the 
half-space defined 
by $\Pi_j$ which contains~$T$ also contains the whole path from~$A$ to~$T$.

Let~$V$ be a neighbour $T_i$ of $T$ with $i>0$. Assume that $V$ is at distance~$h$ 
from~$A$ with $h\leq k$. Consider a shortest path from~$A$ to~$T_i$. As that path and the 
shortest path we considered from~$A$ to~$T$ start both of them from~$A$, there is a 
tile~$B$ on both paths such that the part of those paths from~$A$ to~$B$ is the longest 
common part of the paths from~$A$ to~$T$ and from~$A$ to~$V$. Let~$\{X_i\}$, $\{Y_i\}$ 
constitute the path from $B$ to~$T$, to~$V$ respectively with $X_0$ and $Y_0$ being 
neighbours of~$B$. We know that $X_0$ and~$Y_0$ cannot see each other. Consider $X_1$ 
and~$Y_1$. We claim that if $X_1\not=Y_1$ those tiles cannot see each other which is easy to check on Figure~\ref{fles4}. Accordingly, unless there are few tiles $X_a$, $X_{a+1}$, 
$X_{a+2}$ and $Y_a$, $Y_{a+1}$, $Y_{a+1}$ such that $X_a=Y_a$, $X_{a+1}\not=Y_{a+1}$
and $X_{a+2}=Y_{a+2}$ and as far as $T\not=V$, the conclusion is that $T$ and $V$ 
cannot see each other which is a contradiction with our assumption. That proves the 
lemma. \hfill$\Box$

Note that from the proof of the lemma, we can see that the shortest path from a tile
to another one may not be unique.

The lemma allows us to better see the relation between spheres $\mathcal S$$_k$ of 
radius~$k$ around the same tile. We can see that $\mathcal S$$_{k+2}$ contains 
neighbours of generation~2 of tiles belonging to $\mathcal S$$_k$ and that
$\mathcal S$$_{k+3}$ contains neighbours of generation~3 of tiles belonging 
to~$\mathcal S$$_k$. Accordingly, only spheres $\mathcal S$$_{k+h}$ with $h>3$ have
no contact with $\mathcal S$$_k$.

\section{Main lines of the computation}\label{scenario}

   The first paper about a universal cellular automaton in the pentagrid, the 
tessellation $\{5,4\}$ of the hyperbolic plane, was \cite{fhmmTCS}. That cellular 
automaton was also rotation invariant and, at each step of the computation, the set of 
non quiescent states had infinitely many cycles: we shall say that it is a truly planar 
cellular automaton. That automaton had 22~states. That result was improved by a cellular 
automaton with 9~states in~\cite{mmysPPL}. Recently, it was improved with 5~states, 
see~\cite{mmpenta5st}. A bit later, I proved that in the heptagrid, the tessellation 
$\{7,3\}$ of the hyperbolic plane, there is a weakly universal cellular automaton with 
three states which is rotation invariant and which is truly planar, \cite{mmhepta3st}. 
Later, I improved the result down to two states but the rules are no more rotation 
invariant, see~\cite{mmpaper2st}. Paper \cite{JAC2010} constructs three cellular
automata which are strongly universal and rotation invariant: one in the pentagrid, one 
in the heptagrid, one in the dodecagrid. By strongly universal we mean that the initial 
configuration is finite, {\it i.e.} it lies within a large enough circle or sphere. 
Recently, I succeeded to implement a strongly universal cellular automata in the 
heptagrid, the tessellation $\{7,3\}$ of the hyperbolic plane, which improves that 
latter result: the automaton is rotation invariant and it has seven states, 
see~\cite{mmJAC2021}.

    In the present paper, we tried to follow the construction of~\cite{mmarXiv21a}.
The construction and the rules are the same as in that paper for the tracks and for the
switches but for the convenience of the reader we reproduce it here. But the present
paper clearly differs from~\cite{mmarXiv21a} for what is the implementation of the
operations on a register.

As in previous papers, we simulate a register machine, not necessarily using
the property that two registers are enough to get the strong universality, a result
proved by Coke-Minsky in the sixties, see~\cite{minsky}. 

    The simulation is based on the railway model devised in~\cite{stewart} revisited
by the implementations given in the author's papers, see for instance~\cite{mmpaper2st}.
Sub-section~\ref{railway} describes the main structures of the model. We mainly borrow
its content from previous papers for the sake of the reader. In 
Sub-section~\ref{newrailway} we indicate the new features used in the present simulation.

\subsection{The railway model}\label{railway}

   The railway model of~\cite{stewart} lives in the Euclidean plane. It consists of
{\bf tracks} and {\bf switches} and the configuration of all switches at time~$t$
defines the configuration of the computation at that time. There are three kinds of
switches, illustrated by Figure~\ref{switches}. The changes of the switch configurations
are performed by a locomotive which runs over the circuit defined by the tracks and their
connections organised by the switches.

A switch gathers three tracks $a$, $b$ and~$c$ at a point. In an active crossing,
the locomotive goes from~$a$ to either~$b$ or~$c$. In a passive crossing, it goes
either from~$b$ or~$c$ to~$a$. 

\vskip 10pt
\vtop{
\ligne{\hfill
\includegraphics[scale=0.8]{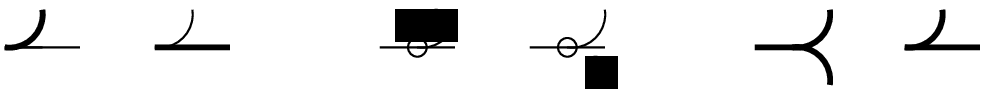}
\hfill}
\begin{fig}\label{switches}
\leurre
The switches used in the railway circuit of the model. To left, the fixed switch, in the
middle, the flip-flop switch, to right the memory switch. In the flip-flop switch, the 
bullet indicates which track has to be taken.
\end{fig}
}

In the fixed switch, the locomotive always goes from~$a$ to~$b$ or always from~$a$ to~$c$.
The passive crossing of the fixed switch is possible. The flip-flop switch is always 
crossed actively only. If the locomotive is sent from~$a$ to~$b$, $c$ by the switch, it 
will be sent to~$c$, $b$ respectively at the next passage. The memory switch can be 
crossed actively or passively. Now, the track taken by the locomotive in an active 
passage is the track taken by the locomotive in the last passive crossing. Of course, at 
the initial time of the computation, for the flip-flop switch and for the memory one, 
the track which will be followed by the locomotive is defined by the implementation.

   As an example, we give here the circuit which stores a one-bit unit of information,
see Figure~\ref{basicelem}. The locomotive may enter the circuit either through the 
gate~$R$ or through the gate~$W$.

\vtop{
\ligne{\hfill
\includegraphics[scale=0.6]{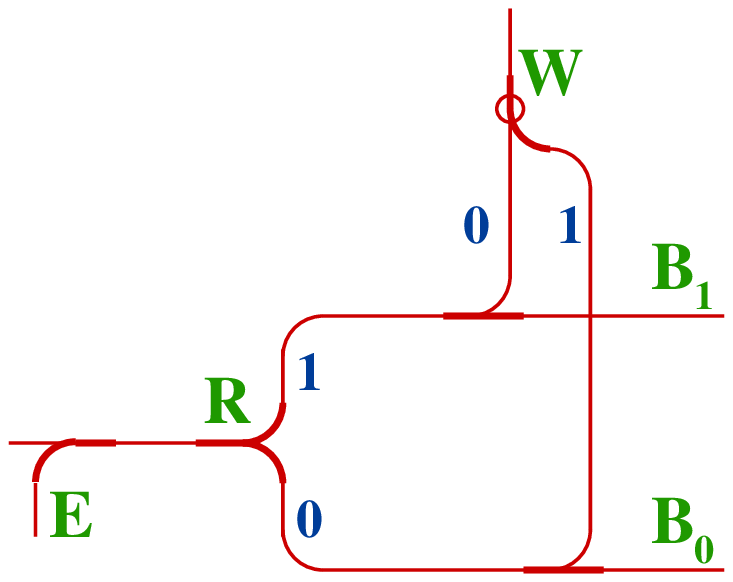}
\hfill}
\begin{fig}\label{basicelem}
\leurre
The basic element containing one bit of information.
\end{fig}
}

  If it enters through the gate~$R$ where a memory switch sits, it goes either through
the track marked with~1 or through the track marked with~0. When it crossed the switch
through track~1, 0, it leaves the unit through the gate~$B_1$, $B_0$ respectively.
Note that on both ways, there are fixed switch sending the locomotive to the appropriate
gate~$B_i$. If the locomotive enters the unit through the gate~$W$, it is sent to the 
gate~$R$, either through track~0 or track~1 from~$W$. Accordingly, the locomotive
arrives to~$R$ where it crosses the switch passively, leaving the unit through the 
gate~$E$ thanks to a fixed switch leading to that latter gate. When the locomotive 
took track~0, 1 from~$W$, the switch after that indicates track~1, 0 respectively and the 
locomotive arrives at~$R$ through track~1, 0 of~$R$. The tracks are numbered according to 
the value stored in the unit. By definition, the unit is~0, 1 when both tracks from~$W$ 
and from~$R$ are~0, 1 respectively. So that, as seen from that study, the entry 
through~$R$ performs a reading of the unit while the entry through~$W$, changes the unit
from~0 to~1 or from~1 to~0: the entry through~$W$ should be used when it is needed to
change the content of the unit and only in that case. The structure works like a memory
which can be read or rewritten. It is the reason to call it the {\bf one-bit memory}.

   We shall see how to combine one-bit memories in the next sub-section as far as we 
introduce several changes to the original setting for the reasons we indicate there.

\subsection{Tuning the railway model}\label{newrailway}

   We start our presentation with a look on the global aspect of the simulation 
illustrated by Figure~\ref{fglobconfig}. On a part of~\HH, we have the program: a green
quadrangle. It contains as many tracks as there are instructions. The return path used by 
the locomotive goes to the next instruction when it incremented a register or when 
it succeeded to decrement it. In the case when it could not decrement the register, the 
return path is different as far as it need going to the appropriate instruction. 

\vskip 10pt
\vtop{
\ligne{\hfill
\includegraphics[scale=0.4]{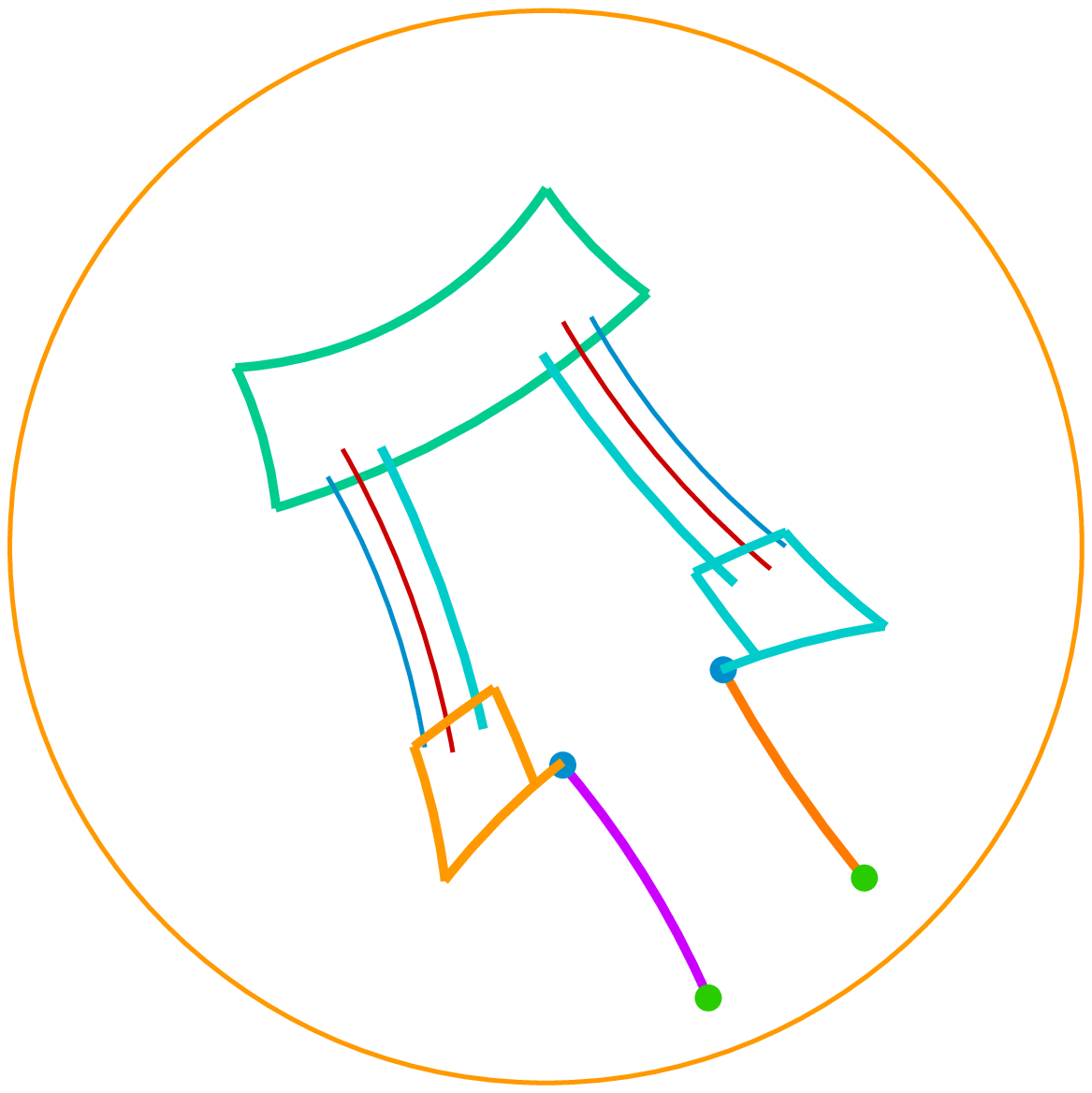}
\hfill}
\begin{fig}\label{fglobconfig}
\leurre
A global view on the simulation.
\end{fig}
}

On another part of the figure, we have two boxes and, attached to each one of them, 
a long segment of line which represents a register. As two registers are enough to 
simulate a Turing machine, see~\cite{minsky}, our illustration contains two registers 
only. At last, and not at all the least, segments of line go in between the program and 
the registers. We can imagine that some of them go in that direction and that the others 
go from the registers to the program. Those segments represent {\bf paths} we study
in the next subsection. They play a key role as far as they convey an information to
the register and go back to the program in order to give a feed back which will determine
the next operation to perform on the registers.

   We first look at the implementation of the tracks in Sub-subsection~\ref{sbbtracks}
and how it is possible to define the crossing of two tracks. In 
Sub-subsection~\ref{sbbswitch} we see how the switches are implemented. Then, in 
Sub-subsection~\ref{sbbunit}, we see how the one-bit memory is implemented in the new 
context and then, in Sub-section~\ref{sbbregdisp}, how we use it in various places. 
At last but not the least, we shall indicate how registers are implemented
in Sub-subsection~\ref{sbbreg}.

\subsubsection{Paths and tracks}\label{sbbtracks}

    We mentioned the key role played by the paths in the computation. Without them,
the locomotive could not perform any operation on the registers. Without them any 
computation is impossible.  Moreover, as can be seen in many papers of the author, that 
one included, it is not an obvious issue which must always be addressed.
  
     In the present paper, 
 we explain the implementation which allowed us to prove
Theorem~\ref{letheo}. The present implementation significantly simplifies what is written 
in~\cite{mmJAC2021}.

   The tracks are one-way. It is useful to reduce the number of states but it raises
a problem as far as a two-way circulation is required in some portions of the circuit.
It is a point where the third dimension comes to help us. In many portions, the
circuit can be implemented on a fixed plane we call \HH, already mentioned in the 
introduction. Roughly speaking, the traffic in one direction will occur in
\HH$_u$ while the reverse running will be performed in \HH$_b$. Occasionally and 
locally, we shall use a plane \VV{} which is orthogonal to~\HH. Note that if \HH{} is 
fixed in all the paper, \VV{} may change according to the context where it is mentioned.

\def\ftt #1 {{\footnotesize\tt#1}}
\def\hhzz{\hskip-0.5pt}
\def\sww{{\ftt W }\hhzz }
\def\sbb{{\ftt B }\hhzz }
\def\srr{{\ftt R }\hhzz }
\def\sgg{{\ftt G }\hhzz }
\def\syy{{\ftt Y }\hhzz }
    A register consists of four sequences of tiles following a line~$\ell$ on~\HH.
In the present paper, the tiles through which the locomotive is running are marked with
what we call {\bf milestones}. All tiles of the dodecagrid are blank, denoted by \sww,
finitely many of them being excepted. Among the non-blank tiles, we have the milestones 
which are blue, denoted by \sbb. Later we say \sbb-, \sww-tile or -cell respectively. 
The blank is the quiescent state of our cellular automaton: if a cell is blank and if 
all its neighbours are blank too, the cell remains blank. We say {\bf tracks} for 
portions of a path which follow a line~$\ell$ defined by a side of the tiles in~\HH. 
Combining several tracks gives rise to paths. On the path, a blue locomotive is running 
which consists of a single blue cell which moves in between blue milestones. 
Figure~\ref{ftracks} illustrates the constitution of the tracks. On the figure, the 
lines are followed by the side~1 of the elements of tracks.

The leftmost picture of the figure shows us that an element of a track consists
of a white dodecahedron $\Delta$ whose face~0 lie on~\HH{} and on the faces~6, 9 and~10
of~$\Delta$ we have three dodecahedrons $\Delta_6$, $\Delta_9$ and $\Delta_{10}$ which
are blue. Those dodecahedrons, close to the face opposite to face~11, opposite to face~0, 
are called the {\bf decoration} of~$\Delta$. In general, the locomotive enters an element 
of the track through its face~5 or~4 and it leaves the element through its face~2. In 
order to allow variations which we further describe, for instance in Figure~\ref{fpaths}, 
the locomotive may also enter through face~1, the exit face being always through face~2.

\vskip 10pt
\vtop{
\ligne{\hfill
\raise 20pt\hbox{\includegraphics[scale=0.4]{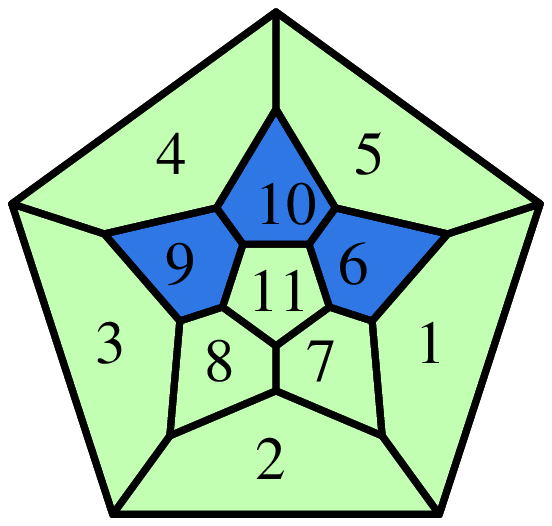}}
\includegraphics[scale=0.8]{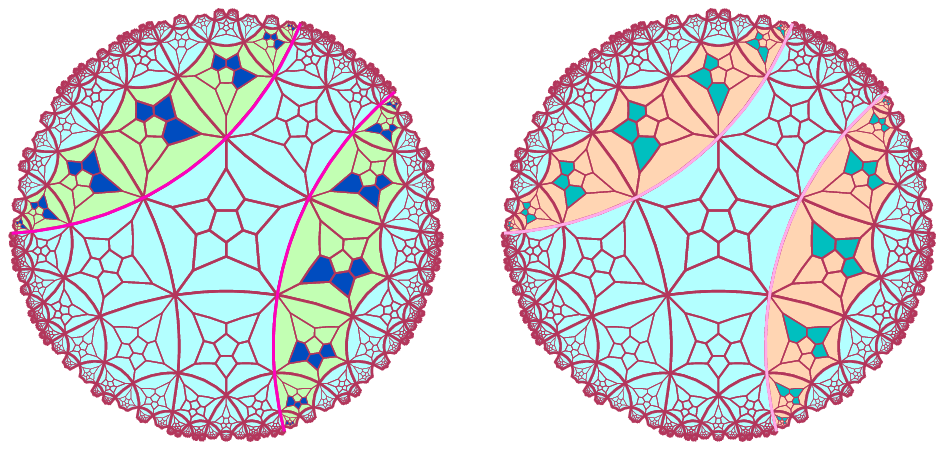}
\hfill}
\begin{fig}\label{ftracks}
\leurre
Leftmost picture: an element of a track. Middle picture: tracks in \HH$_u$.
Rightmost picture, return tracks of the previous ones in \HH$_b$. Note the lines on both
picture, in the middle and to right.
\end{fig}
}

The middle picture of Figure~\ref{ftracks} shows us two tracks which belong to \HH$_u$. 
As clearly seen, the left-hand side track follows a line of~\HH, in mauve in the 
pictures, which contains an edge of the face~1 of all the elements constituting that 
track. A similar remark can be formulated for the right-hand side track. According to the 
convention we indicated on the way a locomotive may cross an element of a track, the 
left-hand side track is going down while the right-hand side one is going up. The 
rightmost picture of the figure shows us tracks which belong to~\HH$_b$, their face~0 
also belonging to~\HH. Imagine that the tracks of the middle picture are removed and 
that we see those tracks of~\HH$_b$ from above~\HH{} as if that plane were translucent. 
We can see that the image we need is an image of~$\Delta$ under a symmetry in the plane 
orthogonal to~\HH{} which contains the opposite edges 1.5 and 8.9 of~$\Delta$. The 
numbering of the faces is thus increasing while counter-clockwise turning around face~0 
while the numbering is clockwise on Figure~\ref{fdodecs}. Note that in~\HH$_b$, the 
direction of the tracks is opposite to that of the tracks of~\HH$_u$ which follow the 
same lines. It is an important feature which allows us to define a two-way path: one 
direction is performed in~\HH$_u$ while the opposite one occurs in~\HH$_b$. Remember
that in both cases, a face of an element of the track is on~\HH, most often face~0.

Moreover, in most parts of the circuit we shall describe a single locomotive is running 
on the circuit. In particular, in the case when a two-way track occurs in some part of 
the circuit, there can never be a locomotive in a track~$\tau$ in~\HH$_u$ and another 
one, at the same time, in the track in~\HH$_b$, below $\tau$.

Figure~\ref{fpaths} illustrates how pieces of tracks indicated in Figure~\ref{ftracks}
can be organised into paths. As long as it will be possible, paths for the locomotive
will follow a straight line of the tiling which is the trace of the dodecagrid on~\HH.
However, as the locomotive goes from the program to the register and back, such a circuit
is organised along a kind of quadrangle $Q$ in the hyperbolic plane as illustrated by 
the top picture of Figure~\ref{fpaths}.

\vskip 10pt
\vtop{
\ligne{\hfill
\raise 5pt\hbox{\includegraphics[scale=0.365]{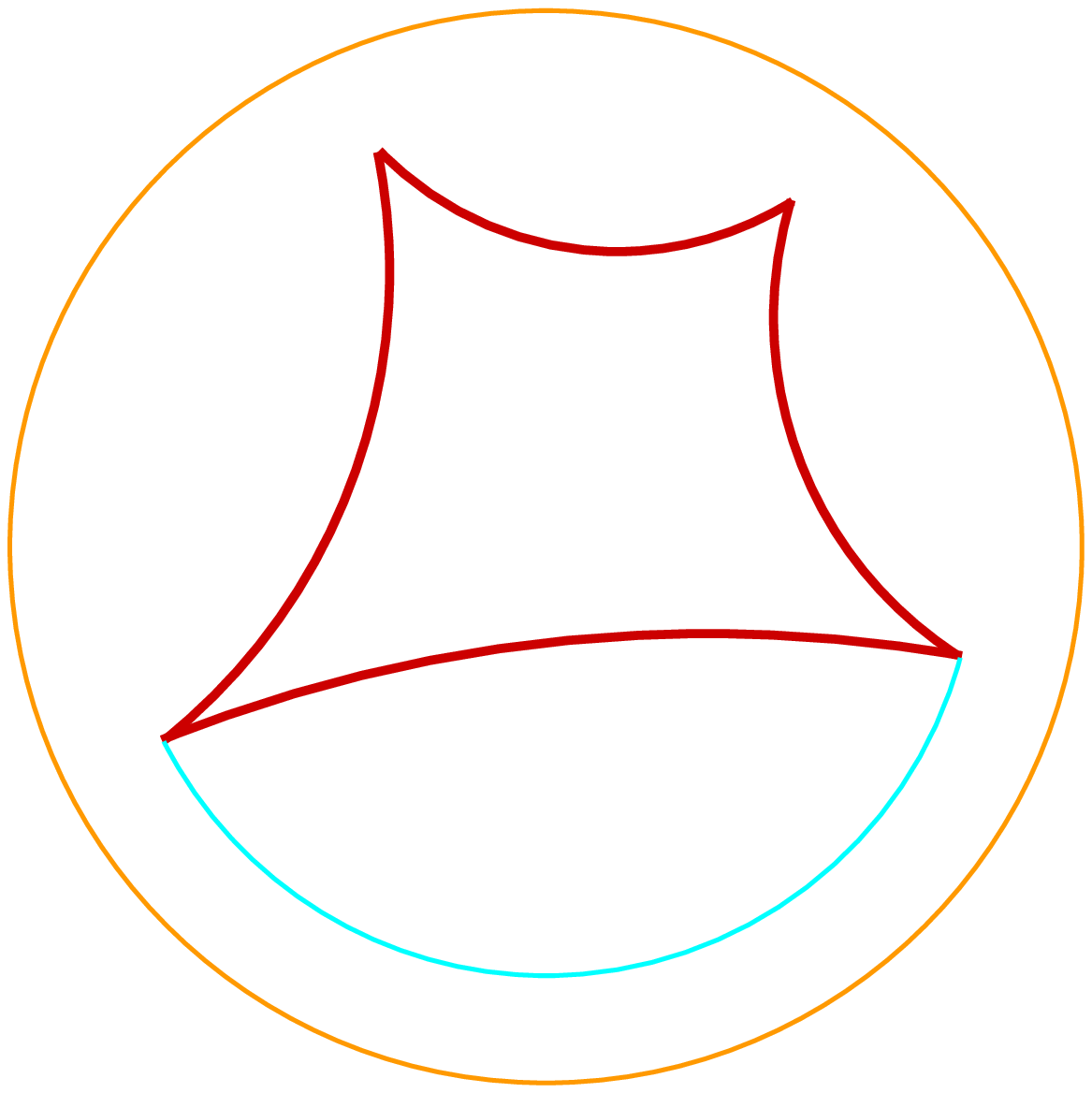}}
\includegraphics[scale=1]{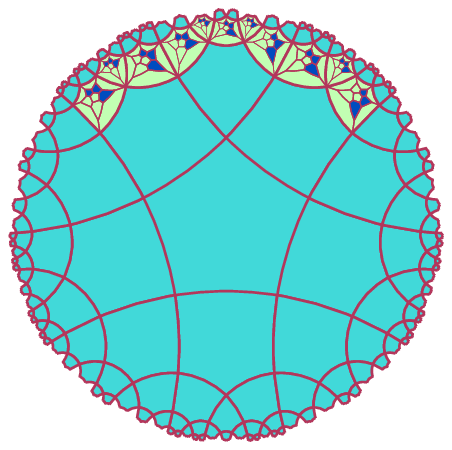}
\hfill}
\vskip-20pt
\begin{fig}\label{fpaths}
\leurre
Basic patterns for the paths. To left, a scheme of a path constituting a quadrangle.
To right, an arc of a circle, long enough to ensure the connection between two tracks.
\end{fig}
}

On that picture, we can see two paths for drawing the bottom side of~$Q$.
The red one is a segment of a straight line in the hyperbolic plane. The blue one is an
arc of a hyperbolic circle. On the model, the arc is longer than the segment. In fact,
the arc is much longer: its length is proportional to an exponential function of the
radius of the circle supporting the arc. The right-hand side picture of the figure shows 
us an arc of a circle. Its length is great enough to ensure the connection of two tracks,
whatever the lines which they follow. By rotating a convenient piece of the arc,
we can see how to connect the sides of~$Q$. The elements with an entry through side~1 
play an important role in the realisation of an arc of circle and on the connection 
between segments of different directions.

   We can see that those constructions are flexible enough, so that we have a relative
freedom in the construction of the circuit.

\subsubsection{The switches}\label{sbbswitch}

The section follows the implementation described in~\cite{mmJAC2021}. We 
reproduce it here for the reader's convenience. The illustration shows us what we
call an {\bf idle configuration}. 
Such a configuration is a configuration where there is no locomotive within a circle of
radius~3 from the central tile: call it the {\bf window}. The left-hand side of 
Figure~\ref{fstab_fxfk} shows us such an idle configuration for the passive fixed switch.
From what we said in Section~\ref{newrailway}, we know that there is no active fixed 
switch, so that the illustration of such a switch concerns the passive one only.

\vskip 10pt
\vtop{
\ligne{\hfill
\includegraphics[scale=1]{hyp3d_fix.ps}
\includegraphics[scale=1]{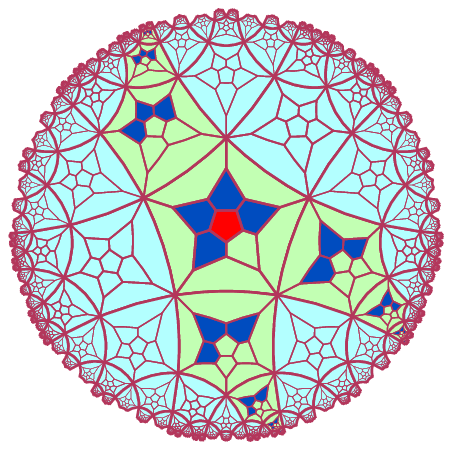}
\hfill}
\vskip-20pt
\begin{fig}\label{fstab_fxfk}
\leurre
Idle configurations, to left of a passive fixed switch, to right of a fork.
\end{fig}
}
\vskip 10pt
We can see that the central tile is an ordinary element of the track. As a locomotive
may enter through its face~5 or through its face~4, the constitution of the switch is
easy: one branch of the switch arrives at face~5{} in the central tile while the other
branch arrives at face~4. We have a single locomotive in the window around the
centre of the fixed switch.

Two important structures are required for implementing the remaining swit-ches :
the fork and the controller. We start with the fork and we examine the case of the
controller after the implementation of the flip-flop switch.

The right-hand side picture of Figure~\ref{fstab_fxfk} illustrates the implementation of 
the {\bf fork} in its idle configuration. That structure receives one locomotive and it 
yields two ones which leave the structure in different directions. As can be seen on the 
figure, the central cell differs from an element of the track: its decoration consists 
of five dodecahedrons which are placed on faces~6, 7, 8, 9 and~11, $\Delta_{11}$ being
red and $\Delta_i$, \hbox{$i\in\{6..9\}$} being blue. Moreover, the 
locomotive enters the tile through its face~2 and the two new locomotives exit through 
faces~4 and~5.

The flip-flop switch and both parts of the memory switch require a much more involved
situation. The global view of an idle configuration is illustrated by 
Figure~\ref{fschflfl}.

\vskip 10pt
\vtop{
\ligne{\hfill
\includegraphics[scale=0.45]{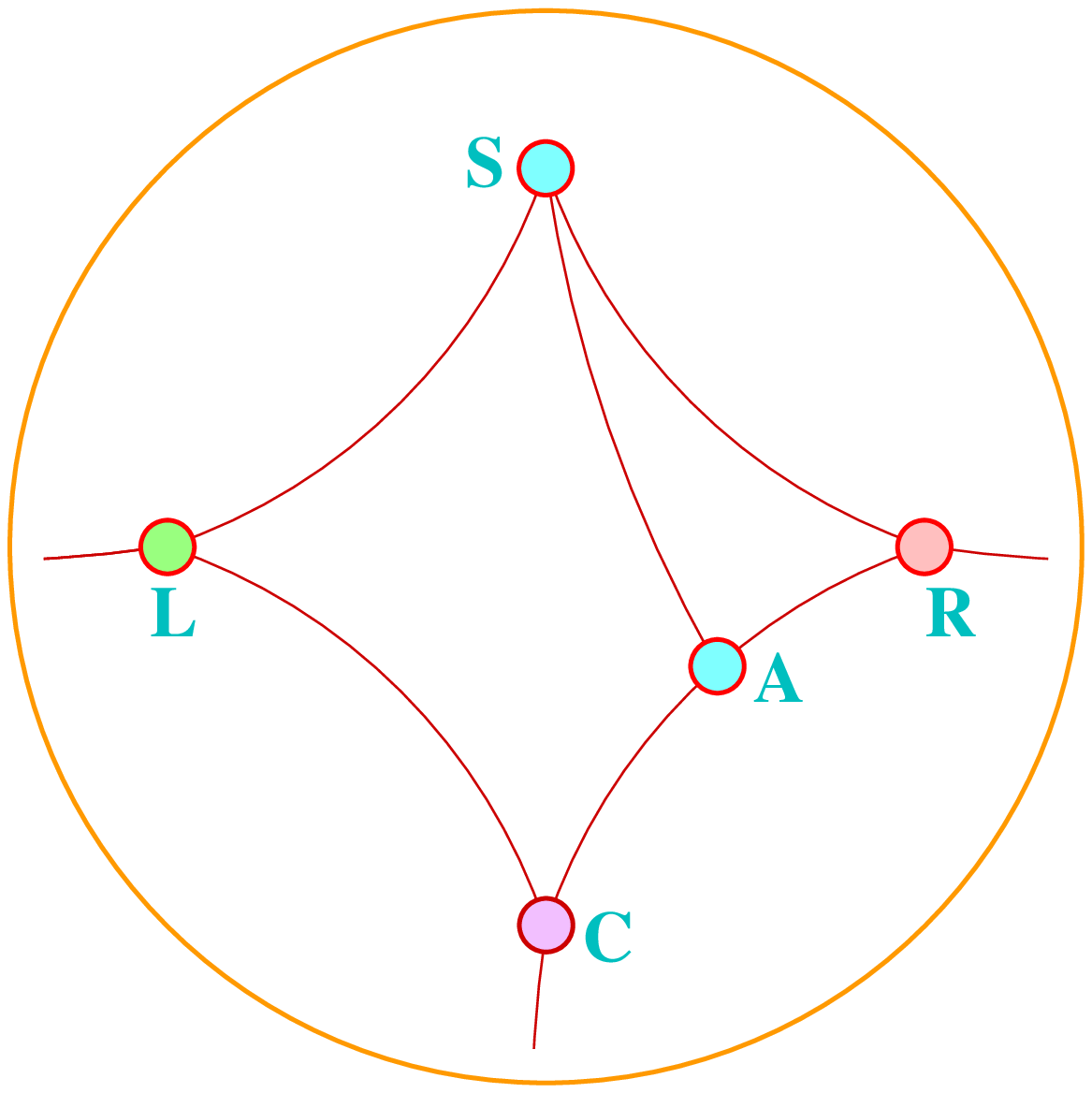}
\raise 50pt\hbox{\includegraphics[scale=0.25]{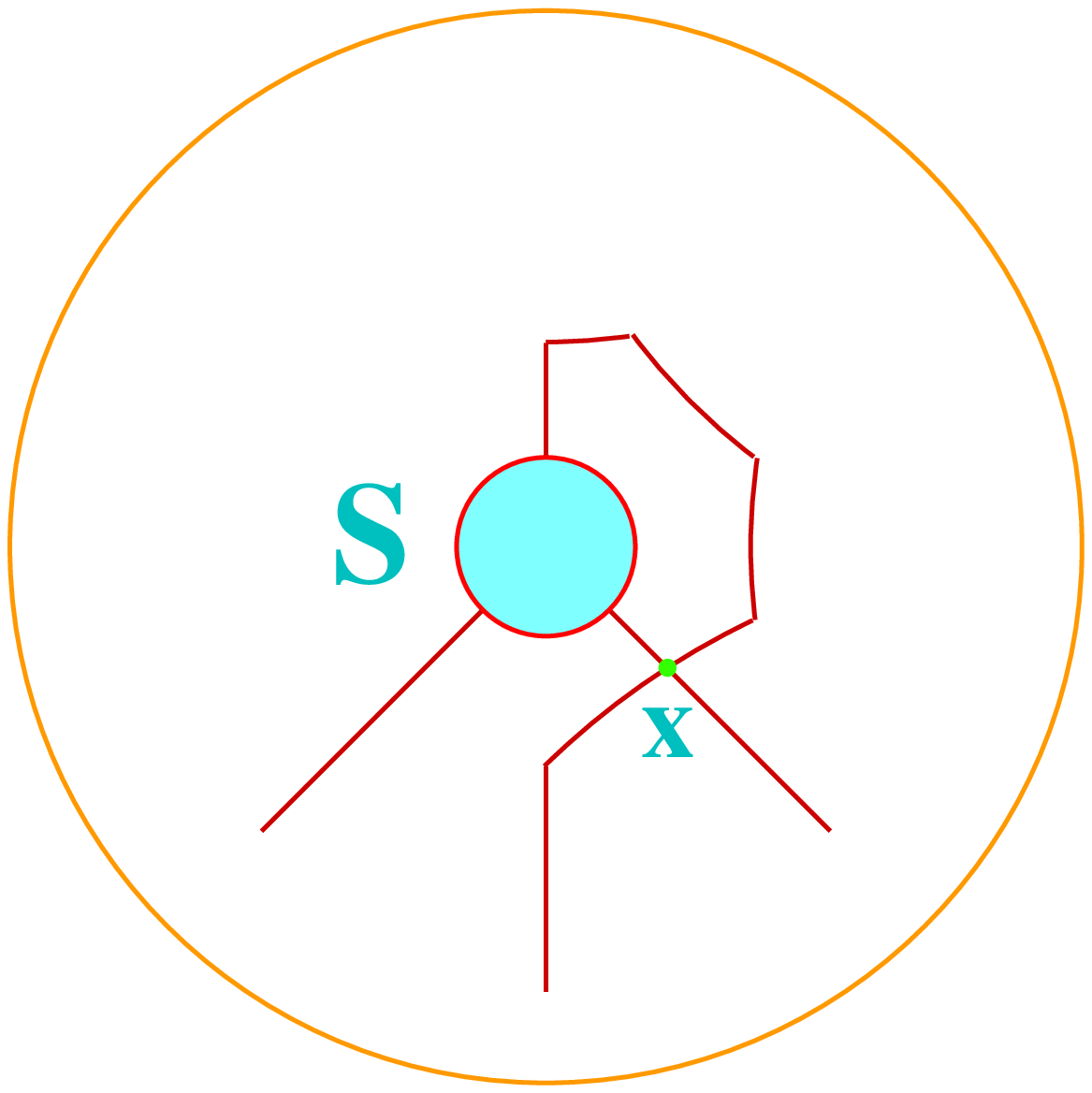}}
\hfill}
\begin{fig}\label{fschflfl}
\leurre
Scheme of the implementation of a flip-flop switch with, to right, a zoom on~{\bf S}.
In the zoom, note the crossing~{\bf x} of a leaving track by the deviated route 
to~{\bf S}.
\end{fig}
}

The locomotive arrives at a segment of a straight line by~$C$ where a fork sits.
Accordingly, two locomotives leave~$C$, one of them towards~$L$, the other towards~$R$.
At~$L$ a controller sits and, on the figure, it let the locomotive go further on the
segment of straight line. Note that the path from~$C$ to~$L$ is also a segment of a
straight line on the figure, which is conformal to the implementation. Now, the structures
which are later involved make the length of that segment to be huge. At~$R$ too a 
controller is sitting but, on the figure, it kills the locomotive which is thus prevented
to go outside the switch. On the way from~$C$ to~$R$ the locomotive meets another fork
at~$A$. The fork sends one of the new locomotives to~$R$ where it is stopped in the 
situation illustrated by the figure and the other is sent to~$S$. There a fork is 
sitting too which sends two locomotives, one of them to~$L$, the other to~$R$. When they
reach their goal, through another face of the controller, the locomotives change the 
configuration of the controller which is sitting there. The controller which let the 
locomotive go will further stop it while the one which stopped it will further let it go.
Accordingly, after the passage of a locomotive and after a certain time, the 
configuration of the flip-flop switch is that we described in Sub-section~\ref{railway}.
We may arrange the circuit so that a new passage of the locomotive at that switch will
happen a long time after the change of function of its controllers is achieved.

We have now to clarify the implementation of the controller. It is illustrated by 
Figure~\ref{fctrl}.

\vskip 10pt
\vtop{
\ligne{\hfill
\includegraphics[scale=1]{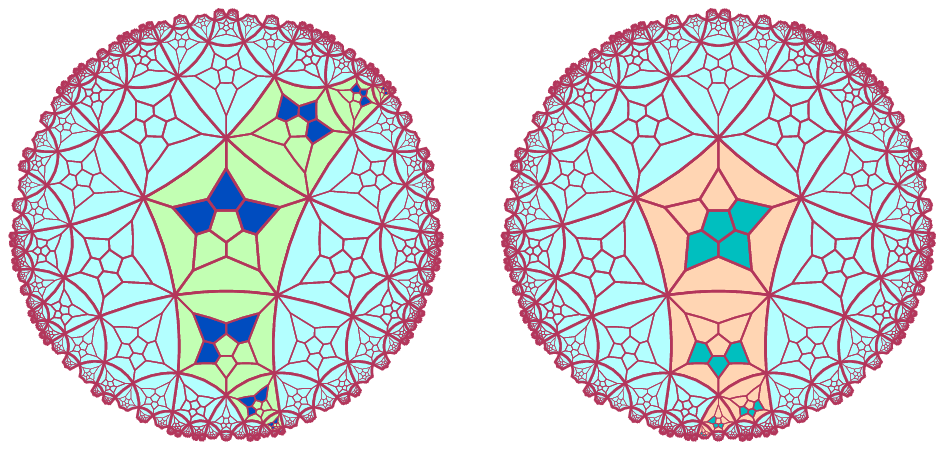}
\hfill}
\begin{fig}\label{fctrl}
\leurre
The idle configuration of the controller. To left, the track in \HH$_u$; to right, 
in~\HH$_b$, the controller and its access by a locomotive.
\end{fig}
}

Here again, the central cell in~\HH$_b$ is not an element of the track. Its decoration
consists of four blue dodecahedrons which are placed on the faces~6, 9, 10 and~11 of
the tile.

   Figure~\ref{fschflfl} requires the explanation of the zoom: three tracks abut the 
point~{\bf S} which is supposed to behave like a fork. However, the configuration of 
the tracks abutting the central tile of a fork is different. It is the reason for which 
the track arriving to~{\bf S} from~{\bf A} is deviated near~{\bf S} as indicated in the
zoom so that the new configuration is conformal to that of a fork. Note that the pieces
constituting the deviation of the track are segments of straight line in Poincar\'e's
disc model. The price to pay is a crossing at~{\bf x} on the picture illustrating the 
zoom. 

   The crossing happens to be a burden in the hyperbolic plane requiring several complex
structures we do not need in the hyperbolic $3D$-space. The third dimension offers
two possible ways to easily realize a crossing: the bridge or the tunnel. In the present
paper, I have chosen the tunnel, illustrated by Figure~\ref{ftunnel}. Two tracks follow
a line: a red one going from right to left on the three pictures and a blue line, from
bottom to top on the left-hand side picture only.

The leftmost picture of the figure is the standard projection of \HH$_u$ on~\HH, where 
each dodecahedron is projected within the face it shares with~\HH. In the picture, the 
faces of the elements of the track which lie on~\HH{} are faces~0. Using the numbering 
of the tiles of~\HH, the elements of the track following the blue line, going from 
bottom to top on the picture are, in this order : sector 3, tiles 21, 8, 3 and~1; the 
central tile, then, in sector~1, tiles~1, 4, 12 and 33. Six other tiles can be seen : in 
sector~2, tiles 3, 8 and 20 and, in sector~5, 33, 12 and~4 as the track goes from
right to left, following the red line~$\ell$. Three tiles are missing for that track: 
tile~1 of sector~5, the central tile and tile~1 of sector~2. Those tiles are not 
in~\HH$_u$ but in \HH$_b$, below~\HH. They are illustrated by the middle picture 
where $\ell$ is again represented. It is the part of the crossing track which passes 
under the track illustrated by the leftmost picture. In the middle picture we can see 
those elements of track as if \HH{} were translucent. Two additional tiles are indicated 
in the middle picture: tile~4 of sector~5 and tile~3 of sector~2. Those tiles correspond 
to the tiles with the same numbers in the same sectors which stand in~\HH$_u$. Each of 
those tiles in \HH$_u$ and in~\HH$_b$ allows a locomotive going upon~\HH{} to go below 
that plane in order to cross the other track. Those elements require the locomotive to 
enter through another face than face~5 or face~4 and to exit through a face~2.

\vskip 10pt
\vtop{
\ligne{\hfill
\includegraphics[scale=0.8]{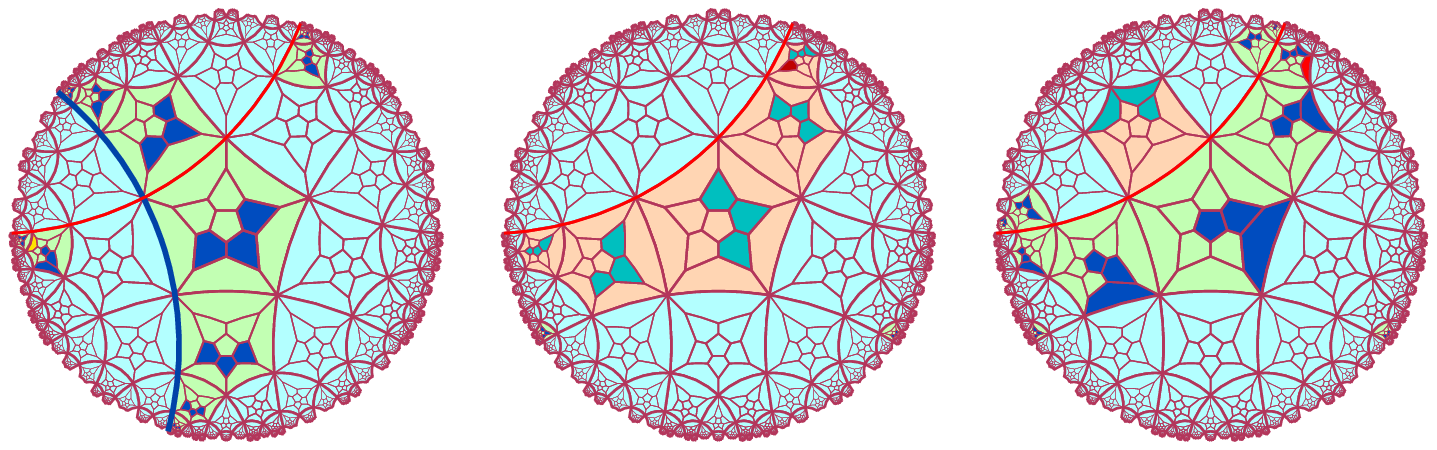}
\hfill}
\begin{fig}\label{ftunnel}
\leurre
The idle configuration of the tunnel. 
\end{fig}
}

The rightmost picture of Figure~\ref{ftunnel} shows us the projection of the tunnel
on the plane~\VV{} which is orthogonal to~\HH, cutting that latter plane along the 
line~$\ell$.  We represent on~\VV{} the tiles which have a face on it only. Those which 
are on the same side of~\VV{} as the central tile in the middle picture are seen in direct
projections. Those which are on the other side are seen as if \VV{} were translucent.

Figure~\ref{fprojs} illustrates the possible projections of an element of the track 
depending on whether it is seen upon~\HH, upon~\VV{} or through those planes by 
transparency, also depending on which face is on~\HH{} and which one on~\VV.

Consider the central tile of the leftmost picture of Figure~\ref{ftunnel}. It is
projected on its face~0 on~\HH{} and the locomotive goes upward, entering through face~5 
and leaving through face~2. Consider the entrance into the tunnel. We can see that
the entry of the locomotive through face~1 and its leaving through face~2 cannot be
used on both ends of the tunnel: the reason is that, at the entry, the neighbour~6 of the
tile below~\HH{} would be below face~0 of the tile of the track arriving at the entry
over~\HH: that would stop the locomotive. That problem does not occur for the exit,
so that we can use that way of working for the exit. For the entry, we need to change
the tile in order to induce rules which would be compatible with the other rules with
respect to the rotations we consider. Accordingly, this time the entry into the tile 
below~\HH{} is through face~7 and the exit still through face~2. That tile differs from 
a tile of the track by the occurrence of red neighbour on its face~3.

Figure~\ref{fprojs} gives various views on the tiles we just mentioned. The first
row of the picture indicates the projections over \HH{} of tiles of the track
depending on the face which lies on ~\HH: face~0, face~1, face~2 and face~7
for pictures~1, 2, 3 and~4 respectively on the first row of the figure. Picture~5
is the projection of the tile of the entry below~\HH, projected on~\VV{} on its face~8,
where \VV{} is the plane orthogonal to~\HH{} with \hbox{\HH~$\cap$~\VV = $\ell$}.
The second row of the figure, pictures~6 up to~10 gives the same tiles viewed through
a translucent \HH. Picture~11, third row of the figure, illustrates an element of the 
tunnel below~\HH{} seen through a translucent \HH. 

\vskip 10pt
\vtop{
\ligne{\hfill
\includegraphics[scale=0.45]{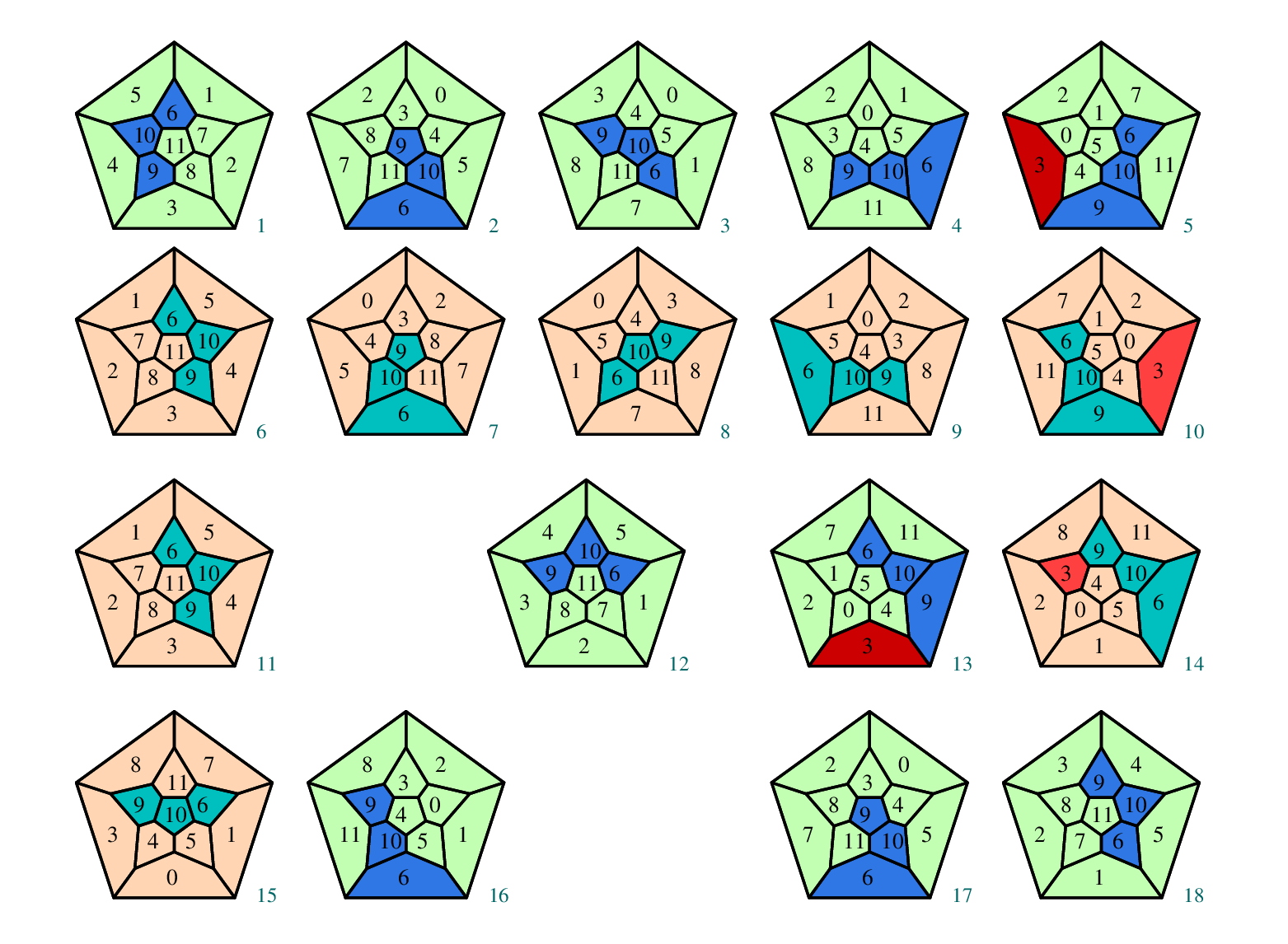}
\hfill}
\begin{fig}\label{fprojs}
\leurre
Projection of various positions of the element track upon~\HH, \VV{} and also of 
elements below \HH.
\end{fig}
}

On the same row, the illustrated tiles
are those for the entry into the tunnel: picture~12 for the tile above~\HH{} and 
pictures~13 and~14 for the entry below~\HH{} when it is projected on~\VV, picture~13, 
on its face~8, and when it is viewed through a translucent~\HH, picture~14, through its 
face~7. The last row of the figure illustrates the tiles for the exit from the tunnel. 
Pictures~15 and~16 illustrate the exit tile below~\HH: viewed through a translucent 
face~2 being on~\HH, while picture~16 shows us the projection on its face~0 over~\VV.
Pictures~17 and~18 illustrate the exit tiles upon~\HH. 

   To conclude with the tunnel, the occurrence of a locomotive in the central tile of
the tunnel will not stop a locomotive on the upper way: a locomotive never
stops in an element of the track and, more over, in such a crossing, there is
a single locomotive in a window around the central point of the crossing: if it is
present on one path, it cannot be present on the other one.
\vskip 10pt
   We are now in position to deal with the memory switch, active and passive parts.

\vskip 10pt
\vtop{
\ligne{\hfill
\includegraphics[scale=0.4]{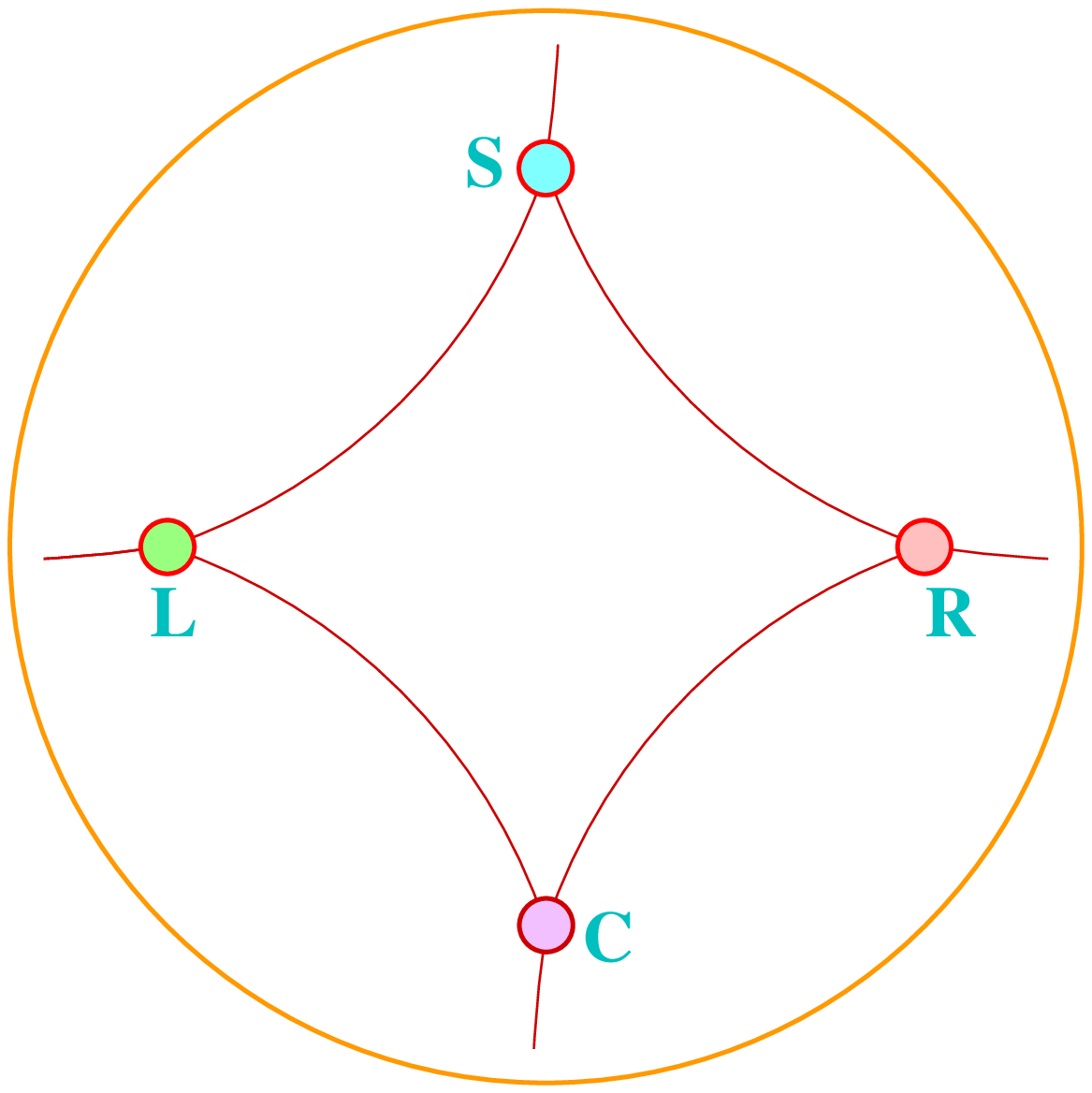}
\includegraphics[scale=0.4]{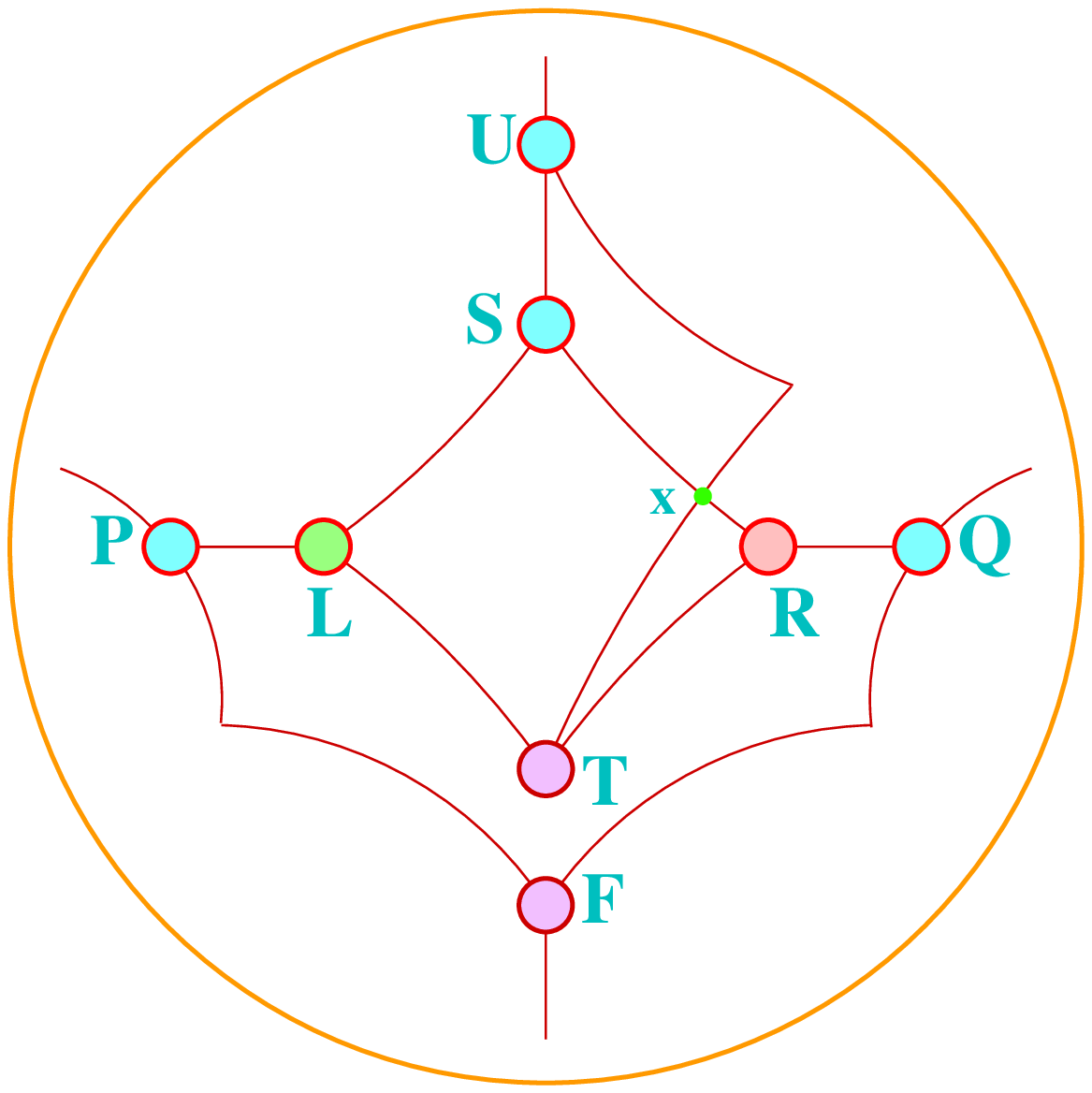}
\hfill}
\begin{fig}\label{fmemo}
\leurre
To left, the active memory switch, to right, the passive one.
\end{fig}
}

   First, we deal with the active part. It looks like the flip-flop switch with this 
difference that there is no fork~$A$ in between the path from the initial fork~$C$
to~$R$, one of the controllers. Figure~\ref{fmemo} illustrates both parts of the 
memory switch: to left of the figure, the active part of the switch, to right, its passive
part.

In the active part of the switch, left-hand side picture of Figure~\ref{fmemo}, 
the locomotive arrives to a fork sitting at~$C$. From there two locomotives are sent,
one to~$L$, the other to~$R$ and the working of the switch at this point is alike that
of a flip-flop switch. The difference lies in the fact that the passage of the locomotive
does not trigger the exchange of the roles between the controllers. That change is 
triggered by the passage of a locomotive through the non-selected track of the passive
part of the switch. When it is the case, a locomotive is sent from the passive part
to the active one. That locomotive arrives at the fork which is sitting at~$S$. The fork
creates two locomotives which are sent to~$L$ and~$R$ in order to change the permissive
controller to a blocking one and to change the blocking one into a permissive one.

Let us look at the working of the passive part. The locomotive arrives to the switch 
through~$P$ or through~$Q$. Assume that it is through~$P$. A fork sitting at~$P$ sends a
locomotive to the fixed switch~$F$ which let the locomotive leave the switch. The other
locomotive sent by~$P$ goes to~$L$. If that side is that of the selected track, the
controller sitting at~$L$ blocks the locomotive so that no change is performed, neither in
the passive switch, nor in the active one. Accordingly, the selected track of the active
switch is controlled by a permissive controller while the corresponding selected track
of the passive switch is controlled by a blocking controller. Presently, assume that
the side of~$L$ is not that of the selected track. It means that $L$ let the locomotive
go to~$T$ where a fixed switch sends the locomotive to a fork at~$U$. That fork
sends a locomotive to the fork~$S$ of the active switch and the other locomotive is sent
to~$S$ of the passive switch. At that point~$S$, a fork sends a locomotive to~$L$ and
another one to~$R$ in order to change the working of both controllers to the
opposite task. As a parallel change occurs in the active switch the selected track
is redefined in both parts of the switch.

   That working raises several remarks. First, the role of the controllers in the active
and in the memory switches are opposite. Nevertheless, in both cases, the same
programmable controller is used exactly because it is programmable in the way we just
described. The second remark is that all the tracks indicated in the pictures are pieces
of straight lines of the hyperbolic plane. A last remark is that we used one crossing, two
fixed switches, at $F$ and at~$T$ and four forks, at~$P$, $Q$, $U$ and~$S$. Each
structure requires some space, at least a disc whose radius is the length of four tiles
aligned along a straight line. Consequently, the passive memory switch requires a huge
amount of tiles. Again, we may assume that a new passage of the locomotive to the
switch happens after the changes has been performed when it is the case they should 
occur.

\subsubsection{The one-bit memory}\label{sbbunit}

\def\WW{{\bf W}}
\def\AA{{\bf A}}
\def\RR{{\bf R}}
\def\EE{{\bf E}}
\def\FF{{\bf F}}
\def\PPP{{\bf P}}
\def\bbz{{\bf b0}}
\def\bbu{{\bf b1}}
\def\zz{{\bf 0}}
\def\uu{{\bf 1}}
It is now time to implement the one-bit memory. Figure~\ref{fonebit} illustrates the
construction for the implementation of Figure~\ref{basicelem} in the dodecagrid.

\vskip 10pt
\vtop{
\ligne{\hfill
\includegraphics[scale=0.5]{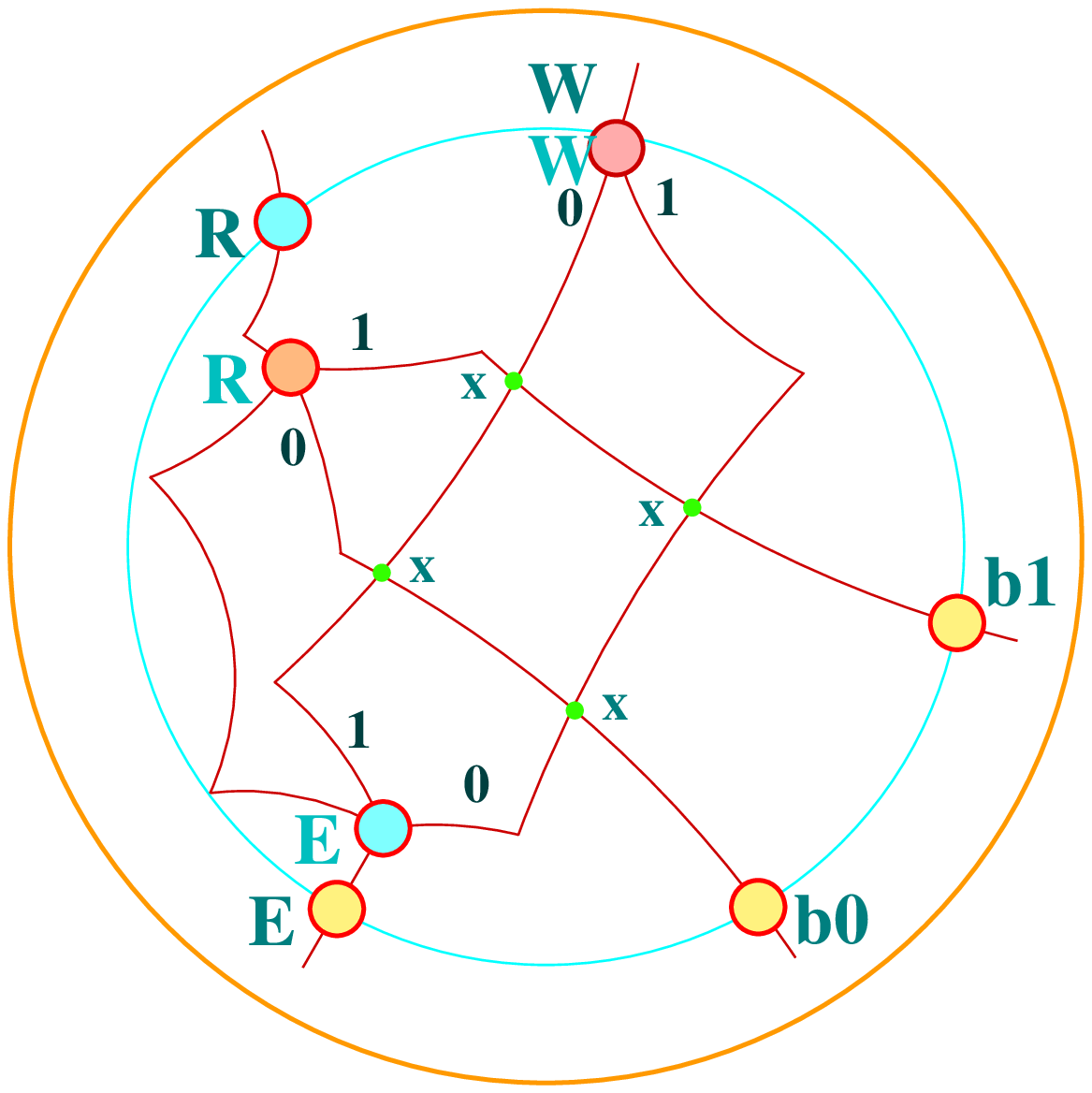}
\hfill}
\begin{fig}\label{fonebit}
\leurre
The idle configuration of the one-bit memory. Note the four crossings in the 
implementation. Note that the connection from~\EE{} to~\RR{} is realized by three 
segments of straight lines.
\end{fig}
}

We can see the active memory switch at~$R$ and the passive one at~$E$. The dark letters
which stand by the blue circle indicate {\bf gates} of the one-bit memory: \WW, \RR, \EE,
\bbz{} and \bbu. We can easily see that if the locomotive enters the unit through 
the gate~\RR, then it leaves the memory through the gate~\bbz{} or through the gate~\bbu{}
depending on the information stored in the memory: that information is provided the unit
by the positions of the switch at~\WW{} and those at~\RR{} and~\EE. Note that the
positions at~\RR{} and at~\EE{} are connected by the path from~\EE{} to~\RR, see the
figure.

When the locomotive enters the memory through the gate~\WW{} where a flip-flop switch is
sitting, it goes to~\RR{} through one of both tracks leaving the switch. If it goes
through the track marked by~\zz, \uu, it arrives to~\EE{} by the track marked with the
opposite symbol, \uu, \zz{} respectively. Indeed, when the locomotive crosses~\WW,
the passing makes the selected track to be changed so, if it went through one track, after
the passage, in particular when the locomotive arrives at~\EE, the new selected track
at~\WW{} is the track through which the locomotive did not pass. So that the
track marked by one symbol at~\WW{} should be marked by the opposite one at~\EE. The
selection observed at~\EE{} is transferred to~\RR{} thanks to the path connecting~\EE{}
to~\RR.

   As the one-bit memory will be used later, we introduce a simplified notation:
in Figure~\ref{fonebit}, the memory structure is enclosed in a blue circle. At its 
circumference the gates are repeated by the same symbols. In the next figures, when a
one-bit memory will be used, we shall simply indicate it by a disc with, at its border, 
the five gates mentioned in Figure~\ref{fonebit}.

\subsubsection{From instructions to registers and back}\label{sbbregdisp}

   As will be explained in Sub-subsection~\ref{sbbreg}, when a locomotive arrives at a 
register in order to performer the operation dictated by the program, it knows only
which instruction to perform, whether it is an incrementation or a decrementation. In
particular, after performing the operation, the locomotive does not know to which point of
the program it must go back. As far as a moving cell which always remains in the same 
state has no memory, that information concerning the return way must be stored somewhere 
in the circuit.

\def\DDI{{${\mathbb D}_I$}}
\def\DDD{{${\mathbb D}_D$}}
   To that goal we define two structures~\DDI {} and \DDD {} for incrementing and 
decrementing instructions respectively. Each structure consists of as many units
as there are instructions of the corresponding type operating on the same register.
Accordingly, each register is dotted with specific \DDI {} and \DDD. We can imagine
that the small orange and blue boxes of Figure~\ref{fglobconfig} contain, each one,
a copy of \DDI {} and a copy of~\DDD.

First, we consider the case of~\DDI. Each unit is based on a one-bit memory which is
illustrated by Figure~\ref{fdispinc}. The working of~\DDI {} is the following. An 
instruction for incrementing the register $R$ is connected through a path to a specific 
unit of~\DDI. The path goes from the program to the gate~\WW{} of that unit. At the 
initial time, the configuration of~\DDI {} is such that all its unit contain the bit~0: 
the switches of the one-bit memory are in a position which, by definition defines bit~0.
Also, each switch at~\AA{} which is a flip-flop switch selects the path leading to the
register. Accordingly, when the locomotive enters the unit, it will change the flip-flop 
at~\WW{} and the memory switches at~\RR{} and \EE{} so that, by definition, the memory 
contains the bit~1. The locomotive leaves the memory through the gate~\EE{} and it meets 
the flip-flop switch at~\AA. The switch sends the locomotive to the register and, just
after it is crossed by the locomotive, it selects the other path, that one which leads
back to the instruction which is the next one after the executed instruction.

When the locomotive completed the incrementation of~$R$ it goes back to the program.
It enters the memory of the first unit through its ~\RR-gate. If it reads~0, it leaves 
the memory through~\bbz{} and the path sends it to the next unit. So that the locomotive
visits the units of the~\DDI {} attached to~$R$ until it finds the unique unit whose 
one-bit memory is set to~1. So that now, we may assume that entering through~\RR, the 
locomotive reads~1. Accordingly it leaves the unit through~\bbu{} which leads it to~\WW. 
Consequently, the locomotive rewrites the bit, turning it to~0 and again exits through 
the gate~\EE. It again meets the flip-flop switch at~\AA{} which sends the locomotive on 
the path leading back to the program. That new visit of the flip-flop switch at~\AA{}
make the switch again select the track leading to~$R$. Accordingly, when the locomotive 
leaves~\DDI {} the structure recovered its initial configuration. Accordingly, that 
scheme allows us to correctly simulate the working of the units of~\DDI.

\vskip 10pt
\vtop{
\ligne{\hfill
\includegraphics[scale=0.5]{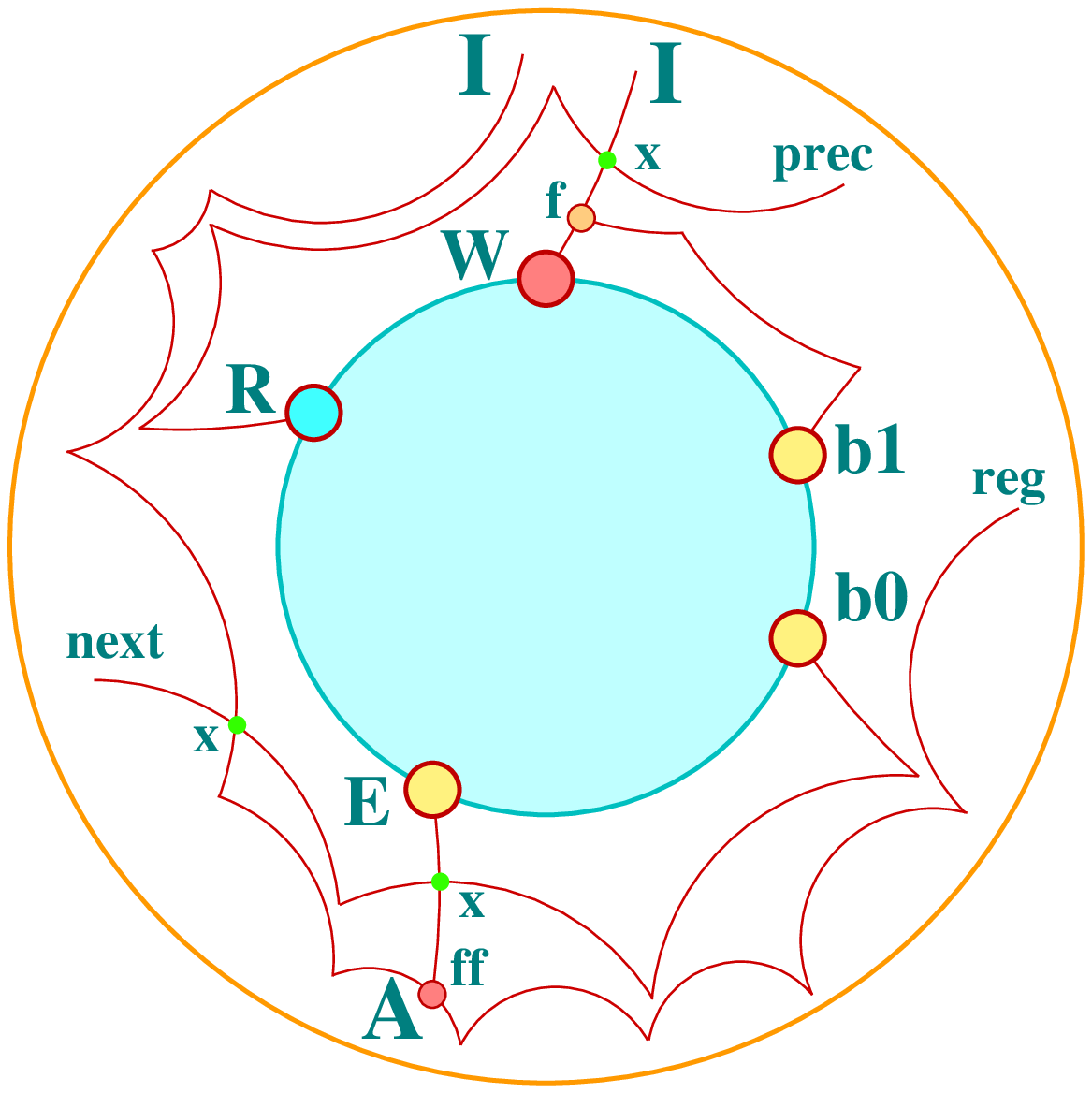}
\hfill}
\begin{fig}\label{fdispinc}
\leurre
The idle configuration of a unit of the structure which memorises the right incrementing
instruction. Note the three crossings in the 
implementation. Also note that the intensive use of sequences of connected segments of
straight lines.
\end{fig}
}

Secondly, we examine the structure of \DDD {} which plays for the 
decrementing instruction the role which \DDI {} plays for the incrementing ones. The 
structure is more complex for the following reason. An incrementing instruction is 
always performed which is not necessarily the case for a decrementing instruction. 
Indeed, if the register contains the value~0, it cannot be decremented. We say that 
the register is empty. In that case, the next instruction to be performed is not the 
next one in the program. It is the reason why the case of an empty register requires to 
be differently dealt with. Concretely, it means that the return track of the locomotive 
depends on whether the register was empty or not at the arrival of the locomotive
at the register.

Accordingly we need two one-bit memories instead of one. More over, as far as the bit~1
locates the unit of~\DDD {} which will send the locomotive to the right place in the 
program, the content of the memories should be the same: 0, if the locomotive did not
visited that unit at its arrival to~\DDD, 1 if it visited the required unit.
There are two return tracks after decrementation: the $D$-track when the 
decrementation could be performed, the $Z$-track when the decrementing locomotive found 
an empty register. We call $D$-, $Z$-{\bf memory} the one-bit memory visited by $D$-, 
$Z$-track respectively. 

The complication entailed by that organisation make a 
representation of a unit in a single window hardly readable.
It is the reason why it was split into three windows as illustrated by 
Figure~\ref{fdispdec}.

\vskip 10pt
\vtop{
\ligne{\hfill
\includegraphics[scale=0.45]{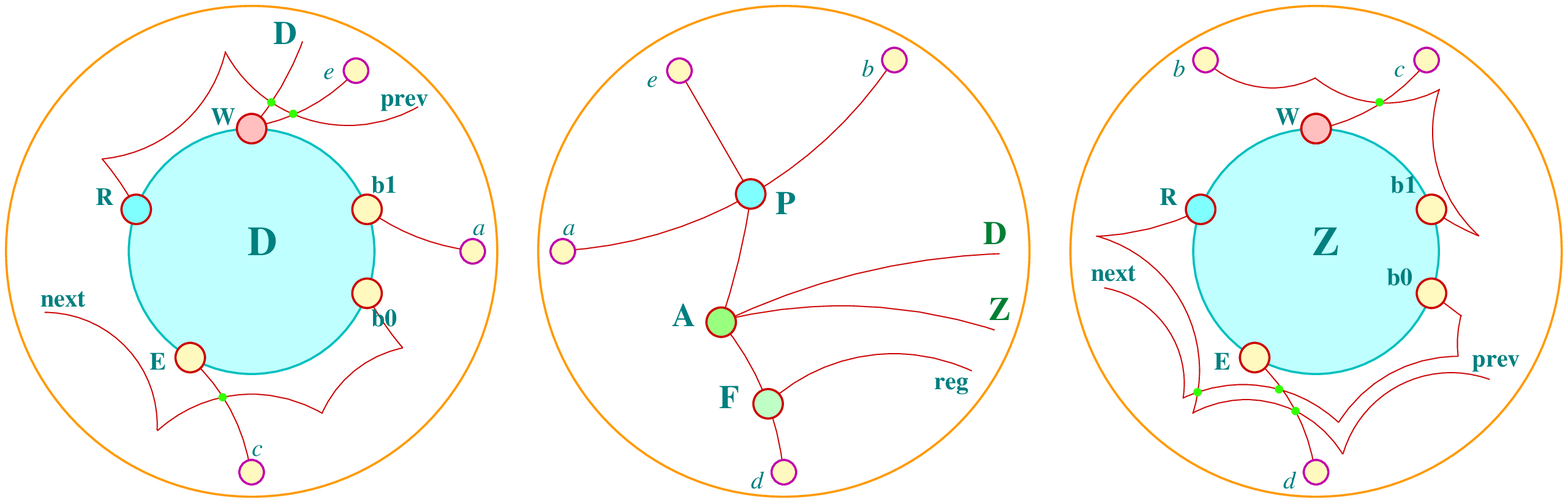}
\hfill}
\begin{fig}\label{fdispdec}
\leurre
The three windows illustrating the idle configuration of a unit of the structure which 
memorises the right decrementing instruction. Note the small discs accompanied with
small italic letters allowing the locomotive to pass from one window to the appropriate 
one. Also note the sketchy representation of the one-bit memories.
\end{fig}
}

   The $D$-memory is represented by the left-hand side picture of
the figure. A locomotive sent by the program for decrementing a register~$R$ arrives
at the appropriate unit of the \DDD {} attached to~$R$ by the track marked by $D$ in 
the picture. As far as the locomotive enters the memory through its gate~\WW, it rewrites 
its content from~0 to~1. Now, it has to mark the $Z$-memory which is represented on 
the right-hand side picture of the figure. To that goal, the track leaving the $D$-memory
through its gate~\EE, goes to the point~$c$, that italic letter marking a small yellow 
disc close to the border of the window. The track is continued from a small yellow disc 
close to the border of the right-hand side window marked with the same letter~$c$. Other 
such discs in the figure are marked by other letters which are pairwise the same in two 
different windows. From that second $c$, the track leads the locomotive to the 
gate~\WW{} of the $Z$-memory whose content, accordingly, will be changed from~0 to~1. 
Leaving the $Z$-memory through its gate~\EE, the locomotive is lead to a small disc~$d$ 
so that the corresponding track is continued by the small disc~$d$ we find in the middle 
window of Figure~\ref{fdispdec}. From there, the locomotive is lead to a flip-flop 
switch sitting at~\FF{} which, in its initial configuration, selects the track leading 
to~$R$. Now, after the switch is passed by the locomotive, its selection is 
changed, indicating the track leading to another switch sitting at~\AA.

   Consider the case when the locomotive returns from the register after a successful 
decrementation. It one by one visits the units of~\DDD, arriving to the unit through its
\RR-gate. If it reads~0, the track issued from~\bbz{} of the $D$-memory leads the 
locomotive to the \RR-gate of the $D$-memory of the next unit. So that such a motion is 
repeated until the locomotive reads~1{} in the $D$-memory. When it is the case, the 
locomotive leaves the memory through its gate~\bbu{} which
leads the locomotive to a small disc~$a$, so that the track is continued from the
small disc~$a$ we can see in the middle picture of Figure~\ref{fdispdec}. From there,
the locomotive arrives to~\PPP, the passive part of the memory switch whose active part
lies at~\AA. The locomotive is sent from~\PPP{} to a small disc~$e$ whose track arriving
there is continued by the track issued from the small disc~$e$ of the left-hand side
picture of the figure. That track leads the locomotive back to the \WW-gate of the
$D$-memory. Consequently, the content~1 of the memory is returned to~0 and leaving
the memory through its gate~\EE, the locomotive again arrives to the \WW-gate of the
$Z$-memory through the small discs~$c$. Again the \EE-gate of the $Z$-memory sends
the locomotive to the small disc~$d$ which, through the other such disc in the middle
window, again arrives at \AA. As far as it took much more time for the locomotive to 
rewrite both memories of the unit than for another locomotive to go from~\PPP{} to~\AA{}
in order to set the selection of the active switch, the switch at~\AA{} indicates the
track corresponding to the branch \PPP$a$ of the passive switch. That track leads to
the right decrementing instruction in the program and after that passage of the 
locomotive, the switch at~\FF{} indicates the path to~$R$: the structure recovered its
initial configuration which does not contain which track is selected at~\AA.

   Presently, consider the case when the locomotive returns from an empty $R$.
It returned through the $Z$-track and it arrives to the~\DDD {} attached to~$R$.
There it visits the units of the structure reading the content of its $Z$-memory through
the track leading to the \RR-gate of that memory in the unit. So that the locomotive 
visits the units until the single one whose $Z$-memory contains~1. When it is the case,
the locomotive leaves the $Z$-memory through its gate~\bbu{} which sends it to the 
\WW-gate of the $D$-memory through the small disc~$b$. But the continuation happens by~$b$
of the middle picture which leads the locomotive to~\PPP. The passive switch sends the 
locomotive to~$e$ and from there to the \WW-gate of the $D$-memory. Accordingly, what
we saw in a visit through a unit thanks to the $D$-track occurs once again. 
Accordingly, the locomotive
rewrites the contents of both the $D$- and the $Z$-memories, returning them from~1 to~0.
Then, the locomotive leaves the $Z$-memory through its \EE-gate so that, via the disc~$d$,
it arrives to~\FF{} where the flip-flop switch sitting there sends it to the switch
sitting at~\AA. Now, during its trip for the \EE-gate of the $Z$-memory to the \WW-gate
of the $D$-memory, the locomotive arrived to~\PPP{} through the branch \PPP$b$ of that
passive switch. Accordingly, the switch selected that track and, in the meanwhile, sent
another locomotive to~\AA{} in order to select the appropriate track, that which leads
to $Z$ in the program. As the locomotive passed for a second time through~\FF, the
switch selects presently the track leading to~$R$, its initial configuration. Accordingly,
the scheme illustrated by the picture performs what is expected.

   At last and not the least, we have to look at what happens when the locomotive arrives
to the program after performing its operation on the register. As already mentioned,
whatever the track arriving to the program, it leads to the appropriate instruction: the 
next one in the program if the just operated one was an incrementation or a successful 
decrementation, a specified one if the decrementation was unsuccessful. We have to 
remember, when we shall discuss the rules that when it leaves a register, whatever the 
return path, the locomotive must be blue.

\subsubsection{Constitution of a register}\label{sbbreg}

   The implementation of the register requires a special examination. As the computation
is supposed to start from a finite configuration, at initial time, each register has
only finitely many non blank tiles but at least a few ones more for a reason which will 
soon be clear. So that incrementation and decrementation necessarily change the length 
of the register which is always finite. 

    Our choice is to make the register grow as long as the computation is not completed.
The operations of incrementing and decrementing can be implemented by a change of colour
at the beginning end of the register: its content is the number of cells of that 
particular colour. The continuation of the register is made at speed 
$\displaystyle{1\over2}$ for a reason we soon indicate. It happens independently of the
operations performed on the register. As far as the length of the track from the register
and back is enormous, the other end of the register is farther and farther from its 
beginning end. So, to stop the register when the stopping instruction is reached, the
program sends a locomotive to all registers. Each locomotive crosses its register at 
speed~1 which, later, sends a stopping signal at speed~1 too. That condition guarantees 
that the stopping signal will eventually reach the growing end of the register so that 
it can stop the continuation.

A register consists of four {\bf strands}, each one being a finite track of the same 
length along a line~$\ell$, all of them starting from the same vertex on~$\ell$.
Two strands are in~\HH$_u$ and the two others are in~\HH$_b$. One strand in~\HH$_u$
contains the content of the register, we denote it by \RR$_c$. That strand is blue, 
denoted by~\sbb {} and the content is blank, denote by~\sww{} as already mentioned. 
The strand below~\RR$_c$ is \RR$_i$. It is red, denoted by~\srr, and it receives the 
locomotive which has to perform an incrementation. The other strand in~\HH$_u$ is red too,
again denoted by~\srr. That strand, denoted by \RR$_d$, receives the locomotive which has
to perform a decrementing instruction. The fourth strand, hence in~\HH$_b$, is blue and 
is the path for the stopping locomotive. We denote it by~\RR$_s$. That strand is also the
returning path for a successful decrementation. The return path for an incrementation
is~\RR$_c$. Note that the just described working is very different from what is described
in~\cite{mmarXiv21a}.

The growing end consists of the 
last tiles of each strands, those tiles being green, denoted by~\sgg{}. The four green
tiles share a side on~$\ell$. We shall see more precisely how it is 
performed when we shall study the rules for implementing the simulation. In order to 
facilitate our explanations, if $\mathcal S$ is one of the four strands we depicted, 
$\mathcal S$(0) denotes its first cell, more generally, $\mathcal S$$(i)$, 
with \hbox{$i\in{\mathbb N}^+$}, denotes the $i^{\rm th}$ element on the strand. As we 
also need to define the cell which has a side on~$\ell$ and which is continued by 
$\mathcal S$, we denote that cell by $\mathcal S$(-1). We also need to consider 
$\mathcal S$(-2) for two strands. We define a numbering for the tiles of $\mathcal S$ as 
illustrated by Figure~\ref{fles4} and as explained in our
discussion about the neighbours of a tile. The motion of a locomotive from a face~5 to a
face~2 allows us to see that on two strands, \RR$_c$ and \RR$_s$, the motion goes 
on $\mathcal S$($n$) by following increasing values of~$n$ while on the two other ones, 
\RR$_i$ and \RR$_d$, it goes by following decreasing values of~$n$. However, on~\RR$_c$
and on~\RR$_s$ we need to implement too the opposite motion of the locomotive.
Clearly, \RR$_c$($n$)
and \RR$_s$ can see \RR$_i$($n$) and \RR$_d$($n$), but \RR$_c$($n$) cannot see 
\RR$_s$($n$) and \RR$_i$($n$) cannot see \RR$_d$($n$). Also, for each $n$ in $\mathbb N$,
$\mathcal S$($n$) can see both $\mathcal S$($n$-1) and $\mathcal S$($n$+1).

\section{The rules}\label{srules}

   Let us remind the reader that cellular automata are a model of massive parallelism.
The base of a cellular automaton is a cell. The set of cells is supposed to be homogeneous
in several aspects: the neighbours of each cell constitute subsets which have the same
structure; the cell changes its state at each tip of a discrete clock according to the
states of its neighbours and its own state. The change is dictated by a finite automaton 
which is the same for each cell. A tessellation is a relevant space for implementing
cellular automata. Let $T$ be a tile and let $N(T)$ be the set of its neighbours. In a 
tessellation, $N(T)$ is the same for any~$T$ and two tiles of the tessellation are 
isomorphic with respect to the geometry of the space on which the tessellation is defined.
The dodecagrid is a tessellation of the hyperbolic $3D$-space. Moreover, there is an 
algorithm to locate the tiles which is cubic in time and in the size of the code 
attached to each tile, see \cite{mmbook2} for instance. However, we do not need that 
location system. The main reason is that we work on the basis of~\HH{} where there is a 
linear algorithm to locate the cells, see~\cite{mmbook2} too, and our incursions in the 
third dimension does not drive us far from~\HH. The tunnel and the growth of 
the registers indicated us how far it can be.

\def\PP{$\mathbb P$}
    The way the automaton manages the change of states can be defined by a finite set
of {\bf rules} we shall call the {\bf program} of the automaton denoted by \PP. We 
organise~\PP{} as a {\bf table} displayed at the end of the paper. The table is organised
according to the role of the corresponding instructions with respect to the place of 
the locomotive in the circuit it performs.
In Sub-section~\ref{rrotinv}
we shall define the format of the rules, what means the invariance we consider in the
present paper and how
we check it. 
In Sub-section~\ref{rtracks} we look at the application of the rules for 
the tracks, including those for the crossings. In the same sub-section
we do the same for the rules managing the working of the switches, mainly their control 
structures.
In Sub-section~\ref{rreg} we study the rules for the register, separately considering 
the growth of the register, its incrementation and its decrementation, the case of an 
empty register being included. Sub-section~\ref{srstop} investigates the rules for 
stopping the computation. The rules are displayed by Table~\ref{trall} at the end of the
paper.

\subsection{Format of the rules and rotation half-invariance}\label{rrotinv}

A cell of the dodecagrid in the cellular automaton we define with the set of rules
studied in this section consists of a tile of the tiling, we call it the {\bf support} 
of the cell, together with the finite automaton defined by the rules of the present 
section. However, we shall use the words cells and tiles as synonyms for the sake of
simplicity. The neighbour of a cell~$c$ whose support is~$\Delta$ is a cell whose support 
is one of the $\Delta_i$, \hbox{$i\in\{0..11\}$}, where $\Delta_i$ shares the face~$i$
of~$\Delta$. Accordingly, $\Delta$ can see each $\Delta_i$ and it is seen by that latter
cell but $\Delta_i$ and $\Delta_j$ with $i\not=j$ cannot see each other, as already known.
We also say that $\Delta_i$ is the {\bf $i$-neighbour} of $\Delta$. The index
of $\Delta_i$ refers to a numbering of the faces. We use the one we described
in Section~\ref{intro} with the help of Figure~\ref{fdodecs}. As long as it will be
possible, cells in~\HH$_u$ or in~\HH$_b$ will have their face~0 on~\HH. We already 
fixed the face~1{} of the strands in the previous section.

   As usual, we call {\bf alphabet} of our cellular automaton the finite set of its 
possible states. The alphabet of our cellular automaton consists of \sww, \sbb, \srr{} 
and \sgg{} we already met. We also call \sww{} the {\bf blank} as far as it is 
the quiescent state of the automaton and we remind the reader that the cells of the 
dodecagrid are blank, except finitely many of them at whichever time. Let $c$ be a 
cell of the dodecagrid whose state is {\ftt S$_o$ }
and whose support is $\Delta$, and let {\ftt S$_i$ } be the state of its $i$-neighbour.
Let {\ftt S$_n$ } be the {\bf new} state of~$c$, {\it i.e.} the state taken by~$c$
when it is the state associated with {\ftt S$_o$ } and the {\ftt S$_i$ } by the
automaton. We write this as follows:

\def\llrule #1 #2 #3 #4 #5 #6 #7 {%
\hbox{\ftt{#1.#2#3#4#5#6#7} }
}
\def\rrrule #1 #2 #3 #4 #5 #6 #7 {%
\hbox{\ftt{#1#2#3#4#5#6.#7} }
}
\vskip 5pt
\ligne{\hfill
\llrule {S$_o$} {S$_0$} {S$_1$} {S$_2$} {S$_3$} {S$_4$} {S$_5$} 
\hskip-4pt\rrrule {S$_6$} {S$_7$} {S$_8$} {S$_9$} {S$_{10}$} {S$_{11}$} {S$_n$} {}
\hfill(\numerrel)\hskip 10pt}
\vskip 5pt
\noindent
and we say that {\ftt S } is a {\bf rule} of the automaton. All rules of the automaton
we give in the present section obey the format defined by~(3). Of course, {\ftt S$_o$ }
and the {\ftt S$_i$ } belong to the alphabet 
\hbox{$\mathbb A = \{$\sww, \sbb, \srr, \sgg$\}$}.
From the remark on $\mathbb A$, we can read a rule as a word on $\mathbb A$: it is 
enough to remove the dots which occur in(3). We also call {\bf neighbourhood} of a rule
the sub-word \ftt{S$_0$S$_1$S$_2$S$_3$S$_4$S$_5$S$_6$S$_7$S$_8$S$_9$S$_{10}$S$_{11}$}
of~(3).

   Let $\mathcal R$ be the set of rotations around a face of~$\Delta$ leaving the 
support of~$c$ globally invariant.
Let $\rho\in\mathcal R$ and let 
\hbox{{\ftt S } $\in$ \PP}. We call {\bf rotated image of {\ftt S } under $\rho$} the 
rule:
\vskip 5pt
\ligne{\hfill 
\llrule {S$_o$} {S$_{\rho(0)}$} {S$_{\rho(1)}$} {S$_{\rho(2)}$} {S$_{\rho(3)}$} 
{S$_{\rho(4)}$} {S$_{\rho(5)}$} 
\hskip-4pt\rrrule {S$_{\rho(6)}$} {S$_{\rho(7)}$} {S$_{\rho(8)}$} {S$_{\rho(9)}$} 
{S$_{\rho(10)}$} {S$_{\rho(11)}$} {S$_n$} {} 
\hfill\hskip 10pt}
\vskip 5pt
\noindent
which we denote \ftt{S}$_\rho$ {} and we also say that \ftt{S}$_\rho$ {} as a word is a 
{\bf rotated form} of \ftt{S}.

A rule {\ftt S } is said {\bf rotation half-invariant} if and only if: 
\vskip 5pt
\ligne{\hfill $\forall \rho\in{\mathcal R}$
({\ftt S } $\in$ \PP{} $\Rightarrow$ {\ftt S }$_\rho$ $\in$ \PP)\hfill
(\numerrel)\hskip 10pt}
\vskip 5pt
We say half-invariance as far as invariance refers to the set of all rotations leaving
$\Delta$ globally invariant.
It is important to note that the rotation half-invariance requires the invariance in 
all rotations around a face of~$\Delta$ leaving the dodecahedron globally invariant. 
In (4), $\mathcal R$ cannot be replaced by a set of generators. For example, we have 
seen that the set of rotations around a face leaving a fixed face of~$\Delta$ invariant 
defines six generators. Now, if {\ftt S } is rotation invariant in those generators, it 
does not involve that it would be invariant under products of those generators. In other 
words, if $g_1$ and $g_2$ are such generators, 
\hbox{{\ftt S }$_{g_1}$, {\ftt S }$_{g_2}$ $\in$ \PP} does not involve
that \hbox{{\ftt S }$_{g_1\circ g_2 }$} should be in \PP{} as far as {\ftt S }$_{g_1}$
is generally formally different from {\ftt S } even if {\ftt S }$_{g_1}$ were in \PP.
Accordingly, rotation half-invariance in the case of a cellular automaton in the 
dodecagrid remains a strong constraint: it requires to check the application of twenty 
five rotations to each rule of its program. By comparison, the rotation invariance in 
the heptagrid, for instance, requires seven conditions only. In the Euclidean frame, 
the rotation invariance for a square grid requires four conditions and for the cube 
twenty four of them. In some sense, rotation half-invariance in the dodecagrid is a
kind of analog of rotation invariance in the cubic tiling of the Euclidean $3D$-space.
This is why the result of Theorem~\ref{letheo} is significant: the reduction from 5 to~4,
despite the relaxation of invariance to half-invariance remains a true effort.

    There is a simple way to check rotation half-invariance for the rules of~\PP. We 
already noticed that a rule in the format (3) can be read as a word on $\mathbb A$ and we
have also read \ftt{S}$_\rho$ {} as a word
for any $\rho$ in $\mathcal R$. If we lexicographically order the rotation forms of
\ftt{S } for all $\rho$ in $\mathcal R$,
we can choose the smallest one as representative of
all the others. Call that word the {\bf minimal form} of the rotated forms of 
{\ftt S } and denote it by min$_{\mathcal R}$({\ftt S }). We get the following lemma:

\begin{lem}\label{minimal}
Let {\ftt U } and {\ftt V } be two rules of \PP. Then {\ftt V } is a rotated image
of {\ftt U } if and only if their minimal forms are identical.
\end{lem}

\noindent
Proof. Denote {\ftt S }$_{\mathcal R}$ the set of all rotated forms of {\ftt S }. Then,
clearly, \hbox{min$_{\mathcal R}${\ftt S } = {\ftt S }$_\rho$} for some 
$\rho\in$~$\mathcal R$.
Also note that for two rotations $\rho_1$ and $\rho_2$ in $\mathcal R$ and for any rule 
{\ftt S }, we have 
\hbox{{\ftt S }$_{{\rho_1}_{\rho_2}}$ = {\ftt S }$_{\rho_1\circ\rho_2}$}. As far as 
$\mathcal R$ is a group, we get that
{\ftt U } $\in$ {\ftt V }$_{\mathcal R}$ if and only if {\ftt V } $\in$ 
{\ftt U }$_{\mathcal R}$ and also
if and only if {\ftt U }$_{\mathcal R}$ = {\ftt V }$_{\mathcal R}$. Which proves the 
lemma.\hfill $\Box$

\def\MM{$\mathbb M$}
\def\TT{$\mathbb T$}

   The condition given by the lemma induces a criterion which is still hardly manageable 
for handy computations. Fortunately, the criterion can easily be performed by an 
algorithm we leave to the reader. We programmed such an algorithm to check the
rotation invariance of the set of rules we define in the paper. Also note that the
lemma gives us the way to simplify the presentation of~\PP: if we strictly apply
(4), as far as for each {\ftt S } in \PP{} and for each $\rho$ in $\mathcal R${} 
{\ftt S }$_\rho$
should also be in \PP, the number of rules in \PP{} is a multiple of twenty five. 
Accordingly, if in \PP{} we replace all rules which are rotations image of each other 
by their minimal forms, we get a new set of rules \MM{} whose number of elements is that 
of~\PP{} divided by twenty five. The set of rules we shall present is not exactly \MM. 
Instead of replacing the rules in {\ftt S }$_{\mathcal R}$ by the minimal form, we take 
a more understandable element which will appear from our investigations. It is that set 
of rules, whose cardinality is that of \MM{} we present and we denote it by \TT. 
Accordingly for two rules {\ftt U } and {\ftt V } in \TT, we have that 
\hbox{min$_{\mathcal R}${\ftt U } $\not=$ min$_{\mathcal R}${\ftt V }}. In that case, 
we say that {\ftt U }
and {\ftt V } are {\bf half-rotationally independent}. If the rules of \TT{} are 
pairwise rotationally independent, we say that \TT{} is {\bf half-rotationally coherent}.
The computation program I devised in establishing \TT{} also checked the half-rotational
coherence of that set of rules, displayed by Table~\ref{trall}.

We often considered what we call an idle configuration. In such a condition, the 
configuration must remain unchanged as long as the locomotive is not in the window which 
defines the configuration. Such rules are called {\bf conservative}. When the locomotive 
falls within the window, some cells are changed while the others remain unchanged. The
rules which change a cell
are called {\bf motion rules}. Among those which do not change a cell, some 
of them have the locomotive in their neighbourhood, in general for one step of the 
computation. Although the cell is not changed, we say that it can see the locomotive so 
that it is a witness of its passage so that we call such rules {\bf witness rules}. 
They appear to be an intermediate class between the conservative rules to which they 
belong as far as the new state is the same as the previous one but their neighbourhood 
is changed so that they are in some sense affected by the motion to which they 
contribute by their conservative function. In our discussion, we shall make the 
distinction between conservative and motion rules as long as it will be needed.
The rules are numbered to facilitate their references in the text.

\ifnum 1=0 {
\subsection{Rules regarding generations~2 and~3}\label{rstopgen}

   In our discussion leading to Proposition~\ref{pneighgen} and to Lemma~\ref{ldistneigh},
we considered neighbours of generation~1, generation~2 and generation~3. Of course, 
the rules of~\TT{} are potentially applied to any cell. However, the question of what 
can be said about that distinction among neighbours is important for establishing a 
simulating program. It is know that for hyperbolic spaces the 
number of tiles within a ball of radius~$k$ is exponential in~$k$. As an example,
the number of cells which are at a distance 10 from a tile is 1528. For each of these
cells at least 6 of their neighbours are not shared with the other cells of the same 
sphere. And that deals with a single one while many configurations involve around ten 
cells, possibly forty eight for the registers.  Such a heavy number is not manageable 
even by a program.  It is desirable to reduce the number of neighbours to be checked. 

In the discussion of Proposition~\ref{pneighgen} and Lemma~\ref{ldistneigh}, we 
considered generations~1, 2 and~3 of a given cell and we considered the connections 
between the spheres around a cell. It is possible to define the notion of neighbour of a cell
of generation~$n$ for any positive~$n$: the neighbours of the generation~$n$+1 of a 
tile~$T$ are the neighbours of those of the generation~$n$ which do not belong to a 
generation~$m$ with \hbox{$m<n$}. We noticed that a neighbour of $T$ of generation~3 has 
only a vertex in common with~$T$. Clearly, the neighbours of generation~4 are separated 
from~$T$ by at least one tile. Repeating the argument, the tiles of generation~$4n$ are 
further and further from~$T$ as $n$ is increasing, a property we already mentioned.

Consider the rules defined by the following patterns:

\vskip 5pt
\ligne{\hfill
$\vcenter{
\vtop{\leftskip 0pt\parindent 0pt\hsize=285pt
\ligne{\hfill
\hbox{$(a)$ \llrule {W} {X} {W} {W} {W} {W} {W} 
\hskip-3.5pt\rrrule {W} {W} {W} {W} {W} {W} {W} },
\hbox{$(b)$ \llrule {W} {X} {Y} {W} {W} {W} {W} 
\hskip-3.5pt\rrrule {W} {W} {W} {W} {W} {W} {W} },
\hbox{$(c)$ \llrule {W} {X} {Y} {Z} {W} {W} {W} 
\hskip-3.5pt\rrrule {W} {W} {W} {W} {W} {W} {W} },
\hfill}
\ligne{\hfill with \hbox{\ftt{X} {}, \syy {}, \ftt{Z} {}  $\in$ 
$\{$\sbb, \srr, \sgg, \syy$\}$}
\hfill}
}}$
\hfill(\numerrel)\hskip 10pt}
\vskip 5pt
We may assume that \hbox{\ftt{X} {} $\leq$ \syy {} $\leq$ \ftt{Z} }. Indeed, with rotation
invariance, assuming \hbox{\sww {} $<$ \ftt{X} {} $<$ \syy}, it is not difficult to see 
that the rules of the forms
\hbox{$(i)$\hskip 5pt\llrule {W} {X} {Y} {W} {W} {W} {W} 
\hskip-3.5pt\rrrule {W} {W} {W} {W} {W} {W} {W} } and
\hbox{$(ii)$\hskip 5pt\llrule {W} {Y} {X} {W} {W} {W} {W} 
\hskip-3.5pt\rrrule {W} {W} {W} {W} {W} {W} {W} } have the same minimal form
\hbox{\llrule {W} {W} {W} {W} {W} {W} {W} 
\hskip-3.5pt\rrrule {W} {W} {W} {W} {X} {Y} {W} }. Indeed, a rotation around face~0
allows us to pass from~$(i)$ to $(ii)$.
For the rules of the form~$(c)$ in (4), 
we can subdivide $(c)$ in five cases assuming 
\hbox{\sww {} $<$ \ftt{X} {} $<$ \syy {} $<$ \ftt{z} }:
\vskip 5pt
\ligne{\hfill
$\vcenter{\vtop{\leftskip 0pt\parindent 0pt\hsize=250pt
\ligne{\hfill
\hbox{$(c_1)$ \llrule {W} {X} {X} {X} {W} {W} {W} 
\hskip-3.5pt\rrrule {W} {W} {W} {W} {W} {W} {W} }\hskip-3pt,
\hbox{$(c_2)$ \llrule {W} {X} {X} {Y} {W} {W} {W} 
\hskip-3.5pt\rrrule {W} {W} {W} {W} {W} {W} {W} }\hskip-3pt,
\hbox{$(c_3)$ \llrule {W} {Y} {Y} {X} {W} {W} {W} 
\hskip-3.5pt\rrrule {W} {W} {W} {W} {W} {W} {W} }\hskip-3pt,
\hfill}
\ligne{\hfill
\hbox{$(c_4)$ \llrule {W} {X} {Y} {Z} {W} {W} {W} 
\hskip-3.5pt\rrrule {W} {W} {W} {W} {W} {W} {W} }\hskip -3pt,
\hbox{$(c_5)$ \llrule {W} {X} {Z} {Y} {W} {W} {W} 
\hskip-3.5pt\rrrule {W} {W} {W} {W} {W} {W} {W} }\hskip -3pt.
\hfill}
}}$
\hfill (\numerrel)\hskip 10pt}
\vskip 5pt
Taking into account rotation invariance,
there are 4 rules of the form~$(a)$, 10 of them of the form~$(b)$. There are 4 rules
for each form $(c_1)$, $(c_4)$ and $(c_5)$ and 6 of them for $(c_2)$ and $(c_3)$.
so that we get 38 rules to which we may append the rule
\hbox{$(w)$ \llrule {W} {W} {W} {W} {W} {W} {W} 
\hskip-3.5pt\rrrule {W} {W} {W} {W} {W} {W} {W} }\hskip -3pt, which simply says that
\sww {} is a quiescent state. As will be seen later, all these rules are not needed.
We need only 19 of them as can be checked in Table~\ref{trall}, the part devoted to
the conservative rules for \sww.

However, the rules deduced from~(4) allow us to prove an interesting result:

\begin{prop}\label{pdecid}
Consider a cellular automaton $A$ whose program is rotationally coherent and which
contains the rules~$(4)$.
	Then, the halting of the computation of~$A$ starting from a 
finite configuration is decidable.
\end{prop}

\noindent
Proof. Under the hypothesis of the proposition, $A$ contains all possible rules of~(4)
modulo rotation invariance.
Fix a non-blank cell $T_0$ of the initial configuration of a computation of~$A$.
Let $\mathcal B$ be the smallest ball around $T_0$ outside which all cells are blank at 
initial time of the computation. Let $k$ be its radius and let $U$ be a tile at 
distance~$k$ of~$T_0$. Consider~$V$ a neighbour of~$U$ which does not belong 
to~$\mathcal B$. By construction, $V$ is at 
distance~$k$+1 and it is blank. The neighbourhood of~$V$ contains at most three 
neighbours which belong to $\mathcal B$. That assertion is entailed by our study of the 
neighbours of generation~1, 2 and~3 around a tile. Accordingly, at least one of the rules
from~(4) apply to~$V$ which remains blank. That argument can be repeated to any
tile at distance~$k$ from $T_0$, so that the cells outside $\mathcal B$ always remain
blank. Accordingly, the computation remains inside $\mathcal B$. Consequently, either
the computation halts at some time or it becomes periodic after a certain time: both
situation are algorithmically detectable.\hfill$\Box$

    Proposition~\ref{pdecid} is not contradictory with Theorem~\ref{letheo}: indeed,
\TT{} does not satisfy Proposition~\ref{pdecid} as far as it rules out a single rule
of~(4) replacing it by a rule
of the form \hbox{\llrule {W} {X} {W} {W} {W} {W} {W} 
\hskip-3.5pt\rrrule {W} {W} {W} {W} {W} {W} {Y} } , where \ftt{X} {} and \syy {} are
non blank, as will further be seen. Accordingly, withdrawing one carefully chosen rule
from~(4) is enough to establish the strong universality of an appropriate computation. 

However, as far as \TT{} contains all rules of~(4) of the forms~$(b)$ and~$(c)$ and all
rules of~$(a)$ but one, we may consider that a blank cell~$V$ which is a neighbour
of generation~2 or~3 of non-blank cells of the configuration remains blank as long 
as~$V$ remains a neighbour of generation~2 or~3 of those cells. That property allows us
to restrict the simulation to neighbours of generation at most~2 of cells of the 
configuration. We shall always check that such an assumption is sound. Later on,
the rules are referred to Table~\ref{trall} by the number they are displayed with there.
} \fi

\newcount\compteregle\compteregle= 0
\def\gzzz{\global\advance\compteregle by 1}
\def\vszz{\vskip-2pt}
\newdimen\largeouille\largeouille=87pt

\subsection{The rules for the tracks and the switches}\label{rtracks}

    We have already met the rule which defines the quiescent state: a state which is 
unchanged if the cell and all its neighbours are in that state. In \TT, \sww {} is
the quiescent state and \TT{} contains the corresponding rule, rule~1{} in 
Table~\ref{trall}.

   We start our study of the rules and the construction of the rules by those which
manage the tracks. The motion rules are simple as they implement the following abstract 
scheme:
\vskip 5pt
\ligne{\hfill\ftt{
   WWW\hskip 20pt LWW\hskip 20pt WLW\hskip 20pt WWL}
\hfill(\numerrel)\hskip 10pt}
\vskip 5pt

To the left of (5), we have the conservative situation, rule~19.
Then, the locomotive enters the window, rule~20, and, in the two 
successive steps, rule~23 and~24 it crosses the window. Then, it witnesses the occurrence
of the locomotive in the next element of the track, rule~24 and, which, at the next step,
the cell recovers the conservative condition requiring rule~19. The implementation of 
those rules in the dodecagrid are easy. Figure~\ref{ftracks} allows us to check the rules
devoted to the tracks in Table~\ref{trall}, namely rules~19 up to~25. 
Note that rules~29 up to~32 deal with that element which allows a locomotive to go
from on tile in~\HH$_u$ into the tile which hangs below it in~\HH$_b$. For that cell, we 
remark the occurrence of state~\srr{} for its neighbour~3.

First, let us look at the conservative rules. They mainly deal with the decorations
of the tiles of a track. A cell belonging to the decorations is \sbb{} an it is 
surrounded by white cells so that rule~16 applies. Also, rules~17 and~18 apply: the 
decoration remains in place during the computation. The rules indicate that the milestone
witnesses the occurrence of the locomotive in the cell. If we look carefully at 
Figure~\ref{ftracks}, tiles on faces~6 and~10 do not see each other but, as known from 
the proof of Proposition~\ref{pneighgen} they have a common neighbour which is 
white and which must remain white: whence rules~2 or~3 are needed. Note at that occasion
that rules~2 and~3, although there is a single~\sbb-cell in their neighbourhood
are not rotation half-invariant. The minimal form of rule~2 is
\hbox{\llrule {W} {W} {W} {W} {W} {W} {W} 
\hskip-3.5pt\rrrule {W} {W} {W} {W} {R} {W} {W} } {} while that of rule~3 is 
\hbox{\llrule {W} {W} {W} {W} {W} {W} {W} 
\hskip-3.5pt\rrrule {W} {W} {W} {W} {W} {R} {W} } {}.
Nonetheless, those rules are rotation invariant: in~\cite{mmarXiv21a}, rule~2 only 
is present.
Note that rule~21 is needed for the passive fixed switch when the locomotive arrives at
it from the left-hand side. Rules~20 and~21 are not rotation half-invariant but
they are rotation invariant. For the tunnel, as far as we need an entry through the face~1
of an element of the track, rule~25 allows that possibility.
The special element of the track with a red cell in its decoration requires rule~29 when
it is idle. The red cell of its decoration requires rule~28 for that cell. It also 
requires rules~4 and~5 for a white neighbour. But rules~12 and~13 are needed too for
a white neighbour which can also see the blue cell on the face~9 of that element of the
track.

\newdimen\svhsize
\newdimen\lalongue\lalongue=170pt
\def\traceline #1 #2 {%
        \svhsize=\hsize
        \hsize=\lalongue
	\ligne{\hbox to 20pt{\hfill\ftt{#1} }\hskip 20pt\hbox{\ftt{#2} }\hfill} {}
        \hsize=\svhsize
\vskip -3pt
}
\vskip 10pt

Rules~33 up to~39 deal with the controller. Rule~33 prevents a locomotive to enter an 
element of the track when the element can see a blue cell through its face~0.
Rule~34 concerns the cell~$\kappa$ which directly leads to the controller: when the 
controller is blue the locomotive enters~$\kappa$. Rule~35: $\kappa$ can see the 
locomotive enter the controller when that one is white. Rules~36 up to~39 deal with the 
controller~$\xi$ itself. Rules~36 and~37 are the conservative rules: 
when $\xi$ is white it remains white, when it is blue, it remains blue.
Rule~38 change~$\xi$ from \sww{} to \sbb, while rule~39 performs the opposite 
transformation. In both cases, the change happens because $\xi$ can see the locomotive
in~$\kappa$ through its face~5. 
Rules~40 and~41 are introduced for technical purposes in the simulating program.

Rules~42 up to~46 deal with the fork. A single locomotive arrives at the fork and two
of them leave it. Rule~43 is the conservative rule for the central cell~$\varphi$ of the
fork. Rule~44 says that a locomotive arrives at~$\varphi$ which can see it through its
face~2. Rule~45 witnesses the occurrence of the locomotive in~$\varphi$. Rule~46
says that two locomotive leave~$\varphi$: one of them is seen from the face~5 of~$\varphi$
the other one is seen through its face~4. Rule~42 also is needed: it deals with 
the \sbb-cell~$\beta$ which is put on face~6 over~$\varphi$. When the locomotive is 
in $\varphi$, $\beta$ which is always\sbb{} can see the locomotive through its face~0
and it also can see a decoration of the element of track which is seen by $\varphi$
through its five~5. Whence rule~42.

   Note that there are no special rules for the switches outside those we indicated 
up to now, even for the fixed switch as far as for that switch, we already mentioned
rule~21 used when the locomotive arrives at the switch from its left-hand side.

  The main features of the switches are the organization of their needed 
circuits in the dodecagrid. We have seen that the elements of the track are enough for 
that purpose. The rules for the fork and for the controller complete the needed set of
rules.

\subsection{The rules for the register}\label{rreg}

   The present part of~\TT{} is the biggest as all the remaining rules are devoted to
the registers: it contains 207 rules from the 251 ones of~\TT.

First, in Sub-subsection~\ref{sbbrstrands} we examine the conservative 
rules for each strand of a register together with the few rules allowing the 
motions in the strand described in Sub-section~\ref{newrailway}. 
Then, in Sub-subsection~\ref{sbbrgrow} we give the rules for the growth of the 
register. Next, we study the decrementation in Sub-section~\ref{sbbrregdec}.
After that, Sub-subsection~\ref{sbbrreginc} manages the incrementation.
At last, Sub-subsection~\ref{srstop} deals with the halting of the
computation which concerns the register as a specific strand is devoted to that
operation.

\subsubsection{The strands of a register}\label{sbbrstrands}

   Before looking at each strand separately, we have to look at the implementation
we give to the strands. 

    We use the notations introduced in Sub-subsection~\ref{sbbreg}. Let us remind the 
reader that the four strands of a register are \RR$_c$, \RR$_i$, \RR$_d$
and \RR$_s$. Two of them, \RR$_c$ and~\RR$_d$ are in~\HH$_u$ while the two others,
\RR$_i$ and~\RR$_s$ are in~\HH$_b$. Each $\mathcal S$($n$) has its face~0 on~\HH{}
and for each of them, its face~1 has a side on a line~$\ell$ of~\HH{} attached to that
register. That fixes the numbering as already mentioned. Note that for strands 
in~\HH$_u$ the numbering is increasing while clockwise turning around the projection of 
the cell. For the strands seen through a translucent \HH{} the numbering is 
counterclockwise increasing. Note that face~0 is the same for \RR$_c$($n$) and for 
\RR$_i$($n$) for any $n$ in $\mathbb N$. Similarly, \RR$_d$($n$) and \RR$_s$($n$) share 
the same face~0 for the same values of~$n$.

A cell has at least four important neighbours : a cell of the same coordinate belonging 
to another strand seen from its faces~0 and~1 and two cells of the same strand seen 
from its faces~5 and~2. We append two decorations to each $\mathcal S$($n$) : a blue 
cell on its faces~4 and~6 for \RR$_i$ and \RR$_d$, on its faces~3 and~7 on \RR$_c$ 
and~\RR$_s$. 

The main reason for those additional decorations is to ensure the rotation 
half-invariance of the rules. Let us now explain why namely those faces are involved. 
To better understand that, it is necessary to describe how the register is growing. We 
know
that \RR$_i$ and \RR$_d$ are \srr{}  and that \RR$_s$ is \sbb. For what is 
\RR$_c$, we know that its beginning is \sww{} unless its value $c$ is~0{} in which case
it is \sbb{} in every cell. However, as already mentioned, as far as the growing process
is constant, the \sbb-part of \RR$_c$ is much bigger than its \sww-beginning, so that
close to its \sgg-end, \RR$_c$ is also \sbb. Let $N_t$ be the coordinate such that, 
at time~$t$, $\mathcal S$$(n)$ is \sww{} if $n>N_t$ and that it is the colour of its 
strand if $n < N_t$. For $n=N_t$, $\mathcal S$$(n)$ is \sgg, as already mentioned. 
At time $t$+1, each neighbour of~$\mathcal S$$(n)$ such that its unique 
non-blank neighbour is \sgg{} becomes~\sbb{} while $\mathcal S$$(n)$ remains \sgg. 
To decide what to do at time~$t$+2, it is necessary to closer look at the situation. 
Two figures help us to see the configuration of the end of the register: 
Figures~\ref{fcut} and~\ref{fregfin}.

The former figure is a cut along a plane~\VV{} which is perpendicular to~$\ell$ and which
contains a face of each $\mathcal S$$(n)$ for some~$n$. On the right-hand side of the
figure, we have the representation on~\HH{} while, on the left-hand side, we have
the projections of the tiles we can see. Remember the condition we have given for 
a neighbour $S_i$ of $\mathcal S$$(n)$: $S_i$ must have eleven white neighbours while
its single non-blank neighbour is $\mathcal S$$(n)$ itself whose colour is~\sgg.
Which faces of $\mathcal S$$(n)$ satisfy such a condition? 

\vskip 10pt
\vtop{
\ligne{\hfill
\includegraphics[scale=1]{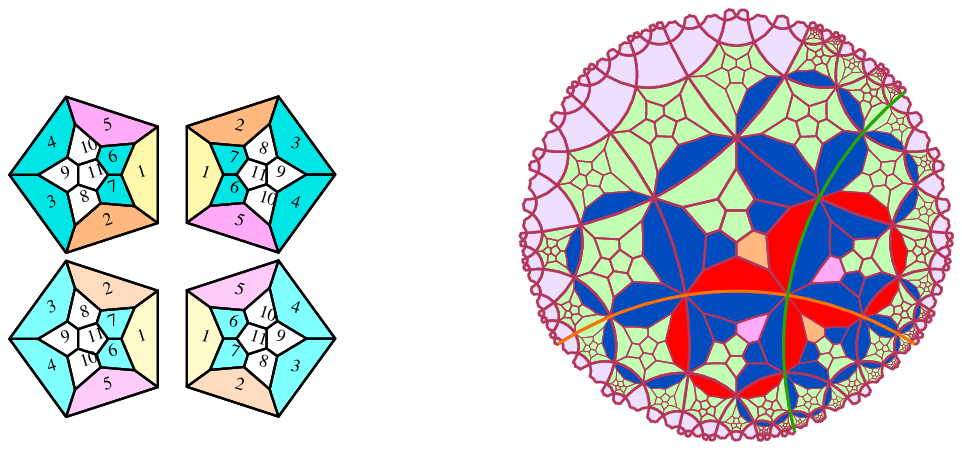}
\hfill}
\begin{fig}\label{fcut}
\leurre
Cut of the strands of a register along a plane~\VV, perpendicular to~$\ell$.
Note the correspondence of colours between the left- and right-hand sides of the figure.
\end{fig}
}
\vskip 5pt
The figure indicates that on all cells, {\it a priori}, we have all faces except faces~0,
and~1, and also face~2 for~\RR$_c$ and \RR$_s$ and face~5 for \RR$_i$ and \RR$_d$. Denote
by $S_i$ the $i$-neighbour of $\mathcal S$($n$). We have that $S_7$ and $S_8$ of each
$\mathcal S$($n$) can see $S_6$ and $S_{10}$ respectively on $\mathcal S$$(n$$-$$1)$ 
for \RR$_i$ and \RR$_d$, on $\mathcal S$$(n$+$1)$ for \RR$_c$ and \RR$_s$. Moreover,
$S_3$ of each $\mathcal S$($n$) can see $S_4$ on $\mathcal S$($n$$-$1) for \RR$_i$
and \RR$_d$ or on $\mathcal S$($n$+1) for \RR$_c$ and \RR$_s$. Moreover, 
$S_3$ and $S_4$ of $\mathcal S$($n$) for \RR$_c$, \RR$_d$ can see the $S_i$ of 
$\mathcal S$($n$) of the same index~$i$ for \RR$_i$, \RR$_s$ respectively. Now, 
on \RR$_c$ and \RR$_s$, $S_7$ of $\mathcal S$($n$) can see $S_6$ of \RR$_d$ and \RR$_i$
respectively for the same coordinate. If we wish to obtain \sgg{} on 
$\mathcal S$($N_t$+1), we must take advantage of the fact that the $S_2$ of \RR$_c$ 
and \RR$_s$ can see the $S_5$ of both \RR$_i$ and \RR$_d$, similarly for the $S_5$ of \RR$_i$ and \RR$_d$
with the $S_2$ of both \RR$_c$ and \RR$_s$. Accordingly, each of those $S_i$'s have 
two \sbb-neighbours and they all can see a \sgg-cell. We can then decide that
in such a case the \sbb-cell becomes \sgg{} and that the former \sgg-cell takes the
colour of its strand. We remain with what to do with the other \sbb-neighbours of
the \sgg-$\mathcal S$$(N_t)$. From what we just remarked, $S_6$ or $S_7$, and also $S_3$ 
or $S_4$, it depends on which strand the cell is, is blue and can see another blue 
neighbour. We can decide that such a \sbb-cell remains \sbb. 

If we do that, what can be said? In that case, we can see that on \RR$_c$ or \RR$_s$,
an $S_6$ neighbour of the \sgg-cell which is white remains white as far as it can see
the \sgg-cell and a \sbb-decoration on $\mathcal S$($N_t$$-$1). Accordingly, that cell
remains white. We have the same argument with $S_4$ for the same cell. A similar argument
holds for $S_7$ and $S_3$ on the \sgg-cell of \RR$_i$ and of \RR$_d$. Accordingly,
at time $t$+1 we can decide that the \sbb-cells which can see their \sgg-neighbour as
their unique non-blank neighbour become white. Accordingly, the new cell of the strand
receives the same decoration as its neighbour $\mathcal S$($N_t$$-$1) on the strand.
All other neighbours are white outside its neighbours~0, 1, 2, 5 and outside the 
decorations, neighbours~3 and~7 on \RR$_c$ and on \RR$_s$, neighbours~4 and~6 on \RR$_i$
and on \RR$_d$.

Those conclusions hold for $\mathcal S$($n$) with $n>0$. For $n=0$ it is important that
the cell knows that it is the first one of the strand. As the cell cannot see the 
decoration of its neighbours, that information of being the first cell must be given
in the decoration. On one hand, \RR$_c$ and \RR$_i$ receive one additional 
blue cell on face~9. On another hand, \RR$_d$ and \RR$_s$ receives an additional red cell
on face~9. 

We gather that information on the neighbours of $\mathcal S$($n$) as follows:
\vskip 10pt
\ligne{\hfill
$\vcenter{\vtop{\leftskip 0pt\parindent 0pt\hsize=225pt
	\ligne{\hfill$\mathcal S$($n$), $n>0$:\hfill}
\ligne{\RR$_c$ : 0: \srr, 1: \syy, 2:\sbb$^*$, 3:\sbb, 5:\sbb$^*$, 7: \sbb,
others: \sww;\hfill}
\ligne{\RR$_i$ : 0: \sbb$^*$, 1: \sbb, 2:\srr, 4:\sbb, 5:\srr, 6: \sbb, 
others: \sww;\hfill}
\ligne{\RR$_d$ : 0: \sbb, 1: \sbb$^*$, 2:\syy, 4:\sbb, 5:\syy, 6: \sbb, 
others: \sww;\hfill}
\ligne{\RR$_s$ : 0: \syy, 1: \srr, 2:\sbb, 3:\sbb, 5:\sbb, 7: \sbb, 
others: \sww;\hfill}
	\ligne{\hfill$\mathcal S$(0):\hfill}
\ligne{\RR$_c$ : 0: \srr, 1: \syy, 2:\sbb$^*$, 3:\sbb, 5:\sww, 7: \sbb,
9: \sbb, others: \sww;\hfill}
\ligne{\RR$_i$ : 0: \sbb$^*$, 1: \sbb, 2:\sww, 4:\sbb, 5:\srr, 6: \sbb, 
9: \sbb, others: \sww;\hfill}
\ligne{\RR$_d$ : 0: \sbb, 1: \sbb$^*$, 2:\syy, 4:\sbb, 5:\syy, 6: \sbb, 
9: \srr, others: \sww;\hfill}
\ligne{\RR$_s$ : 0: \syy, 1: \srr, 2:\sbb, 3:\sbb, 5:\sbb, 7: \sbb, 
9: \srr, others: \sww;\hfill}
}}$
\hfill(\numerrel)\hskip 10pt}
\vskip 10pt
Note that in (6), the asterisk indicates that \sbb{} may be replaced by \sww: remember
that on~\RR$_c$, the value~$v$ stored in the register is indicated by $v$ contiguous
blank cells starting from \RR$_c$(0) which, consequently, is \sbb{} if and only if 
$v=0$. That possibility also concerns \RR$_i$ and \RR$_d$ which can see \RR$_c$ through 
their face~0 and~1 respectively. It does not concern \RR$_s$ which cannot see \RR$_c$.

\vtop{
\ligne{\hfill
\includegraphics[scale=1]{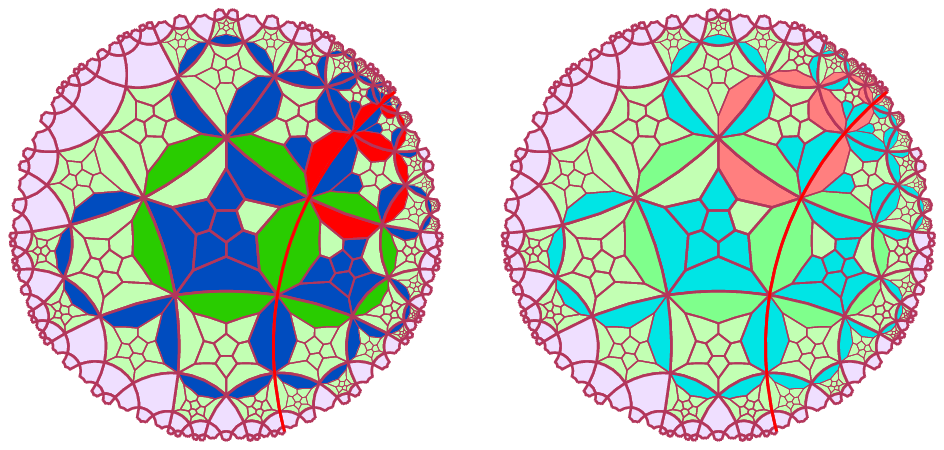}
\hfill}
\begin{fig}\label{fregfin}
\leurre
The idle configuration of the growing end of a register at the time when blue cells cover
the green ones. To left, in~\HH$_u$, to right, in~\HH$_b$ through a translucent~\HH. 
Note the line~$\ell$, in red in the pictures.
\end{fig}
}

Figure~\ref{fregfin} illustrates the configuration of the end of the register
when the \sgg
\ligne{\hfill}

\noindent
-cells are covered by \sbb-neighbours. We can see the faces which are 
covered with a blank cell. The left-hand side part of the figure illustrates \RR$_c$
and \RR$_d$, the strands in~\HH$_u$. The right-hand side part of the figure illustrates
\RR$_i$ and \RR$_s$ which are in~\HH$_b$ and which are viewed through a 
translucent~\HH. The figure applies the convention for colouring the faces of the
dodecahedrons we already defined. It can be seen that the cells of \RR$_c$ and of \RR$_s$ 
are blue, while those of \RR$_i$ and of \RR$_d$ are red.

\vskip 10pt
\vtop{
\ligne{\hfill
\includegraphics[scale=1]{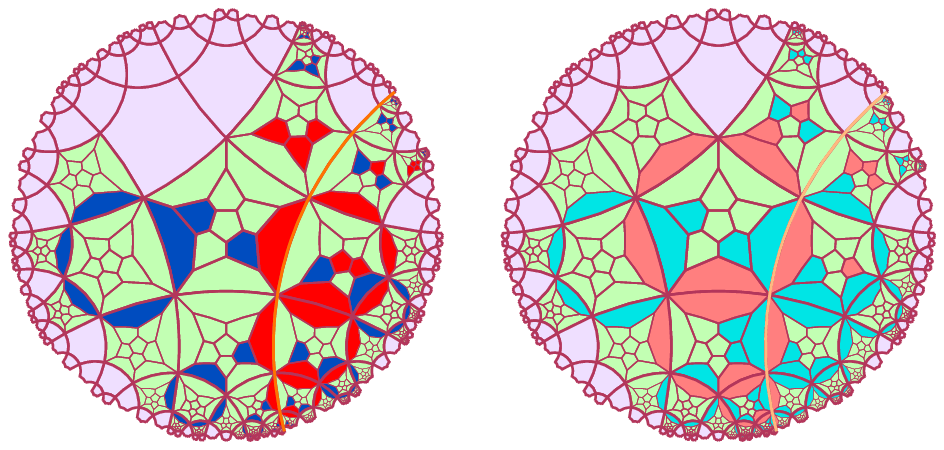}
\hfill}
\begin{fig}\label{fregdeb}
\leurre
The idle configuration of $\mathcal S$$(n)$ for $n\in\{$$-$$1,0..3\}$ for \RR$_c$,
\RR$_i$, \RR$_d$ and \RR$_s$.
To left, the strands in~\HH$_u$, to right, those which
are in~\HH$_b$ viewed through a translucent~\HH. Note the line~$\ell$, in orange in 
the figures.
\end{fig}
}
\vskip 10pt

In the present paper, the strands of a register have a different working from that 
of~\cite{mmarXiv21a}. \RR$_c$ still contains the value~$c$ of the register represented
by $c$ \sww-cells from \RR$_c$(0) up to \RR$_c$($n$-1) when $n>0$. The cells \RR$_c$($n$)
are \sbb{} when $c\geq n$. The locomotive arrives at~\RR$_i$(0) for a decrementation, 
at \RR$_d$(0) for a decrementation. When the operation has been performed, the locomotive
exits through \RR$_c$ in the case of an incrementation, through \RR$_s$ in the case
of a decrementation. It is a main difference with~\cite{mmarXiv21a}. But the stopping
locomotive also arrives at \RR$_s$(0). Accordingly, two tracks abut \RR$_s$(0): one 
is used for stopping the computation,the other is used when the locomotive returns to the
program after a successful decrementation. When the decrementation cannot be performed,
it can be noted by \RR$_d$(0) which can see an arriving locomotive at \RR$_d$(-1) and
\sbb{} in \RR$_c$(0), indicating that $c=0$. Accordingly, two tracks abut \RR$_d$(0) as
well: one for a decrementing locomotive arriving at \RR$_d$, the other for the returning
track when the decrementation cannot be performed. That latter track is the first part
of the $Z$-track leading to the \DDD{} of the register. The first tile of the $Z$-track
is denoted by \RR$_s$(-2)$z$.

That implementation is illustrated by Figure~\ref{fregdeb} on which we also can see the
decoration of the various $\mathcal S$(-1). In (7), following the conventions of~(6),
we summarize what is illustrated by Figure~\ref{fregdeb} as follow, including 
\RR$_s$(-2)$z$:

\vskip 10pt
\ligne{\hfill
$\vcenter{\vtop{\leftskip 0pt\parindent 0pt\hsize=160pt
	\ligne{\hfill$\mathcal S$(-1):\hfill}
\ligne{\RR$_c$ : 2, \sbb$^*$, 6, 9, 10: \syy, others: \sww;\hfill}
\ligne{\RR$_i$ : 2: \srr, 6,9: \sbb, 7,10: \srr, others: \sww;\hfill}
\ligne{\RR$_d$ : 2: \srr, 6,9: \sbb, 10:\srr, others: \sww;\hfill}
\ligne{\RR$_s$ : 4: \sbb, 6,9,10: \sbb, 7: \srr, others: \sww;\hfill}
	\ligne{\hfill$\mathcal S$(-2):\hfill}
\ligne{\RR$_d$(-2)$z$ : 7,8,9,11: \srr, others: \sww;\hfill}
\ligne{\RR$_s$(-2)$s$ : 6,9,10: \sbb, 8: \srr, others: \sww;\hfill}
\ligne{\RR$_s$(-2)$d$ : 6,9,10: \sbb, 11: \srr, others: \sww;\hfill}
}}$
\hfill(\numerrel)\hskip 10pt}
\vskip 10pt
We indicated the cell \RR$_d$(-2)$z$ together with the cells \RR$_s$(-2)$s$ and
\RR$_s$(-2)$d$. In fact, two tracks abut \RR$_d$(-1): one of them is an ordinary track
ending the path coming from the \DDD{} of the register, the other is the first track of 
the $Z$-path whose first cell is \RR$_d$(-2)$z$. Also, two tracks abut \RR$_s$(-1): 
one of them is the last track of the stopping path coming from the program part of the circuit, its last cell is \RR$_s$(-2)$s$; the other is the first track
of the returning path to \DDD{} after a successful decrementation, its first cell
is \RR$_s$(-2)$d$. The justification of the patterns indicated in~(7) lies in the
rules to which we now turn.

Accordingly, let us look at the rules. We start with the conservative ones which allow 
the structure to remain unchanged as long as the locomotive is not present. We also append
the motion rules for the locomotive running on the strands.

\vskip 10pt
In order to check the rules, it is enough to combine the information given in
(6), (7) and (5), knowing that the concerned neighbours are 0, 1, 3, 4, 6, 7 and 9 
for (6), we have to consider also neighbours 8 and 11 for~(7) and, at last, the concerned 
neighbours for~(5) are~2 and~5. Of course, we have to take into account the direction
of the motion, in different directions on \RR$_d$ and on \RR$_s$.

Rules~47 up to~69 deal with $\mathcal S$(-1) and rules~70 up to~76 deal with two cells
of $\mathcal S$(-2), those which belong to \RR$_s$.

Rules~47 up to~52 manage the motion on~\RR$_c$(-1): after the conservative rule~47,
we have the witness one~48 where the cell can see the arriving locomotive in \RR$_i$(-1).
The other rules are also witness rules, except rules~50 and~52 which make the locomotive
get out of the cell, a situation which occurs for locomotive which returns after 
performing an incrementation.

Rules~53 up to~57 deal with \RR$_i$(-1) while rules~58 up to~62 deal with \RR$_d$(-1).
The first group of rules manages the arrival of an incrementing locomotive, the
second one manages the arrival of a decrementing locomotive. Rule~53, 58 are the 
conservative rules for \RR$_i$(-1), \RR$_d$(-1) respectively, in both cases indicating 
that the cell can see that \RR$_i$(0), \RR$_d$(0) respectively is \srr. 

Rules~77 up to~114 deal with $\mathcal S$($n$) when \hbox{$n\in\mathbb N$}. We have rules
both when $n=0$ and when $n>0$. When $n=0$ we can notice the suffixes \ftt{WBWW} {} and 
\ftt{WRWW} for \RR$_c$(0), \RR$_i$(0) and \RR$_d$(0), \RR$_s$(0) respectively in the 
neighbourhood of the corresponding rules. We also notice the prefixes of the same
neighbourhoods: \ftt{RR} for \RR$_c$ and \RR$_s$, \ftt{WB} for \RR$_i$ and \ftt{BW} for
\RR$_d$ when $n<c$, where $c$ is the value stored in \RR$_c$. The prefix is \ftt{BB} for
both \RR$_i$ and \RR$_d$ when $n\geq c$ but we note that property in other rules.

Rules~77 up to~90 deal with \RR$_c$, considering also the motion of an \srr-locomotive
toward \RR$_c$(0) as far as \RR$_c$ is also the return track for a locomotive which
performed an incrementation. Note that there are rules for \RR$_c$(0), from~77 up to~80 
together with rule~90 and rules for \RR$_c$($n$), $n>0$, from~81 up to~89.
Rules~91 up to~98 deal with \RR$_i$ with a similar decomposition, rules~99 up to 108 deal
with \RR$_d$. Rules~109 up to~114 deal with \RR$_s$: they also describe the motion of a
stopping locomotive which is \sgg. It arrives as a \sbb-locomotive at \RR$_s$(-2) and then
at \RR$_s$(-1) but it enters \RR$_s$(0) as a \sgg-locomotive. The reason for that is
simple: the locomotive in \RR$_s$ cannot be \sbb{} as far as the cells of that strand
are already \sbb. It cannot be \srr{} as far as it would interfere with the motion
on \RR$_c$ of a locomotive which returns to~\DDI{} after performing the incrementation
of the register. Accordingly, we remain with a \sgg-locomotive: rule~110 make the cell
becoming \sgg{} as far as it can see the locomotive in \RR$_s$(-1) through its face~5.
Rule~111 make the cell return to its \sbb-state and rules~112 up to~114 deal with the
same motion of the \sgg-locomotive when it crosses \RR$_s$($n$) with $n>0$.

\subsubsection{The growing end of the register}~\label{sbbrgrow}

   We have studied the process in the previous sub-subsection thanks to 
Proposition~\ref{pneighgen} and thanks to Figures~\ref{fregdeb} and~\ref{fregfin}.
We revisit the argument given there on checking the rules devoted to that point.

   The rules for the growth of the registers in Table~\ref{trall} are from~115 up to~126.
Rule~115 is the conservative rule for \RR$_cS$ and for \RR$_s$ while rule~122 is the
conservative rule for \RR$_i$ and for \RR$_d$. Rule~6 says that a white cell with a single
non-blank neighbour which is \sgg{} becomes \sbb. Accordingly, rules~116 and~123 
illustrate that situation for \sgg{} which, at that occasion, takes the state of its 
strand, \sbb{} and \srr{} respectively. Rule~117 says that a \sbb-cell whose unique 
non-blank neighbour is \sgg{} becomes \sww. Rule~118 says that if the cell can see two
\sbb-cells and if its other neighbours are \sww{} the cell remains \sbb. If it can see
one \sgg-neighbours and two \sbb-ones, the others being \sww, then the cell become \sgg:
it said by rules~120 and~121: two rules as far as the place of~\sgg{} among the neighbours
is not the same also with respect to the rotations considered in the paper. At last, 
rule~119 says that if a \sbb-cell has two non-blank neighbours, one of the \sgg{} and the
other~\sbb, then the cell remains \sbb. Moreover, the \sgg{} neighbour is seen through
face~0.

\subsubsection{Decrementation of the register}\label{sbbrregdec}

   That operation is the most difficult to solve for the simulation. We formulate a
general principle which we shall as far as possible implement in all cases. 
A blue locomotive arrives at \RR$_d$(0) and it goes along \RR$_d$($n$) with increasing~$n$
as long as it can see \sww{} in \RR$_c$($n$) through its face~1. When the \sww-cell of
\RR$_c$($c$-1), where $c$ is the value stored on \RR$_c$, can see the locomotive
in \RR$_d$($c$-1), it becomes \sbb{} which performs the decrementation. At the same time,
\RR$_d$($c$) turns from~\srr{} to \sgg{} which triggers a red locomotive on \RR$_s$($c$).
That locomotive goes to \RR$_s$(-1) where it is sent as a blue locomotive on the
return path to \DDD.

We have three different cases: the general case, when the content~$c$ of the register is
large enough, then when $c=2$, when $c=1$ and, not at all the least, when $c=0$.
By large enough we mean that the detection of the first blue cell on \RR$_c$ happens
on a cell which cannot be seen by \RR$_c$(0). We also require a bit more: that all 
possible motion rules of the locomotive could be tested on that white portion of the
register. The case $c=2$ is particular as the detection of the first blue cell
can be seen from \RR$_c$(0). For the same reason, $c=1$ is a particular cell. For what
is $c=0$, it is the case when the decrementation cannot be performed which, of course, 
requires a special treatment. The general principle is depicted by~(8).

\vskip 10pt
\ligne{\hfill
\vtop{\leftskip 0pt\parindent 0pt\hsize=310pt
\ligne{\hfill\small \RR$_d$\hskip 94pt \RR$_c$\hskip 94pt \RR$_s$\hfill}
\vskip 5pt
\setbox110=\vbox{
\ftt{
* 0 1 2 3 4 5 6 7\hskip 30pt * 0 1 2 3 4 5 6 7\hskip 30pt * 0 1 2 3 4 5 6 7\vskip 3pt
0 W W W W W W B B\hskip 30pt 0 B W W R R R R R\hskip 30pt 0 W W W B B B B B\vszz
1 W W W W W W B B\hskip 30pt 1 W B W R R R R R\hskip 30pt 1 W W W B B B B B\vszz
2 W W W W W W B B\hskip 30pt 2 W W B R R R R R\hskip 30pt 2 W W W B B B B B\vszz
3 W W W W W W B B\hskip 30pt 3 W W W B R R R R\hskip 30pt 3 W W W B B B B B\vszz
4 W W W W W W B B\hskip 30pt 4 W W W R B R R R\hskip 30pt 4 W W W B B B B B\vszz
5 W W W W W W B B\hskip 30pt 5 W W W R R B R R\hskip 30pt 5 W W W B B B B B\vszz
6 W W W W W B B B\hskip 30pt 6 W W W R R R G R\hskip 30pt 6 W W W B B B B B\vszz
7 W W W W W B B B\hskip 30pt 7 W W W R R R G R\hskip 30pt 7 W W W B B B R B\vszz
8 W W W W W B B B\hskip 30pt 8 W W W R R R R R\hskip 30pt 8 W W W B B R B B\vszz
9 W W W W W B B B\hskip 30pt 9 W W W R R R R R\hskip 30pt 9 W W W B R B B B\vszz
0 W W W W W B B B\hskip 30pt 0 W W W R R R R R\hskip 30pt 0 W W W R B B B B\vszz
1 W W W W W B B B\hskip 30pt 1 W W W R R R R R\hskip 30pt 1 W W R B B B B B\vszz
2 W W W W W B B B\hskip 30pt 2 W W W R R R R R\hskip 30pt 2 W W W B B B B B\vszz
}
}
\ligne{\hfill$\vcenter{\box110}$
\hfill(\numerrel)\hskip 10pt}
}
\hfill}
\vskip 10pt
In all cases, the locomotive arrives to the register thanks to a path arriving at
\RR$_d$(-1) through its face~4. The decoration of that cell is \ftt{BWWBRW} {} where 
the leftmost occurrence of \sbb{} corresponds to its neighbour~6, as can be seen in~(7).
The motion of the locomotive on~\RR$_d$ is managed by rules~101 up to~103 for \RR$_d$(0)
and by rules~105 up to~107 for \RR$_d$($n$) with \hbox{$<n<c$}. Rules~99 and~100 are
conservative rules for \RR$_d$(0) when $n=0$ and when $n>0$ respectively. Rule~104
is a conservative rule for a cell \RR$_d$($n$) with \hbox{$0<n<c$}. Rule~109 witnesses the
occurrence of the \sgg-locomotive in \RR$_s$(0). Note that \sgg{} is the single state
which can be used at that stage of the computation in order to indicate to \RR$_s$ that
a locomotive must be raised for returning to~\DDD.

    Rules~126 up to~130 deal with \RR$_s$(-2)$d$. The decoration is given by the suffix
 \ftt{BWWBBR} of the neighbourhood the neighbour of those rules. The cell deals
 with the returning locomotive: it is an \srr-one and it leaves the cell as a 
\sbb-locomotive. The rules simply convey the locomotive to the track ,
rule~126 being purely conservative. Rule~60 makes the locomotive enter \RR$_d$(0).
Rules~128 and~129 convey its crossing of \RR$_d$(0). Rule~60 reminds us that the entry
to~\RR$_d$(0) is through the face~4 of the cell and rule~137 tells us that en entry
through face~5 us ruled out.
\vskip 10pt
    Rules~135 and~136 realize the scheme indicated in~(8): rule~136 performs the
decrementation by transforming the last \sww-cell of \RR$_c$ into a \sbb-one; rule~136
triggers the signal which will make the returning locomotive start on~\RR$_s$.
Rule~131 says that \RR$_s$(0) witnesses the \sbb-locomotive in \RR$_d$(0). Rules~132 and 
133 are conservative rules for \RR$_c$(0) and \RR$_c$($n$) respectively with $n>0$ in 
that latter case. Rule~134 witnesses the motion of the returning locomotive in \RR$_d$ 
by \RR$_s$($n$) with $n>0$. Rule~141 triggers the start of the returning locomotive
moving on \RR$_s$. Rule~139 keeps the \sgg-cell unchanged for one step in order that
rule~142 might apply which allows the cell to return to its \sbb-state. The rule indicates
that the \sgg-state in the cell made many \sww-neighbours turn to~\sbb: namely 
neighbours 8, 9, 10 and~11. Rule~143 and 147 manage the motion of the locomotive on
the cells \RR$_s$($n$) with $n>0$. Rule~144 also manage the motion, witnessing the 
occurrence of the \sgg-cell on~\RR$_d$. Rules~149 up to~152 deal with \RR$_s$(0) which
welcomes the locomotive and which witnesses its leaving to~\RR$_s$(-1). The other
rules witness the motion of the locomotive which is seen from \RR$_d$ and also from \RR$_i$
as well. In particular, rule~153 for \RR$_d$(-1) witnesses the occurrence of the still
red locomotive in \RR$_s$(-1).

Two additional rules are needed when $c=2$. As indicated by the \ftt{WRBBW } suffix
of the neighbourhoods in rule~154 deal with the cell \RR$_s$(-2)$s$ which remains~\sww.
Rule~155 deals with \RR$_s$(-1) and witnesses the occurrence of the locomotive on the
first cell of the track leading to~$DDD$.

Three additional rules manage the case $c = 1$. They deal with the occurrence of 
the locomotive in \RR$_i$(0) and of that of \sgg, later, in \RR$_i$(1), rule~157.
they also witness the motion of the locomotive from~\RR$_s$(-1).

At last, the case $c=0$ is different as far as no decrementation is performed: the cell
\RR$_c$(-1) which can see both \sbb{} in \RR$_c$(0) and the \sbb-locomotive in
\RR$_i$(-1) \sgg{} occurs in \RR$_i$(0) remains \sww, rule~49. But \sgg{} occurs
in \RR$_i$(0) as far as the cell also recognizes the impossibility of decrementing, 
so that it becomes~\sgg{} in order that \srr{} occurs in \RR$_d$(-1) which triggers
the start of a new \sbb-locomotive in \RR$_d$(-2)$z$, the firs cell of the $Z$-path to
\DDD. That behaviour is managed by rules~160 up to~166. Rule~167 triggers that 
locomotive in \RR$_d$(-2)$z$. The other rules, from~166 up to~169 are witnessing ones.

\subsubsection{Incrementation of the register}\label{sbbrreginc}

   Presently, we arrive to the implementation of an incrementation of the register.
Rules~170 up to~205 or Table~\ref{trall} manage that operation, eight rules less than
for the decrementation.

   As indicated in Section~\ref{newrailway}, the incrementation is performed on~\RR$_c$ 
although the locomotive arrives at \RR$_i$(0) after visiting \RR$_s$(-1). 
Here too, we follow a general scheme in all cases. It is a bit easier in that case as 
far as the incrementation is always performed. Note that both \RR$_c$ and \RR$_s$ can 
witness the motion of the locomotive which cannot be seen by \RR$_d$ as long as it happens
in \RR$_i$. When the locomotive goes back through \RR$_c$ as an \srr-locomotive, \RR$_s$
can no more witness it but \RR$_i$ goes on that witnessing and \RR$_d$ can now see that
motion. That general scheme is illustrated by~(9). What triggers the incrementation is
the fact that the cell \RR$_c$($c$-1), where $c$ is the value stored in the register,
can see both \sbb{} in \RR$_c$($c$) and the \sbb-locomotive in \RR$_i$($c$-1).

    However, the arrival of the locomotive is different: the path coming from the \DDI{} 
devoted to the register arrives at \RR$_c$(-2)$_a$ which can see \RR$_c$(-1) through its
face~2 while it seen from that latter cell from its face~5.

   Rules~170 up to~188 are devoted to the general case. Rules~170 and~171 witness that 
 \RR$_c($-1) and \RR$_c$(0) respectively can see a locomotive leaving the cell and
passing through \RR$_i$(0) respectively. Rule 172 is the witness rule for \RR$_i$($n$)
when the locomotive is already in \RR$_i$($n$+1) while rule~173 witnesses the locomotive 
in \RR$_i$ through \RR$_s$($0$). Rule~174, rule 175 and rule~176 witness the motion of the
locomotive along \RR$_i$($n$), $n>0$, from \RR$_c$, from \RR$_s$ and from \RR$_d$.
\vskip 10pt
\setbox115=\vtop{\leftskip 0pt\parindent 0pt\hsize=250pt
\ftt{
* 0 1 2 3 4 5 6 7 8\hskip 40pt * 0 1 2 3 4 5 6 7 8\vskip 3pt
0 W W W W W B B B B\hskip 40pt 0 B W R R R R R R R\vszz
1 W W W W W B B B B\hskip 40pt 1 W B R R R R R R R\vszz
2 W W W W W B B B B\hskip 40pt 2 W W B R R R R R R\vszz
3 W W W W W B B B B\hskip 40pt 3 W W R B R R R R R\vszz
4 W W W W R B B B B\hskip 40pt 4 W W R R B R R R R\vszz
5 W W W R R W B B B\hskip 40pt 5 W W R R R G R R R\vszz
6 W W R W W W B B B\hskip 40pt 6 W W R R R R R R R\vszz
7 W B W W W W B B B\hskip 40pt 7 W W R R R R R R R\vszz
8 B W W W W W B B B\hskip 40pt 8 W W R R R R R R R\vszz
} }
\ligne{\hfill
$\vcenter{\hbox{\box115}}$ \hfill (\numerrel)\hskip 10pt}
\vskip 10pt
Rule~177 deals with the cell \RR$_c$($c$-1), where $c$ is the content of the register,
for the time when that cell which can see \sbb{} in \RR$_c$($c$), can also see the 
\sbb-locomotive in \RR$_i$($c$-1). The rule makes that cell change from~\sww{} to
\srr. Rules~178 up to 181 witness the occurrence of \sgg{}
in \RR$_i$($c$) triggered by rule~136 which is viewed from~\RR$_c$, \RR$_i$ and \RR$_s$.
Note that rule~136, defined for the decrementation in the general case, also applies here.
Indeed, the neighbourhood of a cell belonging to \RR$_i$($n$) and to \RR$_d$($n$) is the
same, which explains the application of that rule in both operations, despite their 
opposite working. That difference already occurs with rule~182 which makes \RR$_c$($c$)
turn from \sbb{} to \sww, realizing the incrementation. The \srr{} occurring in 
\RR$_c$($c$-1) triggers the starting of an \srr-locomotive on \RR$_c$($c$-2) moving
on \RR$_c$ towards \RR$_c$(-1) where it becomes a \sbb-locomotive going to \DDI. Rule~82
allows that \srr-locomotive to start its motion on \RR$_c$ which is controlled by the
rules~77 up to~84., rules~77 up to~79 managing the cell \RR$_c$(0). Note that rule~80
witnesses that the \srr-locomotive was changed into a \sbb-one. 

Three additional rules from~189 up to~191 are needed for the case $c = 2$. The change is
produced by the occurrence of~\srr{} in \RR$_c$(1) which is seen by \RR$_c$(0) through
its face~2, that latter cell being \srr{} too, see rule~190.

In the case when $c=1$, managed by rules~192 up to~196, we have the situation when the
reacting cell of \RR$_c$ is \RR$_c$(0) and the reaction occurs in \RR$_i$(1) too,
whence the suffix \ftt{WBWW}, rules~192 up to~195.

Eventually, we arrive to the case $c = 0$, rules~197 up to~205. Rule~197 manages the
cell \RR$_c$(-1) which can see \sbb{} in \RR$_c$(0) and the locomotive in 
\hbox{\RR$_i$(-1)}.
The cell \RR$_i$(0) which can also see the \sbb-locomotive and which can view that
$c = 0$ becomes \sgg, rule 198. Rule~200 keeps \sgg{} for one step in order the 
appropriate rules eliminate the superfluous \sbb-neighbours of~\sgg. Rule~205 returns
\RR$_i$(0) to \srr{} while the previous rules keep \sbb{} in \RR$_c$(0) and start a 
\sbb-locomotive in \RR$_c$(-1) which goes back to the \DDI{} associate with that register.

\subsection{Stopping the computation}\label{srstop}

We arrive to the last Subsection of the present section. It is not the least important
as far as it deals with the end of the computation. We remind the reader that, by
definition, the computation of a cellular automaton has no halting state. In fact, it
is usually considered that when two successive configurations are identical, it should
be considered as the end of the computation. Indeed, nothing new can appear in a 
configuration which is constantly repeated cell by cell. Consequently, stopping the 
computation will consist in realizing a configuration which is endlessly 
identically repeated. Moreover, such a situation is algorithmically detectable.

   We noted that, outside the simulation of the register machine, we decided to construct
the register according a potentially non stopping process. It was the growth of the
register as defined in Sub-subsection~\ref{sbbrgrow}. In that Sub-subsection, we
underlined that the speed of the construction was $\displaystyle{1\over2}$: it needs two
tips of the clock in order to advance the \sgg-cells of the register by one cell forward.
It is enough for stopping the computation. When the locomotive arrives to the halting
instruction of the register machine, the implementation of that instruction consists in
sending a locomotive to all~\RR$_s$ of all registers of the simulated register
machine. If we have $n$ registers, $k$ forks will allow us to do that where
\hbox{$2^{k-1}< n\leq 2^k$}. The locomotive arrives at~\RR$_s$ at some time~$t$.
We may assume that $t$ is the same for all registers. At time~$t$, the \sgg-cells are 
at some distance~$L$ from \RR$_s$(0). As far as 
one step forward of the \sgg-cells requires two tips of the clock, if the locomotive 
in~\RR$_s$ advances at speed~1, it reaches the \sgg-cells at time~$t$+$2L$. Taking
for $L$ the biggest one for all registers, we can then compute the time at which
the computation of the automaton halts when we know the time of the halting of the
register machine. If that latter one does not halt, no signal is sent to any \RR$_s$
so that in that case, the registers are endlessly growing.

   Rules~206 up to~254 of Table~\ref{trall} deal with stopping the computation. 
We also consider rules~109 up to~114 which manage the first steps of the locomotive
on \RR$_s$, starting from \RR$_s$(0). The rules indicate that \sgg{} is immediately 
replaced by \sbb{} so that man \sbb-neighbours of that very temporary \sgg-cell remain
for ever as far as rule~16 applies to them. Consequently, a 'cloud' of immortal \sbb-cells
is raised during that process. Now, as far as the computation is eventually stopped, the
number of \sbb-cells constituting that 'cloud' over \RR$_s$ is finite and it belongs to 
the configuration which is identically endlessly repeated. If the computation is infinite,
that phenomenon does not occur as far as no locomotive is sent to the \RR$_s$ strands
of the registers.

Rules~206 up to~213 deal with cells which can see the \sgg-locomotive in front of them
and of cells of \RR$_i$ and \RR$_d$ which can see it. Remember that the cells of \RR$_c$
cannot see those of \RR$_s$. Note that rules~207 up to 210 together with rules~212 and~213
witness the occurrence of still present \sbb-neighbours of the previously \sgg-cell.
Rule~214 starts a new process. Let $N$ be defined by the fact that, at that time,
\RR$_s$($N$) contains the \sgg-cell involved in the growing process. Then, rule~214 
transforms the \sbb-cell contained in \RR$_s$($N$-1) into \sww. Rule~215 says that
a \sbb-cell before such a \sww-cell remains \sbb{} and that from now on, a created 
\sww-cell remains \sww. Rules~217 and~218 say that the \sbb- and \srr-cells which can see
a \sww-cell behind them also become \sww. Rule~219 says the same for the \sgg-cell 
involved in the growing process of the register. Rules~220 up to~225 deal with
conservative rules for \sbb- and \srr-cells which should not be turned to \sww{} and those
rules also deal with rules changing \sbb- and \srr-cells when they should be transformed.

  That process is illustrated by~(10). Two possible times of arrival of the 
\sgg-locomotive with respect to the \sgg-end of the register are illustrated. In the 
right-hand part, the \sgg-locomotive arrives one time later compared with the left-hand
side part.

\ifnum 1=0 {
\vskip 10pt
\ligne{\hfill
\vtop{\leftskip 0pt\parindent 0pt\hsize=90pt
$\vcenter{\vbox{
\ftt{
B R B B G W W W W\vszz
B B R B G B W W W\vszz
B B B W B G W W W\vszz
B B B W W G B W W\vszz
B B B W W W G W W\vszz
B B B W W W W B W\vszz
}
}}$
}
\hfill(\numerrel)\hskip 10pt}
} \fi
\vskip 10pt
\ligne{\hfill
\def\gzgz{\hskip 30pt{}}
\vtop{\leftskip 0pt\parindent 0pt\hsize=230pt
$\vcenter{\vbox{
\ftt{
- 0 1 2 3 4 5 6 7 8\gzgz - 0 1 2 3 4 5 6 7 8 9\vskip 3pt
0 G B B B G W W W W\gzgz 0 G B B B G B W W W W\vszz
1 B G B B G W W W W\gzgz 1 B G B B G B W W W W\vszz
2 B B G B G B W W W\gzgz 2 B B G B B G W W W W\vszz
3 B B B W B G W W W\gzgz 3 B B B G B G B W W W\vszz
4 B B B W W G B W W\gzgz 4 B B B B W B G W W W\vszz
5 B B B W W W G W W\gzgz 5 B B B B W W G B W W\vszz
6 B B B W W W W B W\gzgz 6 B B B B W W W G W W\vszz
7 B B B W W W W W W\gzgz 7 B B B B W W W W B W\vszz
8 B B B W W W W W W\gzgz 8 B B B B W W W W W W\vszz
}
}}$
}
\hfill(\numerrel)\hskip 10pt}
\vskip 10pt
We can see that the \sgg-locomotive appears in cell~0 at time~0{} in both sides of~(10).
At the same time, the \sgg-end occurs in cell~4 but cell~5 is \sww{} in the left-hand side
while it is \sbb{} in the right-hand side of~(10). Accordingly, the difference corresponds
to different neighbourhood of the \sgg-end. The first \sww-cell appears on cell~3 at 
time~3 on the left-hand side of~(10) while it appears on cell~4 at time~4 on the 
right-hand side of~(10). The computation is stopped at time~7, 8{} in the left-,right-hand
side of~(10) respectively. In fact, it appears a bit later, as far as four strands are
growing in parallel, so that the first occurrence of~\sww{} appears on \RR$_i$, \RR$_d$
with a delay by one tip of the clock and on \RR$_c$ with a delay of two tips of the clock.That delay holds the presence of four \sbb-cells around~$\ell$ a bit longer.

Indeed, when the \sgg-cells is turned to~\sww{} by rule~219 and also further similar
rules as rules~226, 232 and~233 for instance, the growing process is not immediately
stopped. As long as there can be four \sbb-cells surrounding~$\ell$, new \sgg-cells
occur. Rules~226 up to~254 have \sww{} as their new state: it means that whatever their
current state is, their next one is \sww. Clearly, at some point the process arrives to
a configuration where only three \sbb-cells lie around~$\ell$, so that no new \sgg-cell
occur. Let $L$ denote the position of those \sbb-cells. Starting from that moment, the 
replacement of \sbb- and \srr-cells goes on for \RR$_s$($n$) cells with $n>L$ until
the furthest ones are reached by that growing \sww-signal at speed~1. When it is the case,
no new non-blank cell is created so that starting from that time, the configuration is
endlessly repeated identically.

   That last point completes the proof of Theorem~\ref{letheo}. \hfill$\Box$

\section{Conclusion}\label{conclude}

There are 254 rules which are half-rotation pairwise independent. 

We said that the rotation invariance is a huge constraint. The relaxation to invariance
with respect to the rotations leaving a face invariant is still a big constraint, of
course lighter than the invariance with respect to all rotations leaving the dodecahedron
globally invariant. That relaxation allowed us to get a strongly universal cellular 
automaton with four states.

   Probably, the solution given in the paper is not unique. However, it seems to me that
there are less choice under that assumption than under the hypothesis of full rotation
invariance leading to the same issue at the price of one more state. Probably, the fact 
that more flexibility exists in that latter solution explains the reduction to four state
by the help of relaxing the rotation invariance as already mentioned.
Is it still possible to obtain strong universality with less states? I do not know 
presently how to do. That does not mean, of course, that it would be impossible. 
Experience shows that such a reduction requires to change something in the model. 
The present solution did not change the model presented in~\cite{mmarXiv21a}. It
significantly changed the scenario, in particular in the working of the strands of a
register. So, to change something in the model is the way. To change what exactly? That 
is the question.

\compteregle=0
\ligne{\hfill
\vtop{
\begin{tab}\label{trall}
\leurre
Table of the rules for the cellular automaton.
\end{tab}
\vskip-7pt
\ligne{\hfill
\vtop{\leftskip 0pt\parindent 0pt\hsize=\largeouille
\vskip 3pt
\ligne{\hfill\small blanks\hfill}
\ligne{\hfill\gzzz\ftt{\the\compteregle \ }  
\llrule {W} {W} {W} {W} {W} {W} {W} 
\hskip-3.5pt\rrrule {W} {W} {W} {W} {W} {W} {W} {}\hfill}
\vskip-3pt
\ligne{\hfill\gzzz\ftt{\the\compteregle \ }  
\llrule {W} {B} {W} {W} {W} {W} {W} 
\hskip-3.5pt\rrrule {W} {W} {W} {W} {W} {W} {W} {}\hfill}
\vskip-3pt
\ligne{\hfill\gzzz\ftt{\the\compteregle \ }  
\llrule {W} {W} {B} {W} {W} {W} {W} 
\hskip-3.5pt\rrrule {W} {W} {W} {W} {W} {W} {W} {}\hfill}
\vskip-3pt
\ligne{\hfill\gzzz\ftt{\the\compteregle \ }  
\llrule {W} {R} {W} {W} {W} {W} {W} 
\hskip-3.5pt\rrrule {W} {W} {W} {W} {W} {W} {W} {}\hfill}
\vskip-3pt
\ligne{\hfill\gzzz\ftt{\the\compteregle \ }  
\llrule {W} {W} {R} {W} {W} {W} {W} 
\hskip-3.5pt\rrrule {W} {W} {W} {W} {W} {W} {W} {}\hfill}
\vskip-3pt
\ligne{\hfill\gzzz\ftt{\the\compteregle \ }  
\llrule {W} {G} {W} {W} {W} {W} {W} 
\hskip-3.5pt\rrrule {W} {W} {W} {W} {W} {W} {B} {}\hfill}
\vskip-3pt
\ligne{\hfill\gzzz\ftt{\the\compteregle \ }  
\llrule {W} {W} {W} {W} {W} {W} {G} 
\hskip-3.5pt\rrrule {W} {W} {W} {W} {W} {W} {B} {}\hfill}
\vskip-3pt
\ligne{\hfill\gzzz\ftt{\the\compteregle \ }  
\llrule {W} {B} {B} {W} {W} {W} {W} 
\hskip-3.5pt\rrrule {W} {W} {W} {W} {W} {W} {W} {}\hfill}
\vskip-3pt
\ligne{\hfill\gzzz\ftt{\the\compteregle \ }  
\llrule {W} {B} {W} {B} {W} {W} {W} 
\hskip-3.5pt\rrrule {W} {W} {W} {W} {W} {W} {W} {}\hfill}
\vskip-3pt
\ligne{\hfill\gzzz\ftt{\the\compteregle \ }  
\llrule {W} {B} {R} {W} {W} {W} {W} 
\hskip-3.5pt\rrrule {W} {W} {W} {W} {W} {W} {W} {}\hfill}
\vskip-3pt
\ligne{\hfill\gzzz\ftt{\the\compteregle \ }  
\llrule {W} {R} {R} {W} {W} {W} {W} 
\hskip-3.5pt\rrrule {W} {W} {W} {W} {W} {W} {W} {}\hfill}
\vskip-3pt
\ligne{\hfill\gzzz\ftt{\the\compteregle \ }  
\llrule {W} {R} {B} {W} {W} {W} {W} 
\hskip-3.5pt\rrrule {W} {W} {W} {W} {W} {W} {W} {}\hfill}
\vskip-3pt
\ligne{\hfill\gzzz\ftt{\the\compteregle \ }  
\llrule {W} {R} {W} {B} {W} {W} {W} 
\hskip-3.5pt\rrrule {W} {W} {W} {W} {W} {W} {W} {}\hfill}
\vskip-3pt
\ligne{\hfill\gzzz\ftt{\the\compteregle \ }  
\llrule {W} {G} {B} {W} {W} {W} {W} 
\hskip-3.5pt\rrrule {W} {W} {W} {W} {W} {W} {W} {}\hfill}
\vskip-3pt
\ligne{\hfill\gzzz\ftt{\the\compteregle \ }  
\llrule {W} {G} {R} {W} {W} {W} {W} 
\hskip-3.5pt\rrrule {W} {W} {W} {W} {W} {W} {W} {}\hfill}
\vskip 6pt
\hrule height 0.3pt depth 0.3pt width \hsize
\vskip 3pt
\ligne{\hfill\small tracks\hfill}
\ligne{\hfill\gzzz\ftt{\the\compteregle \ }  
\llrule {B} {W} {W} {W} {W} {W} {W} 
\hskip-3.5pt\rrrule {W} {W} {W} {W} {W} {W} {B} {}\hfill}
\vskip-3pt
\ligne{\hfill\gzzz\ftt{\the\compteregle \ }  
\llrule {B} {B} {W} {W} {W} {W} {W} 
\hskip-3.5pt\rrrule {W} {W} {W} {W} {W} {W} {B} {}\hfill}
\vskip-3pt
\ligne{\hfill\gzzz\ftt{\the\compteregle \ }  
\llrule {B} {W} {B} {W} {W} {W} {W} 
\hskip-3.5pt\rrrule {W} {W} {W} {W} {W} {W} {B} {}\hfill}
\vskip-3pt
\ligne{\hfill\gzzz\ftt{\the\compteregle \ }  
\llrule {W} {W} {W} {W} {W} {W} {W} 
\hskip-3.5pt\rrrule {B} {W} {W} {B} {B} {W} {W} {}\hfill}
\vskip-3pt
\ligne{\hfill\gzzz\ftt{\the\compteregle \ }  
\llrule {W} {W} {W} {W} {W} {W} {B} 
\hskip-3.5pt\rrrule {B} {W} {W} {B} {B} {W} {B} {}\hfill}
\vskip-3pt
\ligne{\hfill\gzzz\ftt{\the\compteregle \ }  
\llrule {W} {W} {W} {W} {W} {B} {W} 
\hskip-3.5pt\rrrule {B} {W} {W} {B} {B} {W} {B} {}\hfill}
\vskip-3pt
\ligne{\hfill\gzzz\ftt{\the\compteregle \ }  
\llrule {W} {W} {W} {R} {W} {W} {W} 
\hskip-3.5pt\rrrule {B} {W} {W} {B} {B} {W} {W} {}\hfill}
\vskip-3pt
\ligne{\hfill\gzzz\ftt{\the\compteregle \ }  
\llrule {B} {W} {W} {W} {W} {W} {W} 
\hskip-3.5pt\rrrule {B} {W} {W} {B} {B} {W} {W} {}\hfill}
\vskip-3pt
\ligne{\hfill\gzzz\ftt{\the\compteregle \ }  
\llrule {W} {W} {W} {B} {W} {W} {W} 
\hskip-3.5pt\rrrule {B} {W} {W} {B} {B} {W} {W} {}\hfill}
\vskip-3pt
\ligne{\hfill\gzzz\ftt{\the\compteregle \ }  
\llrule {W} {W} {B} {W} {W} {W} {W} 
\hskip-3.5pt\rrrule {B} {W} {W} {B} {B} {W} {B} {}\hfill}
\vskip-3pt
\ligne{\hfill\gzzz\ftt{\the\compteregle \ }  
\llrule {W} {B} {W} {W} {W} {W} {W} 
\hskip-3.5pt\rrrule {B} {W} {W} {B} {B} {W} {W} {}\hfill}
\vskip-3pt
\ligne{\hfill\gzzz\ftt{\the\compteregle \ }  
\llrule {B} {W} {R} {W} {W} {W} {W} 
\hskip-3.5pt\rrrule {W} {W} {W} {W} {W} {W} {B} {}\hfill}
\vskip-3pt
\ligne{\hfill\gzzz\ftt{\the\compteregle \ }  
\llrule {R} {W} {W} {W} {W} {W} {W} 
\hskip-3.5pt\rrrule {W} {W} {W} {W} {W} {W} {R} {}\hfill}
\vskip-3pt
\ligne{\hfill\gzzz\ftt{\the\compteregle \ }  
\llrule {W} {W} {W} {W} {R} {W} {W} 
\hskip-3.5pt\rrrule {B} {W} {W} {B} {B} {W} {W} {}\hfill}
\vskip-3pt
\ligne{\hfill\gzzz\ftt{\the\compteregle \ }  
\llrule {W} {W} {W} {W} {R} {W} {W} 
\hskip-3.5pt\rrrule {B} {B} {W} {B} {B} {W} {B} {}\hfill}
\vskip-3pt
\ligne{\hfill\gzzz\ftt{\the\compteregle \ }  
\llrule {B} {W} {W} {W} {R} {W} {W} 
\hskip-3.5pt\rrrule {B} {W} {W} {B} {B} {W} {W} {}\hfill}
\vskip-3pt
\ligne{\hfill\gzzz\ftt{\the\compteregle \ }  
\llrule {W} {W} {W} {B} {R} {W} {W} 
\hskip-3.5pt\rrrule {B} {W} {W} {B} {B} {W} {W} {}\hfill}
\ligne{\hfill\ftt{control } \hfill}
\ligne{\hfill\gzzz\ftt{\the\compteregle \ }  
\llrule {W} {B} {W} {W} {W} {W} {B} 
\hskip-3.5pt\rrrule {B} {W} {W} {B} {B} {W} {W} {}\hfill}
\vskip-3pt
\ligne{\hfill\gzzz\ftt{\the\compteregle \ }  
\llrule {W} {W} {W} {B} {W} {W} {B} 
\hskip-3.5pt\rrrule {B} {W} {W} {B} {B} {W} {B} {}\hfill}
\vskip-3pt
\ligne{\hfill\gzzz\ftt{\the\compteregle \ }  
\llrule {B} {W} {W} {B} {W} {W} {W} 
\hskip-3.5pt\rrrule {B} {W} {W} {B} {B} {W} {W} {}\hfill}
\vskip-3pt
\ligne{\hfill\gzzz\ftt{\the\compteregle \ }  
\llrule {W} {W} {W} {W} {W} {W} {W} 
\hskip-3.5pt\rrrule {B} {W} {W} {B} {B} {B} {W} {}\hfill}
\vskip-3pt
\ligne{\hfill\gzzz\ftt{\the\compteregle \ }  
\llrule {B} {W} {W} {W} {W} {W} {W} 
\hskip-3.5pt\rrrule {B} {W} {W} {B} {B} {B} {B} {}\hfill}
\vskip-3pt
\ligne{\hfill\gzzz\ftt{\the\compteregle \ }  
\llrule {W} {W} {W} {W} {W} {W} {B} 
\hskip-3.5pt\rrrule {B} {W} {W} {B} {B} {B} {B} {}\hfill}
\vskip-3pt
\ligne{\hfill\gzzz\ftt{\the\compteregle \ }  
\llrule {B} {W} {W} {W} {W} {W} {B} 
\hskip-3.5pt\rrrule {B} {W} {W} {B} {B} {B} {W} {}\hfill}
\ligne{\hfill\ftt{for computer only } \hfill}
\ligne{\hfill\gzzz\ftt{\the\compteregle \ }  
\llrule {W} {W} {W} {W} {W} {W} {W} 
\hskip-3.5pt\rrrule {B} {R} {W} {B} {B} {W} {W} {}\hfill}
\vskip-3pt
\ligne{\hfill\gzzz\ftt{\the\compteregle \ }  
\llrule {W} {W} {W} {W} {W} {W} {B} 
\hskip-3.5pt\rrrule {B} {R} {W} {B} {B} {W} {W} {}\hfill}
\ligne{\hfill\ftt{fork } \hfill}
\ligne{\hfill\gzzz\ftt{\the\compteregle \ }  
\llrule {B} {B} {B} {W} {W} {W} {W} 
\hskip-3.5pt\rrrule {W} {W} {W} {W} {W} {W} {B} {}\hfill}
\vskip-3pt
\ligne{\hfill\gzzz\ftt{\the\compteregle \ }  
\llrule {W} {W} {W} {W} {W} {W} {W} 
\hskip-3.5pt\rrrule {B} {B} {B} {B} {W} {R} {W} {}\hfill}
\vskip-3pt
\ligne{\hfill\gzzz\ftt{\the\compteregle \ }  
\llrule {W} {W} {W} {B} {W} {W} {W} 
\hskip-3.5pt\rrrule {B} {B} {B} {B} {W} {R} {B} {}\hfill}
\vskip-3pt
\ligne{\hfill\gzzz\ftt{\the\compteregle \ }  
\llrule {B} {W} {W} {W} {W} {W} {W} 
\hskip-3.5pt\rrrule {B} {B} {B} {B} {W} {R} {W} {}\hfill}
\vskip-3pt
\ligne{\hfill\gzzz\ftt{\the\compteregle \ }  
\llrule {W} {W} {W} {W} {W} {B} {B} 
\hskip-3.5pt\rrrule {B} {B} {B} {B} {W} {R} {W} {}\hfill}
\vskip 6pt
\hrule height 0.3pt depth 0.3pt width \hsize
}\hfill
\vtop{\leftskip 0pt\parindent 0pt\hsize=\largeouille
\vskip 3pt
\ligne{\hfill\small $\mathcal S$(-1), \RR$_s$(-2)\hfill}
\vskip 3pt
\ligne{\hfill\ftt{\RR$_c$(-1) } \hfill}
\vskip 3pt
\ligne{\hfill\gzzz\ftt{\the\compteregle \ }  
\llrule {W} {W} {W} {W} {W} {W} {W} 
\hskip-3.5pt\rrrule {R} {W} {W} {R} {R} {W} {W} {}\hfill}
\vskip-3pt
\ligne{\hfill\gzzz\ftt{\the\compteregle \ }  
\llrule {W} {B} {W} {W} {W} {W} {W} 
\hskip-3.5pt\rrrule {R} {W} {W} {R} {R} {W} {W} {}\hfill}
\vskip-3pt
\ligne{\hfill\gzzz\ftt{\the\compteregle \ }  
\llrule {W} {W} {W} {W} {W} {W} {B} 
\hskip-3.5pt\rrrule {R} {W} {W} {R} {R} {W} {W} {}\hfill}
\vskip-3pt
\ligne{\hfill\gzzz\ftt{\the\compteregle \ }  
\llrule {B} {W} {W} {W} {W} {W} {W} 
\hskip-3.5pt\rrrule {R} {W} {W} {R} {R} {W} {W} {}\hfill}
\vskip-3pt
\ligne{\hfill\gzzz\ftt{\the\compteregle \ }  
\llrule {W} {W} {W} {B} {W} {W} {W} 
\hskip-3.5pt\rrrule {R} {W} {W} {R} {R} {W} {W} {}\hfill}
\vskip-3pt
\ligne{\hfill\gzzz\ftt{\the\compteregle \ }  
\llrule {B} {W} {W} {W} {W} {W} {B} 
\hskip-3.5pt\rrrule {R} {W} {W} {R} {R} {W} {W} {}\hfill}
\vskip 3pt
\ligne{\hfill\ftt{\RR$_i$(-1) } \hfill}
\vskip 3pt
\ligne{\hfill\gzzz\ftt{\the\compteregle \ }  
\llrule {W} {W} {W} {R} {W} {W} {W} 
\hskip-3.5pt\rrrule {B} {R} {W} {B} {R} {W} {W} {}\hfill}
\vskip-3pt
\ligne{\hfill\gzzz\ftt{\the\compteregle \ }  
\llrule {W} {B} {W} {R} {W} {W} {W} 
\hskip-3.5pt\rrrule {B} {R} {W} {B} {R} {W} {W} {}\hfill}
\vskip-3pt
\ligne{\hfill\gzzz\ftt{\the\compteregle \ }  
\llrule {W} {W} {W} {R} {W} {W} {B} 
\hskip-3.5pt\rrrule {B} {R} {W} {B} {R} {W} {B} {}\hfill}
\vskip-3pt
\ligne{\hfill\gzzz\ftt{\the\compteregle \ }  
\llrule {B} {W} {W} {R} {W} {W} {W} 
\hskip-3.5pt\rrrule {B} {R} {W} {B} {R} {W} {W} {}\hfill}
\vskip-3pt
\ligne{\hfill\gzzz\ftt{\the\compteregle \ }  
\llrule {W} {W} {W} {B} {W} {W} {W} 
\hskip-3.5pt\rrrule {B} {R} {W} {B} {R} {W} {W} {}\hfill}
\vskip 3pt
\ligne{\hfill\ftt{\RR$_d$(-1) } \hfill}
\vskip 3pt
\ligne{\hfill\gzzz\ftt{\the\compteregle \ }  
\llrule {W} {W} {W} {R} {W} {W} {W} 
\hskip-3.5pt\rrrule {B} {W} {W} {B} {R} {W} {W} {}\hfill}
\vskip-3pt
\ligne{\hfill\gzzz\ftt{\the\compteregle \ }  
\llrule {W} {B} {W} {R} {W} {W} {W} 
\hskip-3.5pt\rrrule {B} {W} {W} {B} {R} {W} {W} {}\hfill}
\vskip-3pt
\ligne{\hfill\gzzz\ftt{\the\compteregle \ }  
\llrule {W} {W} {W} {R} {W} {B} {W} 
\hskip-3.5pt\rrrule {B} {W} {W} {B} {R} {W} {B} {}\hfill}
\vskip-3pt
\ligne{\hfill\gzzz\ftt{\the\compteregle \ }  
\llrule {B} {W} {W} {R} {W} {W} {W} 
\hskip-3.5pt\rrrule {B} {W} {W} {B} {R} {W} {W} {}\hfill}
\vskip-3pt
\ligne{\hfill\gzzz\ftt{\the\compteregle \ }  
\llrule {W} {W} {W} {B} {W} {W} {W} 
\hskip-3.5pt\rrrule {B} {W} {W} {B} {R} {W} {W} {}\hfill}
\vskip 3pt
\ligne{\hfill\ftt{\RR$_s$(-1) } \hfill}
\vskip 3pt
\ligne{\hfill\gzzz\ftt{\the\compteregle \ }  
\llrule {W} {B} {W} {W} {W} {W} {B} 
\hskip-3.5pt\rrrule {B} {R} {W} {B} {B} {W} {W} {}\hfill}
\vskip-3pt
\ligne{\hfill\gzzz\ftt{\the\compteregle \ }  
\llrule {W} {W} {W} {W} {W} {W} {R} 
\hskip-3.5pt\rrrule {B} {R} {W} {B} {B} {W} {R} {}\hfill}
\vskip-3pt
\ligne{\hfill\gzzz\ftt{\the\compteregle \ }  
\llrule {W} {W} {W} {W} {B} {W} {B} 
\hskip-3.5pt\rrrule {B} {R} {W} {B} {B} {W} {B} {}\hfill}
\vskip-3pt
\ligne{\hfill\gzzz\ftt{\the\compteregle \ }  
\llrule {B} {W} {W} {W} {W} {W} {B} 
\hskip-3.5pt\rrrule {B} {R} {W} {B} {B} {W} {W} {}\hfill}
\vskip-3pt
\ligne{\hfill\gzzz\ftt{\the\compteregle \ }  
\llrule {W} {W} {W} {W} {W} {W} {G} 
\hskip-3.5pt\rrrule {B} {R} {W} {B} {B} {W} {W} {}\hfill}
\vskip-3pt
\ligne{\hfill\gzzz\ftt{\the\compteregle \ }  
\llrule {R} {W} {W} {W} {W} {W} {B} 
\hskip-3.5pt\rrrule {B} {R} {W} {B} {B} {W} {W} {}\hfill}
\vskip-3pt
\ligne{\hfill\gzzz\ftt{\the\compteregle \ }  
\llrule {W} {W} {W} {B} {R} {W} {B} 
\hskip-3.5pt\rrrule {B} {R} {W} {B} {B} {W} {W} {}\hfill}
\vskip 3pt
\ligne{\hfill\ftt{\RR$_s$(-2)$_d$ } \hfill}
\vskip 3pt
\ligne{\hfill\gzzz\ftt{\the\compteregle \ }  
\llrule {W} {W} {W} {W} {W} {W} {W} 
\hskip-3.5pt\rrrule {W} {R} {R} {R} {W} {R} {W} {}\hfill}
\vskip-3pt
\ligne{\hfill\gzzz\ftt{\the\compteregle \ }  
\llrule {W} {W} {W} {W} {W} {B} {W} 
\hskip-3.5pt\rrrule {W} {R} {R} {R} {W} {R} {W} {}\hfill}
\vskip-3pt
\vskip 3pt
\ligne{\hfill\ftt{\RR$_s$(-2)$_s$ } \hfill}
\vskip 3pt
\ligne{\hfill\gzzz\ftt{\the\compteregle \ }  
\llrule {W} {W} {W} {W} {W} {W} {W} 
\hskip-3.5pt\rrrule {B} {W} {R} {B} {B} {W} {W} {}\hfill}
\vskip-3pt
\ligne{\hfill\gzzz\ftt{\the\compteregle \ }  
\llrule {W} {W} {W} {W} {W} {W} {B} 
\hskip-3.5pt\rrrule {B} {W} {R} {B} {B} {W} {B} {}\hfill}
\vskip-3pt
\ligne{\hfill\gzzz\ftt{\the\compteregle \ }  
\llrule {B} {W} {W} {W} {W} {W} {W} 
\hskip-3.5pt\rrrule {B} {W} {R} {B} {B} {W} {W} {}\hfill}
\vskip-3pt
\ligne{\hfill\gzzz\ftt{\the\compteregle \ }  
\llrule {W} {W} {W} {B} {W} {W} {W} 
\hskip-3.5pt\rrrule {B} {W} {R} {B} {B} {W} {W} {}\hfill}
\vskip-3pt
\ligne{\hfill\gzzz\ftt{\the\compteregle \ }  
\llrule {R} {W} {W} {W} {W} {W} {W} 
\hskip-3.5pt\rrrule {B} {W} {R} {B} {B} {W} {W} {}\hfill}
\vskip 6pt
\hrule height 0.3pt depth 0.3pt width \hsize
\vskip 3pt
\ligne{\hfill\small register\hfill}
\vskip 3pt
\ligne{\hfill\ftt{motions } \hfill}
\ligne{\hfill\ftt{\RR$_c$ } \hfill}
\ligne{\hfill\gzzz\ftt{\the\compteregle \ }  
\llrule {W} {R} {R} {W} {B} {W} {W} 
\hskip-3.5pt\rrrule {W} {B} {W} {B} {W} {W} {W} {}\hfill}
\vskip-3pt
\ligne{\hfill\gzzz\ftt{\the\compteregle \ }  
\llrule {W} {R} {R} {R} {B} {W} {W} 
\hskip-3.5pt\rrrule {W} {B} {W} {B} {W} {W} {R} {}\hfill}
\vskip-3pt
\ligne{\hfill\gzzz\ftt{\the\compteregle \ }  
\llrule {R} {R} {R} {W} {B} {W} {W} 
\hskip-3.5pt\rrrule {W} {B} {W} {B} {W} {W} {W} {}\hfill}
\vskip-3pt
\ligne{\hfill\gzzz\ftt{\the\compteregle \ }  
\llrule {W} {R} {R} {B} {B} {W} {W} 
\hskip-3.5pt\rrrule {W} {B} {W} {B} {W} {W} {W} {}\hfill}
\ligne{\hfill\gzzz\ftt{\the\compteregle \ }  
\llrule {W} {R} {R} {W} {B} {W} {W} 
\hskip-3.5pt\rrrule {W} {B} {W} {W} {W} {W} {W} {}\hfill}
\vskip-3pt
\ligne{\hfill\gzzz\ftt{\the\compteregle \ }  
\llrule {W} {R} {R} {R} {B} {W} {W} 
\hskip-3.5pt\rrrule {W} {B} {W} {W} {W} {W} {R} {}\hfill}
}
\hfill
\vtop{\leftskip 0pt\parindent 0pt\hsize=\largeouille
\ligne{\hfill\gzzz\ftt{\the\compteregle \ }  
\llrule {R} {R} {R} {W} {B} {W} {W} 
\hskip-3.5pt\rrrule {W} {B} {W} {W} {W} {W} {W} {}\hfill}
\vskip-3pt
\ligne{\hfill\gzzz\ftt{\the\compteregle \ }  
\llrule {W} {R} {R} {W} {B} {W} {R} 
\hskip-3.5pt\rrrule {W} {B} {W} {W} {W} {W} {W} {}\hfill}
\vskip-3pt
\ligne{\hfill\gzzz\ftt{\the\compteregle \ }  
\llrule {W} {R} {R} {W} {B} {W} {B} 
\hskip-3.5pt\rrrule {W} {B} {W} {B} {W} {W} {W} {}\hfill}
\vskip-3pt
\ligne{\hfill\gzzz\ftt{\the\compteregle \ }  
\llrule {W} {R} {R} {W} {B} {W} {B} 
\hskip-3.5pt\rrrule {W} {B} {W} {W} {W} {W} {B} {}\hfill}
\vskip-3pt
\ligne{\hfill\gzzz\ftt{\the\compteregle \ }  
\llrule {W} {R} {R} {B} {B} {W} {W} 
\hskip-3.5pt\rrrule {W} {B} {W} {W} {W} {W} {W} {}\hfill}
\vskip-3pt
\ligne{\hfill\gzzz\ftt{\the\compteregle \ }  
\llrule {B} {R} {R} {B} {B} {W} {W} 
\hskip-3.5pt\rrrule {W} {B} {W} {W} {W} {W} {B} {}\hfill}
\vskip-3pt
\ligne{\hfill\gzzz\ftt{\the\compteregle \ }  
\llrule {B} {R} {R} {B} {B} {W} {B} 
\hskip-3.5pt\rrrule {W} {B} {W} {W} {W} {W} {B} {}\hfill}
\vskip-3pt
\ligne{\hfill\gzzz\ftt{\the\compteregle \ }  
\llrule {B} {R} {R} {B} {B} {W} {W} 
\hskip-3.5pt\rrrule {W} {B} {W} {B} {W} {W} {B} {}\hfill}
\ligne{\hfill\ftt{\RR$_i$ } \hfill}
\ligne{\hfill\gzzz\ftt{\the\compteregle \ }  
\llrule {R} {W} {B} {W} {W} {B} {R} 
\hskip-3.5pt\rrrule {B} {W} {W} {B} {W} {W} {R} {}\hfill}
\vskip-3pt
\ligne{\hfill\gzzz\ftt{\the\compteregle \ }  
\llrule {R} {B} {B} {W} {W} {B} {R} 
\hskip-3.5pt\rrrule {B} {W} {W} {B} {W} {W} {R} {}\hfill}
\vskip-3pt
\ligne{\hfill\gzzz\ftt{\the\compteregle \ }  
\llrule {R} {W} {B} {B} {W} {B} {R} 
\hskip-3.5pt\rrrule {B} {W} {W} {B} {W} {W} {B} {}\hfill}
\vskip-3pt
\ligne{\hfill\gzzz\ftt{\the\compteregle \ }  
\llrule {B} {W} {B} {W} {W} {B} {R} 
\hskip-3.5pt\rrrule {B} {W} {W} {B} {W} {W} {R} {}\hfill}
\vskip-3pt
\ligne{\hfill\gzzz\ftt{\the\compteregle \ }  
\llrule {R} {W} {B} {W} {W} {B} {B} 
\hskip-3.5pt\rrrule {B} {W} {W} {B} {W} {W} {R} {}\hfill}
\vskip-3pt
\ligne{\hfill\gzzz\ftt{\the\compteregle \ }  
\llrule {R} {W} {B} {R} {W} {B} {R} 
\hskip-3.5pt\rrrule {B} {W} {W} {W} {W} {W} {R} {}\hfill}
\vskip-3pt
\ligne{\hfill\gzzz\ftt{\the\compteregle \ }  
\llrule {R} {W} {B} {B} {W} {B} {R} 
\hskip-3.5pt\rrrule {B} {W} {W} {W} {W} {W} {B} {}\hfill}
\vskip-3pt
\ligne{\hfill\gzzz\ftt{\the\compteregle \ }  
\llrule {B} {W} {B} {R} {W} {B} {R} 
\hskip-3.5pt\rrrule {B} {W} {W} {W} {W} {W} {R} {}\hfill}
\ligne{\hfill\ftt{\RR$_d$ } \hfill}
\ligne{\hfill\gzzz\ftt{\the\compteregle \ }  
\llrule {R} {B} {B} {W} {W} {B} {R} 
\hskip-3.5pt\rrrule {B} {W} {W} {R} {W} {W} {R} {}\hfill}
\vskip-3pt
\ligne{\hfill\gzzz\ftt{\the\compteregle \ }  
\llrule {R} {B} {W} {W} {W} {B} {R} 
\hskip-3.5pt\rrrule {B} {W} {W} {R} {W} {W} {R} {}\hfill}
\vskip-3pt
\ligne{\hfill\gzzz\ftt{\the\compteregle \ }  
\llrule {R} {B} {W} {B} {W} {B} {R} 
\hskip-3.5pt\rrrule {B} {W} {W} {R} {W} {W} {B} {}\hfill}
\vskip-3pt
\ligne{\hfill\gzzz\ftt{\the\compteregle \ }  
\llrule {B} {B} {W} {W} {W} {B} {R} 
\hskip-3.5pt\rrrule {B} {W} {W} {R} {W} {W} {R} {}\hfill}
\vskip-3pt
\ligne{\hfill\gzzz\ftt{\the\compteregle \ }  
\llrule {R} {B} {W} {W} {W} {B} {B} 
\hskip-3.5pt\rrrule {B} {W} {W} {R} {W} {W} {R} {}\hfill}
\vskip-3pt
\ligne{\hfill\gzzz\ftt{\the\compteregle \ }  
\llrule {R} {B} {W} {R} {W} {B} {R} 
\hskip-3.5pt\rrrule {B} {W} {W} {W} {W} {W} {R} {}\hfill}
\vskip-3pt
\ligne{\hfill\gzzz\ftt{\the\compteregle \ }  
\llrule {R} {B} {W} {B} {W} {B} {R} 
\hskip-3.5pt\rrrule {B} {W} {W} {W} {W} {W} {B} {}\hfill}
\vskip-3pt
\ligne{\hfill\gzzz\ftt{\the\compteregle \ }  
\llrule {B} {B} {W} {R} {W} {B} {R} 
\hskip-3.5pt\rrrule {B} {W} {W} {W} {W} {W} {R} {}\hfill}
\vskip-3pt
\ligne{\hfill\gzzz\ftt{\the\compteregle \ }  
\llrule {R} {B} {W} {R} {W} {B} {B} 
\hskip-3.5pt\rrrule {B} {W} {W} {W} {W} {W} {R} {}\hfill}
\vskip-3pt
\ligne{\hfill\gzzz\ftt{\the\compteregle \ }  
\llrule {R} {G} {B} {W} {W} {B} {R} 
\hskip-3.5pt\rrrule {B} {W} {W} {R} {W} {W} {R} {}\hfill}
\ligne{\hfill\ftt{\RR$_s$ } \hfill}
\ligne{\hfill\gzzz\ftt{\the\compteregle \ }  
\llrule {B} {R} {R} {B} {B} {W} {W} 
\hskip-3.5pt\rrrule {W} {B} {W} {R} {W} {W} {B} {}\hfill}
\vskip-3pt
\ligne{\hfill\gzzz\ftt{\the\compteregle \ }  
\llrule {B} {R} {R} {B} {B} {W} {B} 
\hskip-3.5pt\rrrule {W} {B} {W} {R} {W} {W} {G} {}\hfill}
\vskip-3pt
\ligne{\hfill\gzzz\ftt{\the\compteregle \ }  
\llrule {G} {R} {R} {B} {B} {W} {W} 
\hskip-3.5pt\rrrule {W} {B} {W} {R} {W} {W} {B} {}\hfill}
\vskip-3pt
\ligne{\hfill\gzzz\ftt{\the\compteregle \ }  
\llrule {B} {R} {R} {B} {B} {W} {G} 
\hskip-3.5pt\rrrule {W} {B} {W} {W} {W} {W} {G} {}\hfill}
\vskip-3pt
\ligne{\hfill\gzzz\ftt{\the\compteregle \ }  
\llrule {G} {R} {R} {B} {B} {W} {B} 
\hskip-3.5pt\rrrule {W} {B} {W} {W} {W} {W} {B} {}\hfill}
\vskip-3pt
\ligne{\hfill\gzzz\ftt{\the\compteregle \ }  
\llrule {B} {R} {R} {G} {B} {W} {B} 
\hskip-3.5pt\rrrule {W} {B} {W} {W} {W} {W} {B} {}\hfill}
\ligne{\hfill\small growth\hfill}
\ligne{\hfill\ftt{ \RR$_c$, \RR$_s$ } \hfill}
\ligne{\hfill\gzzz\ftt{\the\compteregle \ }  
\llrule {G} {G} {G} {W} {W} {W} {B} 
\hskip-3.5pt\rrrule {W} {W} {W} {W} {W} {W} {G} {}\hfill}
\vskip-3pt
\ligne{\hfill\gzzz\ftt{\the\compteregle \ }  
\llrule {G} {G} {G} {B} {B} {W} {B} 
\hskip-3.5pt\rrrule {W} {B} {B} {B} {B} {B} {B} {}\hfill}
\vskip-3pt
\ligne{\hfill\gzzz\ftt{\the\compteregle \ }  
\llrule {B} {G} {W} {W} {W} {W} {W} 
\hskip-3.5pt\rrrule {W} {W} {W} {W} {W} {W} {W} {}\hfill}
\vskip-3pt
\ligne{\hfill\gzzz\ftt{\the\compteregle \ }  
\llrule {B} {B} {W} {B} {W} {W} {W} 
\hskip-3.5pt\rrrule {W} {W} {W} {W} {W} {W} {B} {}\hfill}
\vskip-3pt
\ligne{\hfill\gzzz\ftt{\the\compteregle \ }  
\llrule {B} {G} {W} {B} {W} {W} {W} 
\hskip-3.5pt\rrrule {W} {W} {W} {W} {W} {W} {B} {}\hfill}
\vskip-3pt
\ligne{\hfill\gzzz\ftt{\the\compteregle \ }  
\llrule {B} {B} {B} {W} {W} {W} {G} 
\hskip-3.5pt\rrrule {W} {W} {W} {W} {W} {W} {G} {}\hfill}
\vskip-3pt
\ligne{\hfill\gzzz\ftt{\the\compteregle \ }  
\llrule {B} {B} {B} {G} {W} {W} {W} 
\hskip-3.5pt\rrrule {W} {W} {W} {W} {W} {W} {G} {}\hfill}
\ligne{\hfill\ftt{ \RR$_i$, \RR$_d$ } \hfill}
\ligne{\hfill\gzzz\ftt{\the\compteregle \ }  
\llrule {G} {G} {G} {R} {W} {W} {W} 
\hskip-3.5pt\rrrule {W} {W} {W} {W} {W} {W} {G} {}\hfill}
\vskip-3pt
\ligne{\hfill\gzzz\ftt{\the\compteregle \ }  
\llrule {G} {G} {G} {R} {W} {B} {B} 
\hskip-3.5pt\rrrule {B} {W} {B} {B} {B} {B} {R} {}\hfill}
\vskip-3pt
\ligne{\hfill\gzzz\ftt{\the\compteregle \ }  
\llrule {B} {R} {W} {B} {W} {W} {W} 
\hskip-3.5pt\rrrule {W} {W} {W} {W} {W} {W} {B} {}\hfill}
\vskip-3pt
\ligne{\hfill\gzzz\ftt{\the\compteregle \ }  
\llrule {R} {B} {B} {R} {W} {B} {G} 
\hskip-3.5pt\rrrule {B} {W} {W} {W} {W} {W} {R} {}\hfill}
\vskip 6pt
\hrule height 0.3pt depth 0.3pt width \hsize
}
\hfill}
}
\hfill}

\ligne{\hfill
\vtop{
\ligne{\hfill
\vtop{\leftskip 0pt\parindent 0pt\hsize=\largeouille
\vskip 3pt
\ligne{\hfill\small decrementation\hfill}
\ligne{\hfill\ftt{general case } \hfill}
\ligne{\hfill\ftt{return strand } \hfill}
\ligne{\hfill\gzzz\ftt{\the\compteregle \ }  
\llrule {W} {W} {W} {W} {W} {W} {W} 
\hskip-3.5pt\rrrule {B} {W} {W} {B} {B} {R} {W} {}\hfill}
\vskip-3pt
\ligne{\hfill\gzzz\ftt{\the\compteregle \ }  
\llrule {W} {W} {W} {W} {W} {W} {B} 
\hskip-3.5pt\rrrule {B} {W} {W} {B} {B} {R} {W} {}\hfill}
\vskip-3pt
\ligne{\hfill\gzzz\ftt{\the\compteregle \ }  
\llrule {W} {W} {W} {W} {W} {W} {R} 
\hskip-3.5pt\rrrule {B} {W} {W} {B} {B} {R} {B} {}\hfill}
\vskip-3pt
\ligne{\hfill\gzzz\ftt{\the\compteregle \ }  
\llrule {B} {W} {W} {W} {W} {W} {W} 
\hskip-3.5pt\rrrule {B} {W} {W} {B} {B} {R} {W} {}\hfill}
\vskip-3pt
\ligne{\hfill\gzzz\ftt{\the\compteregle \ }  
\llrule {W} {W} {W} {B} {W} {W} {W} 
\hskip-3.5pt\rrrule {B} {W} {W} {B} {B} {R} {W} {}\hfill}
\ligne{\hfill\ftt{$\mathcal S$($n$), $n\geq0$ } \hfill}
\ligne{\hfill\gzzz\ftt{\the\compteregle \ }  
\llrule {B} {B} {R} {B} {B} {W} {W} 
\hskip-3.5pt\rrrule {W} {B} {W} {R} {W} {W} {B} {}\hfill}
\vskip-3pt
\ligne{\hfill\gzzz\ftt{\the\compteregle \ }  
\llrule {W} {R} {B} {W} {B} {W} {W} 
\hskip-3.5pt\rrrule {W} {B} {W} {B} {W} {W} {W} {}\hfill}
\vskip-3pt
\ligne{\hfill\gzzz\ftt{\the\compteregle \ }  
\llrule {W} {R} {B} {W} {B} {W} {W} 
\hskip-3.5pt\rrrule {W} {B} {W} {W} {W} {W} {W} {}\hfill}
\vskip-3pt
\ligne{\hfill\gzzz\ftt{\the\compteregle \ }  
\llrule {B} {B} {R} {B} {B} {W} {B} 
\hskip-3.5pt\rrrule {W} {B} {W} {W} {W} {W} {B} {}\hfill}
\vskip-3pt
\ligne{\hfill\gzzz\ftt{\the\compteregle \ }  
\llrule {W} {R} {B} {B} {B} {W} {W} 
\hskip-3.5pt\rrrule {W} {B} {W} {W} {W} {W} {B} {}\hfill}
\vskip-3pt
\ligne{\hfill\gzzz\ftt{\the\compteregle \ }  
\llrule {R} {B} {B} {B} {W} {B} {R} 
\hskip-3.5pt\rrrule {B} {W} {W} {W} {W} {W} {G} {}\hfill}
\vskip-3pt
\ligne{\hfill\gzzz\ftt{\the\compteregle \ }  
\llrule {W} {R} {R} {G} {B} {W} {W} 
\hskip-3.5pt\rrrule {W} {B} {W} {W} {W} {W} {W} {}\hfill}
\vskip-3pt
\ligne{\hfill\gzzz\ftt{\the\compteregle \ }  
\llrule {B} {R} {G} {B} {B} {W} {B} 
\hskip-3.5pt\rrrule {W} {B} {W} {W} {W} {W} {B} {}\hfill}
\vskip-3pt
\ligne{\hfill\gzzz\ftt{\the\compteregle \ }  
\llrule {G} {B} {B} {R} {W} {B} {R} 
\hskip-3.5pt\rrrule {B} {W} {W} {W} {W} {W} {G} {}\hfill}
\vskip-3pt
\ligne{\hfill\gzzz\ftt{\the\compteregle \ }  
\llrule {R} {B} {B} {G} {W} {B} {R} 
\hskip-3.5pt\rrrule {B} {W} {W} {W} {W} {W} {R} {}\hfill}
\vskip-3pt
\ligne{\hfill\gzzz\ftt{\the\compteregle \ }  
\llrule {B} {G} {R} {B} {B} {W} {B} 
\hskip-3.5pt\rrrule {W} {B} {W} {W} {W} {W} {R} {}\hfill}
\vskip-3pt
\ligne{\hfill\gzzz\ftt{\the\compteregle \ }  
\llrule {G} {R} {B} {R} {W} {B} {R} 
\hskip-3.5pt\rrrule {B} {W} {B} {B} {B} {B} {R} {}\hfill}
\vskip-3pt
\ligne{\hfill\gzzz\ftt{\the\compteregle \ }  
\llrule {B} {R} {R} {R} {B} {W} {B} 
\hskip-3.5pt\rrrule {W} {B} {W} {W} {W} {W} {R} {}\hfill}
\vskip-3pt
\ligne{\hfill\gzzz\ftt{\the\compteregle \ }  
\llrule {R} {G} {R} {B} {B} {W} {B} 
\hskip-3.5pt\rrrule {W} {B} {W} {W} {W} {W} {B} {}\hfill}
\vskip-3pt
\ligne{\hfill\gzzz\ftt{\the\compteregle \ }  
\llrule {R} {R} {B} {R} {W} {B} {R} 
\hskip-3.5pt\rrrule {B} {W} {W} {W} {W} {W} {R} {}\hfill}
\vskip-3pt
\ligne{\hfill\gzzz\ftt{\the\compteregle \ }  
\llrule {B} {R} {R} {B} {B} {W} {R} 
\hskip-3.5pt\rrrule {W} {B} {W} {W} {W} {W} {B} {}\hfill}
\vskip-3pt
\ligne{\hfill\gzzz\ftt{\the\compteregle \ }  
\llrule {R} {R} {R} {B} {B} {W} {B} 
\hskip-3.5pt\rrrule {W} {B} {W} {W} {W} {W} {B} {}\hfill}
\vskip-3pt
\ligne{\hfill\gzzz\ftt{\the\compteregle \ }  
\llrule {R} {R} {W} {R} {W} {B} {R} 
\hskip-3.5pt\rrrule {B} {W} {W} {W} {W} {W} {R} {}\hfill}
\vskip-3pt
\ligne{\hfill\gzzz\ftt{\the\compteregle \ }  
\llrule {B} {R} {R} {R} {B} {W} {W} 
\hskip-3.5pt\rrrule {W} {B} {W} {R} {W} {W} {R} {}\hfill}
\vskip-3pt
\ligne{\hfill\gzzz\ftt{\the\compteregle \ }  
\llrule {R} {R} {W} {W} {W} {B} {R} 
\hskip-3.5pt\rrrule {B} {W} {W} {R} {W} {W} {R} {}\hfill}
\vskip-3pt
\ligne{\hfill\gzzz\ftt{\the\compteregle \ }  
\llrule {R} {R} {R} {B} {B} {W} {W} 
\hskip-3.5pt\rrrule {W} {B} {W} {R} {W} {W} {B} {}\hfill}
\vskip-3pt
\ligne{\hfill\gzzz\ftt{\the\compteregle \ }  
\llrule {B} {R} {R} {B} {B} {W} {R} 
\hskip-3.5pt\rrrule {W} {B} {W} {R} {W} {W} {B} {}\hfill}
\vskip-3pt
\ligne{\hfill\gzzz\ftt{\the\compteregle \ }  
\llrule {W} {R} {W} {R} {W} {W} {W} 
\hskip-3.5pt\rrrule {B} {W} {W} {B} {R} {W} {W} {}\hfill}
\ligne{\hfill\ftt{$c = 2$ } \hfill}
\ligne{\hfill\gzzz\ftt{\the\compteregle \ }  
\llrule {W} {W} {W} {R} {W} {W} {W} 
\hskip-3.5pt\rrrule {B} {W} {R} {B} {B} {W} {W} {}\hfill}
\vskip-3pt
\ligne{\hfill\gzzz\ftt{\the\compteregle \ }  
\llrule {W} {W} {W} {B} {W} {W} {W} 
\hskip-3.5pt\rrrule {B} {B} {R} {W} {B} {B} {W} {}\hfill}
\ligne{\hfill\ftt{$c = 1$ } \hfill}
\ligne{\hfill\gzzz\ftt{\the\compteregle \ }  
\llrule {W} {R} {B} {B} {B} {W} {W} 
\hskip-3.5pt\rrrule {W} {B} {W} {B} {W} {W} {B} {}\hfill}
\vskip-3pt
\ligne{\hfill\gzzz\ftt{\the\compteregle \ }  
\llrule {R} {B} {B} {W} {W} {B} {G} 
\hskip-3.5pt\rrrule {B} {W} {W} {R} {W} {W} {R} {}\hfill}
\vskip-3pt
\ligne{\hfill\gzzz\ftt{\the\compteregle \ }  
\llrule {R} {R} {B} {W} {W} {B} {R} 
\hskip-3.5pt\rrrule {B} {W} {W} {R} {W} {W} {R} {}\hfill}
\ligne{\hfill\ftt{$c = 0$ } \hfill}
\ligne{\hfill\gzzz\ftt{\the\compteregle \ }  
\llrule {R} {B} {B} {B} {W} {B} {R} 
\hskip-3.5pt\rrrule {B} {W} {W} {R} {W} {W} {G} {}\hfill}
\vskip-3pt
\ligne{\hfill\gzzz\ftt{\the\compteregle \ }  
\llrule {B} {R} {G} {B} {B} {W} {W} 
\hskip-3.5pt\rrrule {W} {B} {W} {B} {W} {W} {B} {}\hfill}
\vskip-3pt
\ligne{\hfill\gzzz\ftt{\the\compteregle \ }  
\llrule {G} {B} {B} {W} {W} {B} {R} 
\hskip-3.5pt\rrrule {B} {W} {W} {R} {W} {W} {G} {}\hfill}
\vskip-3pt
\ligne{\hfill\gzzz\ftt{\the\compteregle \ }  
\llrule {W} {W} {W} {G} {W} {W} {W} 
\hskip-3.5pt\rrrule {B} {W} {W} {B} {R} {W} {R} {}\hfill}
\vskip-3pt
\ligne{\hfill\gzzz\ftt{\the\compteregle \ }  
\llrule {B} {G} {R} {B} {B} {W} {W} 
\hskip-3.5pt\rrrule {W} {B} {W} {R} {W} {W} {B} {}\hfill}
\vskip-3pt
\ligne{\hfill\gzzz\ftt{\the\compteregle \ }  
\llrule {G} {B} {B} {R} {B} {B} {R} 
\hskip-3.5pt\rrrule {B} {W} {W} {R} {B} {B} {R} {}\hfill}
\vskip-3pt
\ligne{\hfill\gzzz\ftt{\the\compteregle \ }  
\llrule {R} {W} {W} {G} {W} {W} {W} 
\hskip-3.5pt\rrrule {B} {W} {W} {B} {R} {W} {W} {}\hfill}
\vskip-3pt
\ligne{\hfill\gzzz\ftt{\the\compteregle \ }  
\llrule {W} {R} {W} {W} {W} {W} {B} 
\hskip-3.5pt\rrrule {B} {R} {W} {B} {B} {W} {W} {}\hfill}
\vskip-3pt
\ligne{\hfill\gzzz\ftt{\the\compteregle \ }  
\llrule {W} {W} {W} {W} {W} {R} {W} 
\hskip-3.5pt\rrrule {W} {R} {R} {R} {W} {R} {B} {}\hfill}
\vskip-3pt
\ligne{\hfill\gzzz\ftt{\the\compteregle \ }  
\llrule {W} {W} {W} {R} {W} {W} {B} 
\hskip-3.5pt\rrrule {B} {W} {W} {B} {R} {W} {W} {}\hfill}
\vskip-3pt
\ligne{\hfill\gzzz\ftt{\the\compteregle \ }  
\llrule {B} {W} {W} {W} {W} {W} {W} 
\hskip-3.5pt\rrrule {W} {R} {R} {R} {W} {R} {W} {}\hfill}
\vskip 6pt
\hrule height 0.3pt depth 0.3pt width \hsize
}
\hfill
\vtop{\leftskip 0pt\parindent 0pt\hsize=\largeouille
\vskip 3pt
\ligne{\hfill\small incrementation\hfill}
\ligne{\hfill\ftt{general case } \hfill}
\ligne{\hfill\gzzz\ftt{\the\compteregle \ }  
\llrule {W} {W} {W} {B} {W} {W} {W} 
\hskip-3.5pt\rrrule {W} {R} {R} {R} {W} {R} {W} {}\hfill}
\vskip-3pt
\ligne{\hfill\gzzz\ftt{\the\compteregle \ }  
\llrule {W} {B} {R} {W} {B} {W} {W} 
\hskip-3.5pt\rrrule {W} {B} {W} {B} {W} {W} {W} {}\hfill}
\vskip-3pt
\ligne{\hfill\gzzz\ftt{\the\compteregle \ }  
\llrule {R} {B} {B} {R} {W} {B} {R} 
\hskip-3.5pt\rrrule {B} {W} {W} {W} {W} {W} {R} {}\hfill}
\vskip-3pt
\ligne{\hfill\gzzz\ftt{\the\compteregle \ }  
\llrule {B} {R} {B} {B} {B} {W} {W} 
\hskip-3.5pt\rrrule {W} {B} {W} {R} {W} {W} {B} {}\hfill}
\vskip-3pt
\ligne{\hfill\gzzz\ftt{\the\compteregle \ }  
\llrule {W} {B} {R} {W} {B} {W} {W} 
\hskip-3.5pt\rrrule {W} {B} {W} {W} {W} {W} {W} {}\hfill}
\vskip-3pt
\ligne{\hfill\gzzz\ftt{\the\compteregle \ }  
\llrule {B} {R} {B} {B} {B} {W} {B} 
\hskip-3.5pt\rrrule {W} {B} {W} {W} {W} {W} {B} {}\hfill}
\vskip-3pt
\ligne{\hfill\gzzz\ftt{\the\compteregle \ }  
\llrule {R} {W} {B} {R} {W} {B} {B} 
\hskip-3.5pt\rrrule {B} {W} {W} {W} {W} {W} {R} {}\hfill}
\vskip-3pt
\ligne{\hfill\gzzz\ftt{\the\compteregle \ }  
\llrule {W} {B} {R} {B} {B} {W} {W} 
\hskip-3.5pt\rrrule {W} {B} {W} {W} {W} {W} {R} {}\hfill}
\vskip-3pt
\ligne{\hfill\gzzz\ftt{\the\compteregle \ }  
\llrule {R} {W} {B} {R} {W} {B} {G} 
\hskip-3.5pt\rrrule {B} {W} {W} {W} {W} {W} {R} {}\hfill}
\vskip-3pt
\ligne{\hfill\gzzz\ftt{\the\compteregle \ }  
\llrule {R} {R} {R} {B} {B} {W} {W} 
\hskip-3.5pt\rrrule {W} {B} {W} {W} {W} {W} {R} {}\hfill}
\vskip-3pt
\ligne{\hfill\gzzz\ftt{\the\compteregle \ }  
\llrule {R} {R} {B} {R} {W} {B} {G} 
\hskip-3.5pt\rrrule {B} {W} {W} {W} {W} {W} {R} {}\hfill}
\vskip-3pt
\ligne{\hfill\gzzz\ftt{\the\compteregle \ }  
\llrule {R} {B} {R} {R} {W} {B} {R} 
\hskip-3.5pt\rrrule {B} {W} {W} {W} {W} {W} {R} {}\hfill}
\vskip-3pt
\ligne{\hfill\gzzz\ftt{\the\compteregle \ }  
\llrule {B} {G} {R} {B} {B} {W} {R} 
\hskip-3.5pt\rrrule {W} {B} {W} {W} {W} {W} {W} {}\hfill}
\vskip-3pt
\ligne{\hfill\gzzz\ftt{\the\compteregle \ }  
\llrule {R} {R} {R} {R} {B} {W} {W} 
\hskip-3.5pt\rrrule {W} {B} {W} {W} {W} {W} {W} {}\hfill}
\vskip-3pt
\ligne{\hfill\gzzz\ftt{\the\compteregle \ }  
\llrule {R} {R} {R} {W} {B} {W} {R} 
\hskip-3.5pt\rrrule {W} {B} {W} {W} {W} {W} {W} {}\hfill}
\vskip-3pt
\ligne{\hfill\gzzz\ftt{\the\compteregle \ }  
\llrule {W} {G} {R} {B} {B} {W} {R} 
\hskip-3.5pt\rrrule {W} {B} {W} {W} {W} {W} {W} {}\hfill}
\vskip-3pt
\ligne{\hfill\gzzz\ftt{\the\compteregle \ }  
\llrule {G} {W} {B} {R} {W} {B} {R} 
\hskip-3.5pt\rrrule {B} {W} {B} {B} {B} {B} {R} {}\hfill}
\vskip-3pt
\ligne{\hfill\gzzz\ftt{\the\compteregle \ }  
\llrule {R} {R} {B} {W} {W} {B} {R} 
\hskip-3.5pt\rrrule {B} {W} {W} {B} {W} {W} {R} {}\hfill}
\vskip-3pt
\ligne{\hfill\gzzz\ftt{\the\compteregle \ }  
\llrule {R} {B} {R} {W} {W} {B} {R} 
\hskip-3.5pt\rrrule {B} {W} {W} {R} {W} {W} {R} {}\hfill}
\ligne{\hfill\ftt{$c = 2$ } \hfill}
\ligne{\hfill\gzzz\ftt{\the\compteregle \ }  
\llrule {W} {W} {W} {W} {W} {W} {R} 
\hskip-3.5pt\rrrule {R} {W} {W} {R} {R} {W} {B} {}\hfill}
\vskip-3pt
\ligne{\hfill\gzzz\ftt{\the\compteregle \ }  
\llrule {R} {R} {R} {R} {B} {W} {W} 
\hskip-3.5pt\rrrule {W} {B} {W} {B} {W} {W} {W} {}\hfill}
\vskip-3pt
\ligne{\hfill\gzzz\ftt{\the\compteregle \ }  
\llrule {W} {W} {W} {R} {B} {W} {W} 
\hskip-3.5pt\rrrule {B} {W} {W} {B} {R} {W} {W} {}\hfill}
\ligne{\hfill\ftt{$c = 1$ } \hfill}
\ligne{\hfill\gzzz\ftt{\the\compteregle \ }  
\llrule {W} {B} {R} {B} {B} {W} {W} 
\hskip-3.5pt\rrrule {W} {B} {W} {B} {W} {W} {R} {}\hfill}
\vskip-3pt
\ligne{\hfill\gzzz\ftt{\the\compteregle \ }  
\llrule {R} {R} {R} {B} {B} {W} {W} 
\hskip-3.5pt\rrrule {W} {B} {W} {B} {W} {W} {W} {}\hfill}
\vskip-3pt
\ligne{\hfill\gzzz\ftt{\the\compteregle \ }  
\llrule {R} {R} {B} {W} {W} {B} {G} 
\hskip-3.5pt\rrrule {B} {W} {W} {B} {W} {W} {R} {}\hfill}
\vskip-3pt
\ligne{\hfill\gzzz\ftt{\the\compteregle \ }  
\llrule {R} {W} {B} {W} {W} {B} {G} 
\hskip-3.5pt\rrrule {B} {W} {W} {B} {W} {W} {R} {}\hfill}
\vskip-3pt
\ligne{\hfill\gzzz\ftt{\the\compteregle \ }  
\llrule {W} {G} {R} {B} {B} {W} {W} 
\hskip-3.5pt\rrrule {W} {B} {W} {W} {W} {W} {W} {}\hfill}
\ligne{\hfill\ftt{$c = 0$ } \hfill}
\ligne{\hfill\gzzz\ftt{\the\compteregle \ }  
\llrule {W} {B} {W} {W} {W} {W} {B} 
\hskip-3.5pt\rrrule {R} {W} {W} {R} {R} {W} {B} {}\hfill}
\vskip-3pt
\ligne{\hfill\gzzz\ftt{\the\compteregle \ }  
\llrule {R} {B} {B} {B} {W} {B} {R} 
\hskip-3.5pt\rrrule {B} {W} {W} {B} {W} {W} {G} {}\hfill}
\vskip-3pt
\ligne{\hfill\gzzz\ftt{\the\compteregle \ }  
\llrule {W} {B} {W} {G} {W} {W} {W} 
\hskip-3.5pt\rrrule {B} {R} {W} {B} {R} {W} {W} {}\hfill}
\vskip-3pt
\ligne{\hfill\gzzz\ftt{\the\compteregle \ }  
\llrule {B} {G} {R} {B} {B} {W} {B} 
\hskip-3.5pt\rrrule {W} {B} {W} {B} {W} {W} {W} {}\hfill}
\vskip-3pt
\ligne{\hfill\gzzz\ftt{\the\compteregle \ }  
\llrule {G} {B} {B} {W} {W} {B} {R} 
\hskip-3.5pt\rrrule {B} {W} {W} {B} {W} {W} {G} {}\hfill}
\vskip-3pt
\ligne{\hfill\gzzz\ftt{\the\compteregle \ }  
\llrule {B} {R} {G} {B} {B} {W} {W} 
\hskip-3.5pt\rrrule {W} {B} {W} {R} {W} {W} {B} {}\hfill}
\vskip-3pt
\ligne{\hfill\gzzz\ftt{\the\compteregle \ }  
\llrule {W} {W} {W} {G} {W} {W} {W} 
\hskip-3.5pt\rrrule {B} {R} {W} {B} {R} {W} {W} {}\hfill}
\vskip-3pt
\ligne{\hfill\gzzz\ftt{\the\compteregle \ }  
\llrule {W} {G} {R} {B} {B} {W} {W} 
\hskip-3.5pt\rrrule {W} {B} {W} {B} {W} {W} {W} {}\hfill}
\vskip-3pt
\ligne{\hfill\gzzz\ftt{\the\compteregle \ }  
\llrule {G} {W} {B} {W} {B} {B} {R} 
\hskip-3.5pt\rrrule {B} {W} {W} {B} {B} {B} {R} {}\hfill}
\vskip 6pt
\hrule height 0.3pt depth 0.3pt width \hsize
\vskip 3pt
\ligne{\hfill\small halting\hfill}
\ligne{\hfill\gzzz\ftt{\the\compteregle \ }  
\llrule {R} {G} {B} {R} {W} {B} {R} 
\hskip-3.5pt\rrrule {B} {W} {W} {W} {W} {W} {R} {}\hfill}
\vskip-3pt
\ligne{\hfill\gzzz\ftt{\the\compteregle \ }  
\llrule {B} {R} {R} {G} {B} {B} {W} 
\hskip-3.5pt\rrrule {W} {B} {B} {R} {B} {B} {B} {}\hfill}
\vskip-3pt
\ligne{\hfill\gzzz\ftt{\the\compteregle \ }  
\llrule {B} {R} {R} {B} {B} {B} {W} 
\hskip-3.5pt\rrrule {W} {B} {B} {R} {B} {B} {B} {}\hfill}
\vskip-3pt
\ligne{\hfill\gzzz\ftt{\the\compteregle \ }  
\llrule {B} {R} {R} {G} {B} {W} {B} 
\hskip-3.5pt\rrrule {W} {B} {B} {B} {W} {B} {B} {}\hfill}
\vskip-3pt
\ligne{\hfill\gzzz\ftt{\the\compteregle \ }  
\llrule {B} {R} {R} {B} {B} {W} {B} 
\hskip-3.5pt\rrrule {W} {B} {B} {B} {W} {B} {B} {}\hfill}
\vskip-3pt
\ligne{\hfill\gzzz\ftt{\the\compteregle \ }  
\llrule {R} {B} {G} {R} {W} {B} {R} 
\hskip-3.5pt\rrrule {B} {W} {W} {W} {W} {W} {R} {}\hfill}
\vskip-3pt
\ligne{\hfill\gzzz\ftt{\the\compteregle \ }  
\llrule {B} {R} {R} {G} {B} {B} {B} 
\hskip-3.5pt\rrrule {B} {B} {B} {B} {B} {B} {B} {}\hfill}
\vskip-3pt
\ligne{\hfill\gzzz\ftt{\the\compteregle \ }  
\llrule {B} {R} {R} {B} {B} {B} {B} 
\hskip-3.5pt\rrrule {B} {B} {B} {B} {B} {B} {B} {}\hfill}
\vskip-3pt
\ligne{\hfill\gzzz\ftt{\the\compteregle \ }  
\llrule {B} {R} {R} {G} {B} {W} {G} 
\hskip-3.5pt\rrrule {W} {B} {W} {W} {W} {W} {W} {}\hfill}
\vskip-3pt
\ligne{\hfill\gzzz\ftt{\the\compteregle \ }  
\llrule {B} {R} {R} {W} {B} {W} {B} 
\hskip-3.5pt\rrrule {W} {B} {B} {B} {W} {B} {B} {}\hfill}
\vskip-3pt
\ligne{\hfill\gzzz\ftt{\the\compteregle \ }  
\llrule {W} {R} {R} {B} {B} {W} {B} 
\hskip-3.5pt\rrrule {W} {B} {W} {W} {W} {W} {W} {}\hfill}
}
\hfill
\vtop{\leftskip 0pt\parindent 0pt\hsize=\largeouille
\ligne{\hfill\gzzz\ftt{\the\compteregle \ }  
\llrule {B} {R} {R} {G} {B} {W} {W} 
\hskip-3.5pt\rrrule {W} {B} {W} {W} {W} {W} {W} {}\hfill}
\vskip-3pt
\ligne{\hfill\gzzz\ftt{\the\compteregle \ }  
\llrule {R} {B} {W} {R} {W} {B} {G} 
\hskip-3.5pt\rrrule {B} {W} {W} {W} {W} {W} {W} {}\hfill}
\vskip-3pt
\ligne{\hfill\gzzz\ftt{\the\compteregle \ }  
\llrule {G} {G} {G} {B} {B} {W} {W} 
\hskip-3.5pt\rrrule {W} {B} {B} {B} {B} {B} {W} {}\hfill}
\vskip-3pt
\ligne{\hfill\gzzz\ftt{\the\compteregle \ }  
\llrule {R} {B} {B} {R} {W} {B} {W} 
\hskip-3.5pt\rrrule {B} {W} {W} {W} {W} {W} {R} {}\hfill}
\vskip-3pt
\ligne{\hfill\gzzz\ftt{\the\compteregle \ }  
\llrule {B} {W} {R} {B} {B} {W} {B} 
\hskip-3.5pt\rrrule {W} {B} {W} {W} {W} {W} {W} {}\hfill}
\vskip-3pt
\ligne{\hfill\gzzz\ftt{\the\compteregle \ }  
\llrule {W} {B} {W} {R} {W} {B} {R} 
\hskip-3.5pt\rrrule {B} {W} {W} {W} {W} {W} {W} {}\hfill}
\vskip-3pt
\ligne{\hfill\gzzz\ftt{\the\compteregle \ }  
\llrule {R} {B} {W} {W} {W} {B} {G} 
\hskip-3.5pt\rrrule {B} {W} {W} {W} {W} {W} {W} {}\hfill}
\vskip-3pt
\ligne{\hfill\gzzz\ftt{\the\compteregle \ }  
\llrule {B} {R} {R} {W} {B} {W} {B} 
\hskip-3.5pt\rrrule {W} {B} {W} {W} {W} {W} {B} {}\hfill}
\vskip-3pt
\ligne{\hfill\gzzz\ftt{\the\compteregle \ }  
\llrule {W} {R} {W} {W} {B} {W} {B} 
\hskip-3.5pt\rrrule {W} {B} {W} {W} {W} {W} {W} {}\hfill}
\ligne{\hfill\ftt{last \sgg{}'s } \hfill}
\ligne{\hfill\gzzz\ftt{\the\compteregle \ }  
\llrule {G} {G} {G} {W} {W} {W} {W} 
\hskip-3.5pt\rrrule {W} {W} {W} {W} {W} {W} {W} {}\hfill}
\vskip-3pt
\ligne{\hfill\gzzz\ftt{\the\compteregle \ }  
\llrule {W} {W} {R} {B} {B} {W} {B} 
\hskip-3.5pt\rrrule {W} {B} {W} {W} {W} {W} {W} {}\hfill}
\vskip-3pt
\ligne{\hfill\gzzz\ftt{\the\compteregle \ }  
\llrule {W} {W} {W} {R} {W} {B} {W} 
\hskip-3.5pt\rrrule {B} {W} {W} {W} {W} {W} {W} {}\hfill}
\vskip-3pt
\ligne{\hfill\gzzz\ftt{\the\compteregle \ }  
\llrule {R} {W} {W} {R} {W} {B} {R} 
\hskip-3.5pt\rrrule {B} {W} {W} {W} {W} {W} {W} {}\hfill}
\vskip-3pt
\ligne{\hfill\gzzz\ftt{\the\compteregle \ }  
\llrule {B} {W} {R} {G} {B} {W} {W} 
\hskip-3.5pt\rrrule {W} {B} {W} {W} {W} {W} {W} {}\hfill}
\vskip-3pt
\ligne{\hfill\gzzz\ftt{\the\compteregle \ }  
\llrule {W} {B} {W} {W} {W} {B} {G} 
\hskip-3.5pt\rrrule {B} {W} {W} {W} {W} {W} {W} {}\hfill}
\vskip-3pt
\ligne{\hfill\gzzz\ftt{\the\compteregle \ }  
\llrule {G} {G} {W} {W} {W} {B} {B} 
\hskip-3.5pt\rrrule {B} {W} {B} {B} {B} {B} {W} {}\hfill}
\vskip-3pt
\ligne{\hfill\gzzz\ftt{\the\compteregle \ }  
\llrule {G} {W} {G} {R} {W} {B} {B} 
\hskip-3.5pt\rrrule {B} {W} {B} {B} {B} {B} {W} {}\hfill}
\vskip-3pt
\ligne{\hfill\gzzz\ftt{\the\compteregle \ }  
\llrule {W} {R} {W} {W} {B} {W} {W} 
\hskip-3.5pt\rrrule {W} {B} {W} {W} {W} {W} {W} {}\hfill}
\vskip-3pt
\ligne{\hfill\gzzz\ftt{\the\compteregle \ }  
\llrule {W} {G} {G} {B} {B} {W} {W} 
\hskip-3.5pt\rrrule {W} {B} {B} {B} {B} {B} {W} {}\hfill}
\vskip-3pt
\ligne{\hfill\gzzz\ftt{\the\compteregle \ }  
\llrule {W} {W} {W} {W} {B} {W} {B} 
\hskip-3.5pt\rrrule {W} {B} {W} {W} {W} {W} {W} {}\hfill}
\vskip-3pt
\ligne{\hfill\gzzz\ftt{\the\compteregle \ }  
\llrule {W} {W} {W} {R} {W} {B} {R} 
\hskip-3.5pt\rrrule {B} {W} {W} {W} {W} {W} {W} {}\hfill}
\vskip-3pt
\ligne{\hfill\gzzz\ftt{\the\compteregle \ }  
\llrule {W} {W} {R} {B} {B} {W} {W} 
\hskip-3.5pt\rrrule {W} {B} {W} {W} {W} {W} {W} {}\hfill}
\vskip-3pt
\ligne{\hfill\gzzz\ftt{\the\compteregle \ }  
\llrule {W} {W} {W} {W} {W} {B} {W} 
\hskip-3.5pt\rrrule {B} {W} {W} {W} {W} {W} {W} {}\hfill}
\vskip-3pt
\ligne{\hfill\gzzz\ftt{\the\compteregle \ }  
\llrule {R} {W} {W} {W} {W} {B} {W} 
\hskip-3.5pt\rrrule {B} {W} {W} {W} {W} {W} {W} {}\hfill}
\vskip-3pt
\ligne{\hfill\gzzz\ftt{\the\compteregle \ }  
\llrule {B} {W} {W} {G} {B} {W} {W} 
\hskip-3.5pt\rrrule {W} {B} {W} {W} {W} {W} {W} {}\hfill}
\vskip-3pt
\ligne{\hfill\gzzz\ftt{\the\compteregle \ }  
\llrule {W} {W} {B} {R} {W} {B} {G} 
\hskip-3.5pt\rrrule {B} {W} {W} {W} {W} {W} {W} {}\hfill}
\vskip-3pt
\ligne{\hfill\gzzz\ftt{\the\compteregle \ }  
\llrule {G} {G} {B} {W} {W} {W} {W} 
\hskip-3.5pt\rrrule {W} {W} {W} {W} {W} {W} {W} {}\hfill}
\vskip-3pt
\ligne{\hfill\gzzz\ftt{\the\compteregle \ }  
\llrule {G} {B} {G} {W} {W} {W} {W} 
\hskip-3.5pt\rrrule {W} {W} {W} {W} {W} {W} {W} {}\hfill}
\vskip-3pt
\ligne{\hfill\gzzz\ftt{\the\compteregle \ }  
\llrule {W} {W} {W} {B} {B} {W} {W} 
\hskip-3.5pt\rrrule {W} {B} {B} {B} {B} {B} {W} {}\hfill}
\vskip-3pt
\ligne{\hfill\gzzz\ftt{\the\compteregle \ }  
\llrule {B} {G} {G} {W} {W} {W} {W} 
\hskip-3.5pt\rrrule {W} {W} {W} {W} {W} {W} {W} {}\hfill}
\vskip-3pt
\ligne{\hfill\gzzz\ftt{\the\compteregle \ }  
\llrule {W} {W} {W} {G} {B} {W} {W} 
\hskip-3.5pt\rrrule {W} {B} {W} {W} {W} {W} {W} {}\hfill}
\vskip-3pt
\ligne{\hfill\gzzz\ftt{\the\compteregle \ }  
\llrule {G} {W} {W} {B} {B} {W} {W} 
\hskip-3.5pt\rrrule {W} {B} {B} {B} {B} {B} {W} {}\hfill}
\vskip-3pt
\ligne{\hfill\gzzz\ftt{\the\compteregle \ }  
\llrule {W} {G} {W} {W} {W} {B} {B} 
\hskip-3.5pt\rrrule {B} {W} {B} {B} {B} {B} {W} {}\hfill}
\vskip-3pt
\ligne{\hfill\gzzz\ftt{\the\compteregle \ }  
\llrule {W} {W} {G} {W} {W} {B} {B} 
\hskip-3.5pt\rrrule {B} {W} {B} {B} {B} {B} {W} {}\hfill}
\vskip-3pt
\ligne{\hfill\gzzz\ftt{\the\compteregle \ }  
\llrule {W} {W} {W} {W} {B} {W} {W} 
\hskip-3.5pt\rrrule {W} {B} {B} {B} {B} {B} {W} {}\hfill}
\vskip-3pt
\ligne{\hfill\gzzz\ftt{\the\compteregle \ }  
\llrule {W} {W} {W} {W} {W} {B} {B} 
\hskip-3.5pt\rrrule {B} {W} {B} {B} {B} {B} {W} {}\hfill}
\vskip-3pt
\ligne{\hfill\gzzz\ftt{\the\compteregle \ }  
\llrule {G} {B} {B} {W} {W} {W} {W} 
\hskip-3.5pt\rrrule {W} {W} {W} {W} {W} {W} {W} {}\hfill}
\vskip-3pt
\ligne{\hfill\gzzz\ftt{\the\compteregle \ }  
\llrule {B} {W} {G} {W} {W} {W} {W} 
\hskip-3.5pt\rrrule {W} {W} {W} {W} {W} {W} {W} {}\hfill}
\vskip-3pt
}
\hfill }
}
\hfill}

\end{document}